\documentclass[%
reprint,
 amsmath,amssymb,
 aps,
]{revtex4-2}

\usepackage{graphicx}
\usepackage{dcolumn}
\usepackage{bm}
\usepackage{hyperref}
\usepackage{color} 

\begin{document}


\title{Optically tunable photoluminescence and upconversion lasing on a chip}

\author{Christiaan J. Bekker}
 \email{c.bekker@uq.edu.au}
\author{Christopher G. Baker}%
\author{Warwick P. Bowen}%
\affiliation{%
School of Mathematics and Physics\\ The University of Queensland, Brisbane, Australia 4072}%

\date{\today}

\begin{abstract}
The ability to tune the wavelength of light emission on a silicon chip is important for scalable photonic networks, distributed photonic sensor networks and next generation computer architectures. Here we demonstrate light emission in a chip-scale optomechanical device, with wide tunablity provided by a combination of radiation pressure and photothermal effects. To achieve this, we develop an optically active double-disk optomechanical system through implantation of erbium ions. We observe frequency tuning of photoluminescence in the telecommunications band with a wavelength range of 520 pm, green upconversion lasing with a threshold of $340\pm 70 \; \mu$W, and optomechanical self-pulsing caused by the interplay of radiation pressure and thermal effects. These results provide a path towards widely-tunable micron-scale lasers for photonic networks.
\end{abstract}
\maketitle

\section{Introduction}

Active microcavities are a promising technology for the development of on-chip light sources for integrated photonic networks.
The addition of a gain medium to a microcavity allows redistribution of power between optical modes and the ability to generate lasing in the cavity through population inversion. The strong optical confinement and small mode volumes of whispering gallery mode (WGM) systems in particular are useful in applications that require large photonic networks with multiplexing and many components~\cite{atabaki_integrating_2018,elshaari_-chip_2017}. WGM cavities have been demonstrated as on-chip laser light sources~\cite{fujita_microgear_2002} and stabilisation elements for external laser pumps~\cite{yang_fiber-coupled_2003}. A variety of active WGM geometries exist, with gain media introduced through active sol-gel ~\cite{yang_fiber-coupled_2003,lu_-chip_2009,mehrabani_blue_2013, park_titanium_2014, yang_tunable_2017} or silicon-rich oxide~\cite{verbert_efficient_2005} coatings, rare-earth ion implantation~\cite{polman_ultralow-threshold_2004,kalkman_erbium-implanted_2006,kippenberg_demonstration_2006,min_erbium-implanted_2004} or doping~\cite{sandoghdar_very_1996,von_klitzing_very_2000,zhu_all-optical_2018}, embedded quantum wells~\cite{fujita_microgear_2002,mccall_whisperinggallery_1998, princepe_self-sustained_2018} and fluorescent dyes~\cite{lin_characteristics_1986,kuwata-gonokami_polymer_1995,knight_core-resonance_1992,siegle_split-disk_2017}. Individually controllable \textit{in-situ} frequency tuning has been reported in many active photonic systems, including in bottle resonators~\cite{yang_tunable_2017,zhu_all-optical_2018,ward_glass--glass_2016}, semiconductor quantum well cavities~\cite{zhang_simple_2013,perahia_electrostatically_2010}, optical cavity- or strain-coupled quantum dots~\cite{weis_surface_2016,sun_strain_2013,moczala-dusanowska_strain-tunable_2019,kremer_strain-tunable_2014,alegre_optomechanical_2010} and Bragg grating-based erbium-doped optical fiber cavities~\cite{paterno_highly_2007,ball_continuously_1992,cheng_single-longitudinal-mode_2008}. However it has not yet been achieved for on-chip active WGM systems, which either have a fixed emission frequency set by the device geometry (See e.g.~\cite{yang_fiber-coupled_2003,kippenberg_demonstration_2006,fujita_microgear_2002}), are dynamically swept through mechanical self-oscillation~\cite{princepe_self-sustained_2018} or involve chip-scale tuning which cannot be applied to individual devices~\cite{siegle_split-disk_2017}.In the absence of individually controllable frequency tuning, imperfections arising from device fabrication, particularly significant for reflown microtoroids~\cite{baker_high_2016}, make it difficult to interface multiple devices in a  network~\cite{bekker_free_2018}, precluding applications such as large-scale integrated photonic circuits~\cite{minzioni_roadmap_2019,thomson_roadmap_2016}.

In this work, we demonstrate an active double-disk optomechanical system, using implanted erbium ions as the gain medium. The very large optomechanical interaction between the disks provides a degree of control to tune the photoluminescence from the cavity. Combined with the high mechanical compliance of the double-disk geometry, this allows an optically driven tuning range of nearly 7\% of the device free spectral range (FSR), which is demonstrated in the telecommunications band. Furthermore, we demonstrate green ($\sim 550$ nm) upconversion lasing with a threshold of $340\pm 70 \; \mu$W in our double-disk geometry, which arises as a natural characteristic of the implanted erbium~\cite{luo_compact_2016}. The lasing frequency is tunable by the same mechanism as the telecom photoluminescence. Our results provide a pathway towards widely tunable lasers for photonic circuits, with applications such as distributed and precision photonic sensors~\cite{iqbal_label-free_2010,washburn_multiplexed_2016,stern_battery-operated_2018,delhaye_optical_2007}, and photonic buses in next generation computer architectures~\cite{atabaki_integrating_2018,kuramochi_large-scale_2014}.

\begin{figure*}[t]
\centering
\includegraphics[width=0.9\textwidth]{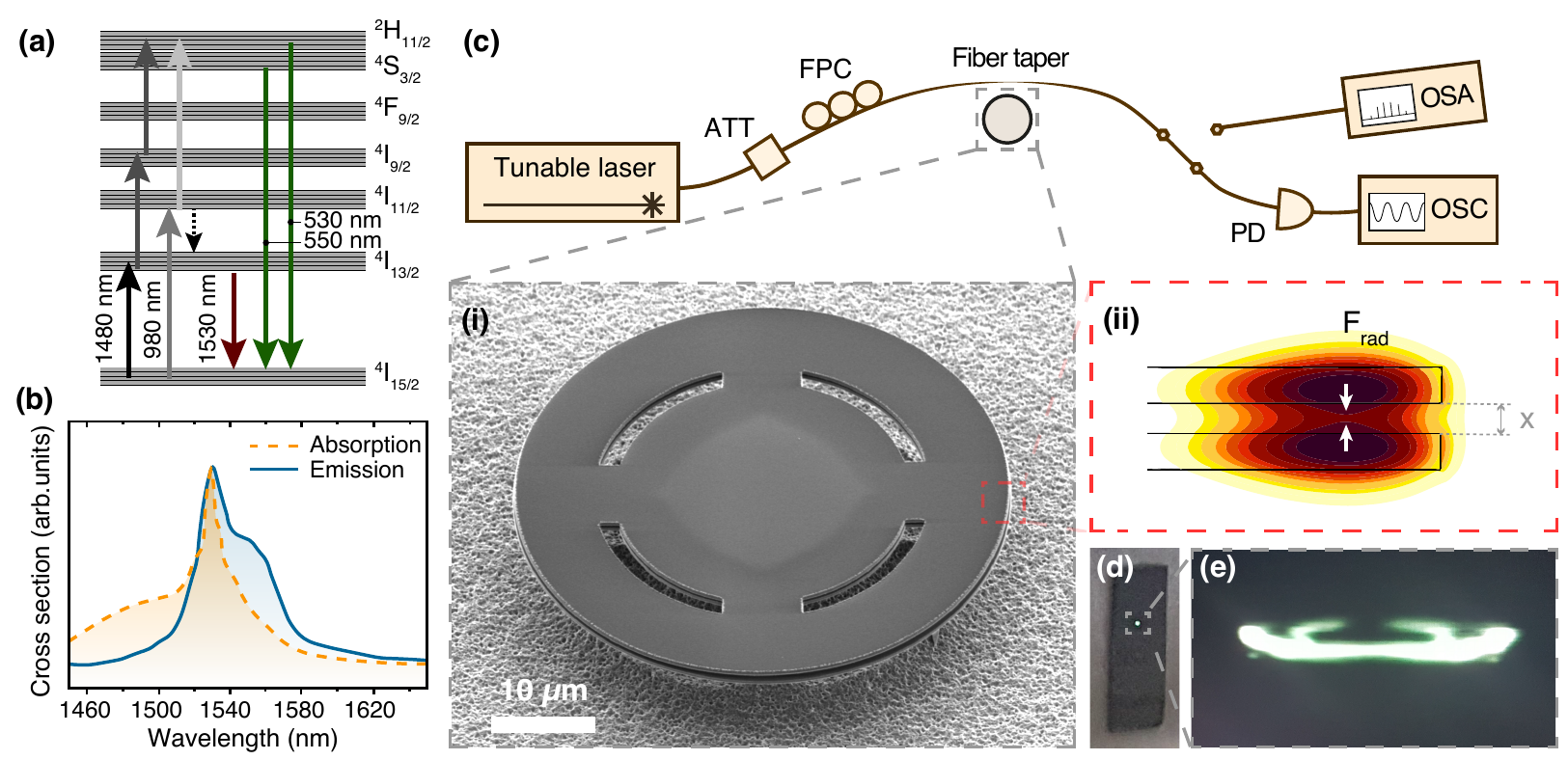}
\caption{\label{Fig:Fig1} (a) Energy levels of erbium implanted in a silica matrix~\cite{polman_erbium_2001,kik_erbium-doped_1998}. Each level is broadened through the Stark effect. Common pumping (left) and emission (right) pathways are denoted by arrows, including multi-photon and non-radiative relaxation (dashed) processes. (b) Absorption and emission cross-sections for erbium at telecommunications wavelengths in a silica matrix. Adapted from data in Ref.~\cite{becker_erbium-doped_1999}. (c) Schematic of the experimental setup. ATT: Attenuator, FPC: fiber Polarization Controller, PD: Photodetector, OSA: Optical Spectrum Analyzer and OSC: Oscilloscope. Insets: i) SEM image of an erbium-doped slotted double-disk device. ii) COMSOL Multiphysics simulation of the fundamental bonding optical WGM, color denotes intensity of electromagnetic field. The frequency of the mode is sensitive to displacements which change the air gap $x$ between the disks, and tuning can be accomplished through the radiation pressure force $F_\text{rad}$. (d) Photograph of green emission from device visible with the naked eye. (e) Microscope camera side-view of device emitting green light. }
\end{figure*}

\section{Active double-disk resonator}

While less widespread than microtoroids or microdisks, double-disk optomechanical systems have been developed by several groups using silica~\cite{lin_mechanical_2009,jiang_high-q_2009,rosenberg_static_2009,bekker_free_2018} and silicon nitride~\cite{lee_silicon_2010,wiederhecker_controlling_2009,wiederhecker_broadband_2011}, with more recent implementations in silicon carbide~\cite{lu_silicon_2019}, lithium niobate~\cite{zheng_high-q_2019,fang_fabrication_2017} and double-layer silicon-on-insulator~\cite{dehghannasiri_integrated_2018}. The geometry consist of two disk cavities stacked vertically such that the evanescent fields of the disks overlap. This causes the optical modes to hybridize into `supermodes' existing in both disks, with a significant proportion of the mode residing in the gap between the disks (See insets of Fig. \ref{Fig:Fig1}(c)). This causes the effective refractive indices $n_\text{eff}$ of the optical modes to be extremely sensitive to the size of the gap. 

The resonance condition for WGM optical cavities is given by:

\begin{equation}
    2\pi R n_\text{eff} = m \lambda_{0,m} \label{Eq:Resonancecondition}
\end{equation}

\noindent where $R$ denotes the radius of the device, $m$ the azimuthal mode number and $\lambda_{0,m}$ the resonance wavelength of the cavity for a given mode number. Therefore, as the effective refractive index of the modes are changed with changing disk spacing $x$, the resonance frequencies of the cavity $\omega_{0,m} = \frac{2\pi c}{\lambda_{0,m}}$ are accordingly shifted. This provides double-disk systems with a very large optomechanical coupling strength, namely $G = \partial \omega _{0,m}/ \partial x \sim 10$~GHz/nm~\cite{bekker_free_2018}, which is to first order dependent on the disks' thickness and vertical (out-of-plane) separation instead of the radius $R$ of the device~\cite{bekker_free_2018}, as is the case in single disk resonators where $G\sim \omega/R$~\cite{ding_high_2010}.  This allows the resonance frequency of the WGM modes to be widely tuned with extremely weak forces, such as radiation pressure~\cite{wiederhecker_broadband_2011, rosenberg_static_2009} and capacitive actuation~ through the use of integrated interdigitated electrodes~\cite{bekker_free_2018}, and for the fabrication of larger devices to increase mechanical compliance without degrading the optomechanical coupling strength. These characteristics are rare among WGM systems, and make double-disk cavities strong candidates for use as on-chip tunable light sources.

\subsection{Erbium luminescence pathways}

The rare-earth element erbium is often used as a gain medium for lasing due to its natural operation in the telecommunications frequency band ($\lambda\sim 1550$~nm). Figure \ref{Fig:Fig1}(a) shows the energy levels of erbium in a silica matrix. Each level is split into a manifold due to the Stark effect~\cite{polman_erbium_1997}, so that the absorption spectrum of the erbium consists of bands with broad absorption ranges.Figure \ref{Fig:Fig1}(b) shows the absorption and emission cross-sections of erbium ions in silica in the telecommunications band ($\lambda \sim 1550$~nm)~\cite{becker_erbium-doped_1999}. The main driving pathway is typically to excite ions to the upper states in the $^4$I$_{15/2}$ band with a laser of wavelength around 1480~nm. The ions decay non-radiatively to the lowest state in the band, and emit a photon of approximately 1530~nm wavelength to return to the ground state. Alternatively, a 980 nm pump source can be used to excite ions to the $^4$I$_{11/2}$ band, as they spontaneously decay from there to the $^4$I$_{15/2}$ band~\cite{yang_tunable_2017}. 

In addition to the driving mechanisms above, absorption of multiple pump photons can excite the erbium to the $^2$H$_{11/2}$ or $^4$S$_{3/2}$ bands, with relaxation to the ground state accompanied by the emission of an upconverted green photon ($\lambda = 550$~nm or 530~nm).

\subsection{Device design and fabrication}

The active double-disks implemented here are 50~$\mu$m in diameter and designed with a slot and four tethers to support the outer annuli, as shown in Fig. \ref{Fig:Fig1}(c)(i). The slot is included to reduce warping of the disks due to internal material stress gradients~\cite{bekker_free_2018} and to allow greater under-etching of the disks. 

The fabrication process for these devices is similar to that presented in Ref.~\cite{bekker_free_2018}, with the addition of an erbium implantation step. A SiO$_2$/$\alpha$-Si/SiO$_2$ stack with nominal thickness 350/300/350 nm is deposited on a crystalline Si substrate using PECVD. The top disk is subsequently implanted with erbium ions using a 1.7 MV tandem Pelletron\textsuperscript{\textregistered} accelerator, with a total ion fluence of $4\times 10^{15}$~ions/cm$^2$. The chip is annealed to repair implantation damage at a temperature of 900$^\circ$C for a period of five hours (See Appendix \ref{Sec:Annealing} for discussion of the annealing process). The double-disk stack is defined through electron-beam lithography with ma-N 2410 negative resist (Micro Resist technology GmbH.) and three successive reactive ion dry-etching steps. Each layer is etched using an optimised etch chemistry consisting of SF$_6$, CHF$_3$ and CF$_4$ gases. After removal of the resist with an oxygen plasma process the silicon substrate and amorphous silicon sacrificial layer are released with an isotropic xenon difluoride dry etch.

\subsection{Measurement Setup}

A schematic of the experimental setup is shown in Fig. \ref{Fig:Fig1}(c). Optical spectroscopy is performed using a tunable continuous wave diode laser (Yenista T100S-HP) in the 1500-1600 nm wavelength band. Input light is coupled into the double-disk through a tapered optical fiber, and the transmitted optical intensity is detected on a high-speed photodetector. Optical quality factors on the order $10^5$ are typically observed in these devices, similar to the values observed in other silica double-disk optomechanical systems~\cite{bekker_free_2018, rosenberg_static_2009}. Alternatively, the transmitted light can be analyzed on an optical spectrum analyzer (OSA) to detect photoluminescence from the implanted erbium. 

While driving at 1480 nm is optimal due to the small emission cross-section of the ions~\cite{kippenberg_demonstration_2006} as shown in Fig. \ref{Fig:Fig1}(b), effective excitation can be achieved up to a wavelength of $\sim 1510$~nm. This allows driving of the device in our experiments using an optical resonance situated around 1506~nm. The resulting photoluminescence is strongest around 1535 nm, and drops off for wavelengths below 1500~nm and above 1570~nm. Section \ref{Sec:Emission} outlines the experimental observation of photoluminescence in the 1550~nm band in our device.

 We additionally observe significant green emission from our devices while pumping in the telecommunications band, both with the naked eye (Fig. \ref{Fig:Fig1}(d)) and with an optical microscope camera (Fig. \ref{Fig:Fig1}(e)). This green emission was not guided in the optical fiber used to pump the cavity, but could be captured with a CCD microscope camera in order to extract the intensity of the emission and reveal upconversion lasing~\cite{lu_-chip_2009}, as discussed in Section \ref{Sec:Green}.

\section{Photoluminescence at telecommunication frequencies}
\label{Sec:Emission}

\begin{figure}[t]
\centering\includegraphics[width=\linewidth]{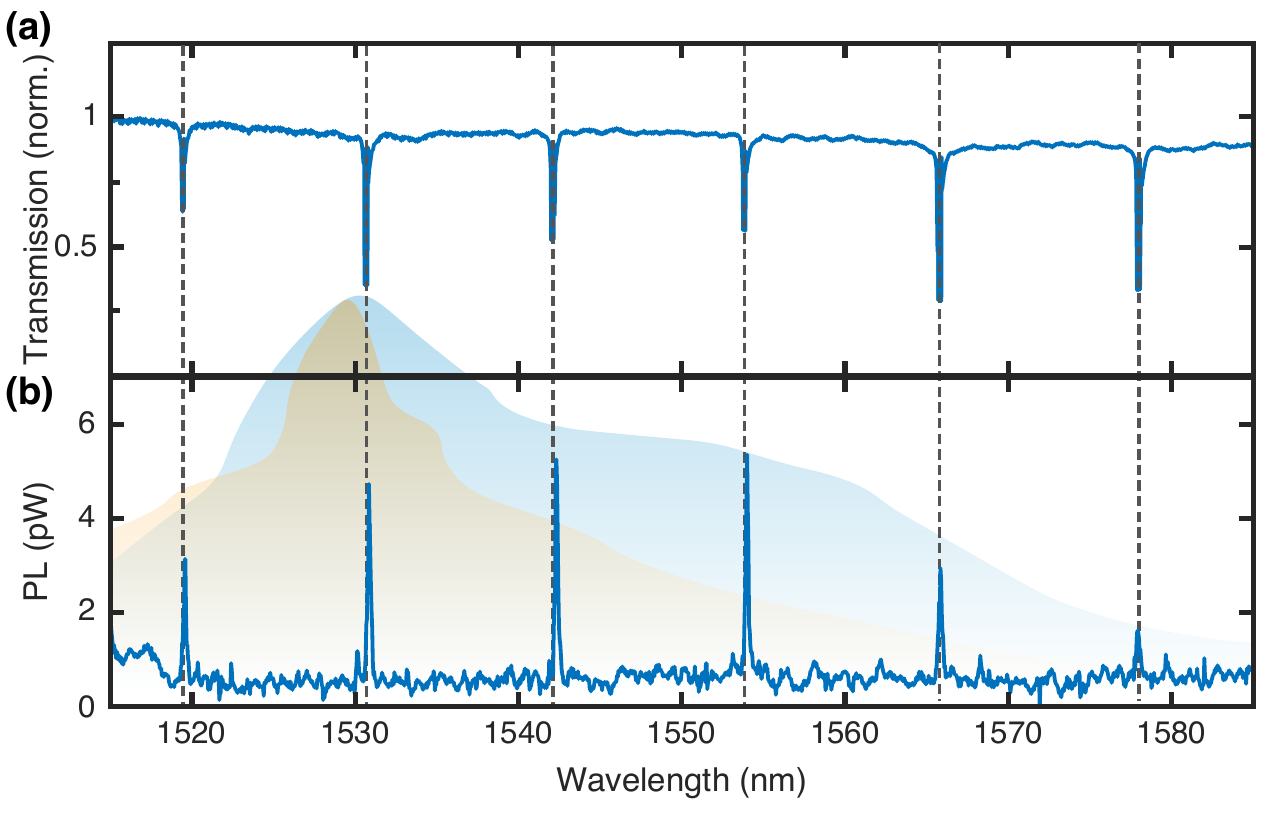}
\caption{\label{Fig:OSAmodescomparison} Comparison between optical mode spectrum and photoluminescence emission peaks in the near infrared band. a) Normalized optical transmission spectrum showing a set of WGMs separated by the device FSR. b) OSA photoluminescence (PL) spectrum acquired while the cavity is pumped using a WGM mode at about 1506~nm showing emission peaks at wavelengths corresponding to device WGMs. The peaks are not visible above the noise floor outside of this band, confirming that the emission is indeed due to the implanted erbium. The shaded areas in (b) denote the emission (blue) and absorption (orange) cross-sections of the erbium ions as a function of wavelength. Note that while emission peaks at 1530 nm, the absorption of the ions is also largest at this point, resulting in a reduced photoluminescence peak height.}
\end{figure}

When erbium ions are implanted in a WGM cavity and the system is pumped in a region of strong erbium absorption, Purcell enhancement~\cite{romeira_purcell_2018} causes the photoluminescence from the ions to preferentially occur at frequencies corresponding to optical cavity modes in the emission region. Hence, when the cavity is optically pumped, spectrally narrow peaks separated by the FSR of the optical cavity are expected across the entire erbium emission range~\cite{jager_whispering_2009, kippenberg_demonstration_2006}. Optical spectroscopy of the device is first performed by sweeping the frequency of the input laser and detecting the transmitted light on a high-speed photodetector. This yields the optical spectrum of the double-disk shown in Fig. \ref{Fig:OSAmodescomparison}(a), where one family of WGM modes is visible. 

In order to detect photoluminescence from the erbium, the pump laser is detuned to the side of an undercoupled optical resonance of the double-disk with a wavelength of 1506.4~nm. The transmitted light is sent to an optical spectrum analyzer (OSA, Yokogawa AQ~6374), and the resulting photoluminescence is plotted in Fig. \ref{Fig:OSAmodescomparison}(b).

Figure \ref{Fig:OSAmodescomparison} confirms the expectation that photoluminescence from the erbium predominantly occurs at the WGM resonances of the double disk. The dashed vertical lines between the plots emphasise the shared frequency of each WGM resonance and emission peak. The comb of photoluminescence peaks extends over the range of 1520-1580~nm, as shown in Fig. \ref{Fig:OSAmodescomparison}(b), and drops below the level of the measurement noise outside of these bounds. This spectrum is consistent with what would be expected for optical emission from the erbium within the WGM cavity~\cite{hsu_ultra-low-threshold_2009, lu_-chip_2009,kippenberg_demonstration_2006, jager_whispering_2009}.

\subsection{Photoluminescence tuning}
\label{Sec:Tuning}

\begin{figure} [t]
\centering
\includegraphics[width=\linewidth]{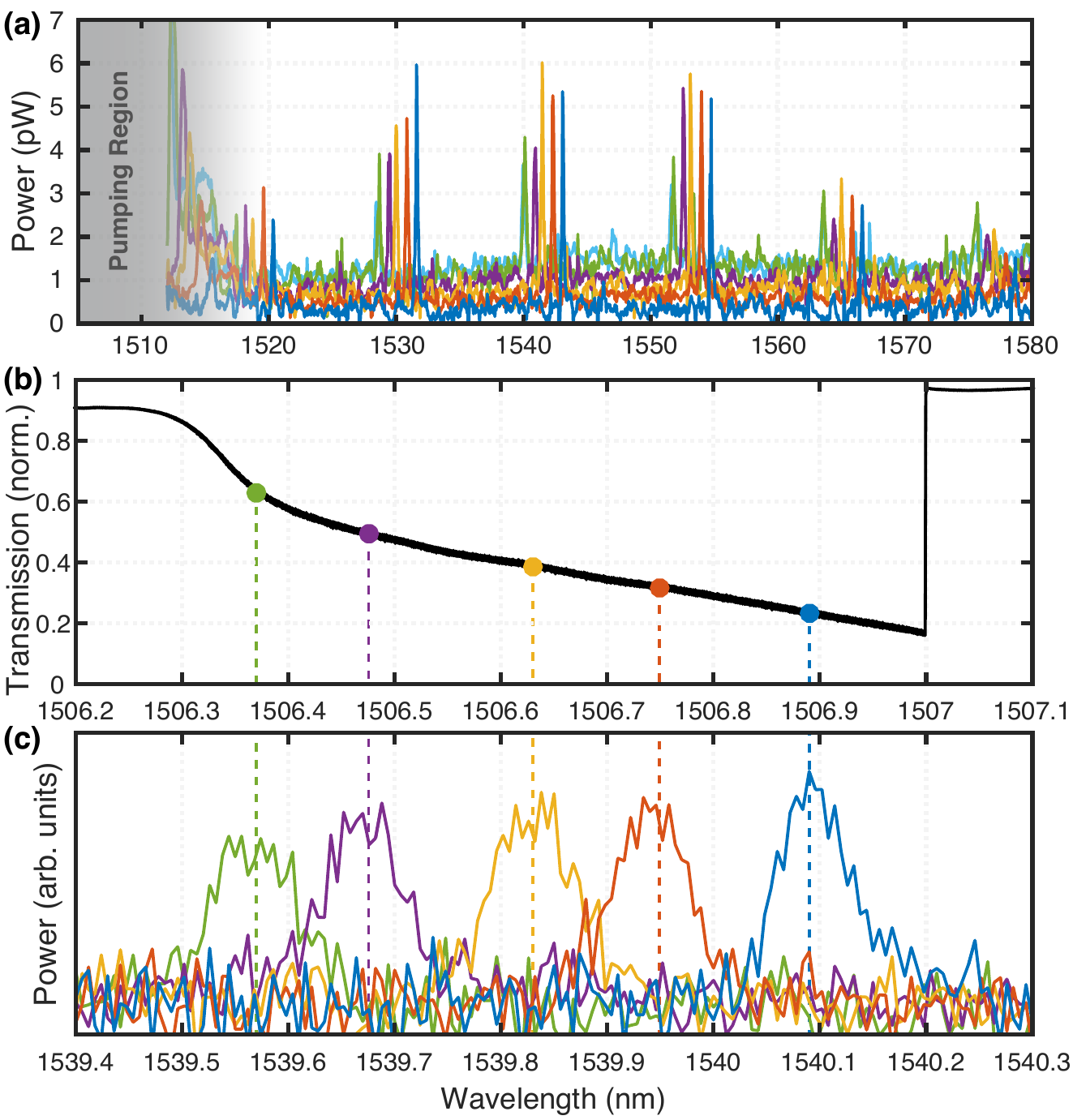}
\caption{\label{Fig:tuningemission} Photoluminescence from the device as a function of the detuning of the pumping laser. a) Optical emission spectrum of multiple modes for different dropped powers (green, lowest; to blue, highest). At frequencies corresponding to the pumped optical mode the detector is saturated, and no data can be collected (gray region). Instead, optical tuning is measured using an adjacent peak corresponding to a mode of the same family. Note that all modes are identically tuned as the dropped power is increased. b) Normalized optical transmission trace of the pumped mode as the input laser is swept through it at a speed of 10~nm/s with a power of 10~mW. The triangular shape is characteristic of optomechanical pulling, and indicates the optical tuning range. Subsequently the laser was set to a constant detuning at each of the positions marked by a colored circle, and the optical emission was measured on an OSA. c) Resulting photoluminescence for each detuning of the input laser from a mode in the same family as the pumped mode. The luminescence peaks clearly demonstrate the photoluminescence frequency tuning, with peak shape indicating the equilibrium spectral shape of the mode.}
\end{figure}

Since the frequencies of the erbium photoluminescence emission peaks are directly linked to the device's WGM resonance frequencies, they can be directly controlled by tuning the optical modes. Figure \ref{Fig:tuningemission}(a) shows the emission spectrum across the full erbium emission range with varied dropped laser power into an optical resonance around 1506~nm. As the dropped power increases we observe an overall red-shift of the spectrum. We note that measurements were not possible close to the pump wavelength because of saturation of the optical spectrum analyser due to the presence of the pump. Fig. \ref{Fig:tuningemission}(b) shows the transmission past the WGM at fixed input power, as the laser is swept across the 1506~nm resonance. The non-Lorentzian profile results from the optical tuning of the resonance frequency. Fig. \ref{Fig:tuningemission}(c) shows the observed tuning of the photoluminescence emitted into an adjacent WGM resonance, which has a nominal wavelength of 1540~nm, due to this optical tuning. Here, the fixed input power results in a smaller tuning frequency of 520~pm as opposed to the $\sim3$~nm achieved in Fig. \ref{Fig:tuningemission}(a) by also modifying the power, but allows for greater consistency and direct comparison between the optical resonance and erbium emission peaks.

Three effects contribute to the observed tuning, namely: the thermo-optic effect, where the refractive index of the silica is modified due to temperature changes; mechanical expansion from photothermal heating; and radiation pressure, where the optical gradient force between the double disks changes the size of the air gap (See Appendix \ref{Sec:Tuningmechanism} for calculations and discussion of the tuning mechanisms present in this system). As the laser is detuned closer to the cavity resonance frequency more optical power enters the cavity, increasing the strength of each these effects to modify the optical path length, either through mechanical deformation or modifying the material refractive index, and hence shifting the optical resonance frequency. This leads to the triangular mode shape in Fig. \ref{Fig:tuningemission}(b).

The photoluminescence tuning accompanying optical resonance frequency tuning is shown in Fig. \ref{Fig:tuningemission}(c), where a photoluminescence peak is tuned as the pumping laser frequency is varied. The central frequency of the photoluminescence peak indicates the tuned resonance frequency of the optical mode and the width of the peak its spectral shape.  Optical resonance frequency tuning over a full FSR using radiation pressure has been demonstrated in similar devices~\cite{wiederhecker_broadband_2011}, allowing photoluminescence at any arbitrary frequency within the material transparency window. Large tuning using capacitive actuation is also possible~\cite{bekker_free_2018,perahia_electrostatically_2010}, and can provide a mechanism to tune photoluminescence independently of the optical pump.

In the literature, strain-tuning of the emission frequencies of quantum dots~\cite{weis_surface_2016,sun_strain_2013,moczala-dusanowska_strain-tunable_2019,kremer_strain-tunable_2014,tao_-chip_2020}, spin defects~\cite{lee_strain_2016} and erbium ions~\cite{feng_enhancing_2010,zhang_single_2019} has been demonstrated. The strain changes the local electromagnetic environment of the active ions in the material, which changes the local electronic wavefunctions around the emitters. Strain tuning of this sort is negligible in our devices both because the material lattice is only affected at the anchor points of the annuli, far from the optically pumped ions; and because the emission spectrum of erbium is far wider than the WGM resonance linewidth, so that the emission wavelengths are predominantly determined by the WGM resonances.

\section{Green upconversion lasing}
\label{Sec:Green}

As the pump power is increased, we expect to see an increase in the photoluminescence intensity in the telecommunications band. However even at low input powers ($< 1$~mW) saturation of the telecommunications band photoluminescence occurs, and we instead observe an increase in the emission of green light from the device arising from upconversion of the input light~\cite{lu_-chip_2009,he_whispering_2013}. Green light emission is visible from the device even with the naked eye, as shown in Fig. \ref{Fig:Fig1}. 

\begin{figure} [t]
\centering
\includegraphics[width=\linewidth]{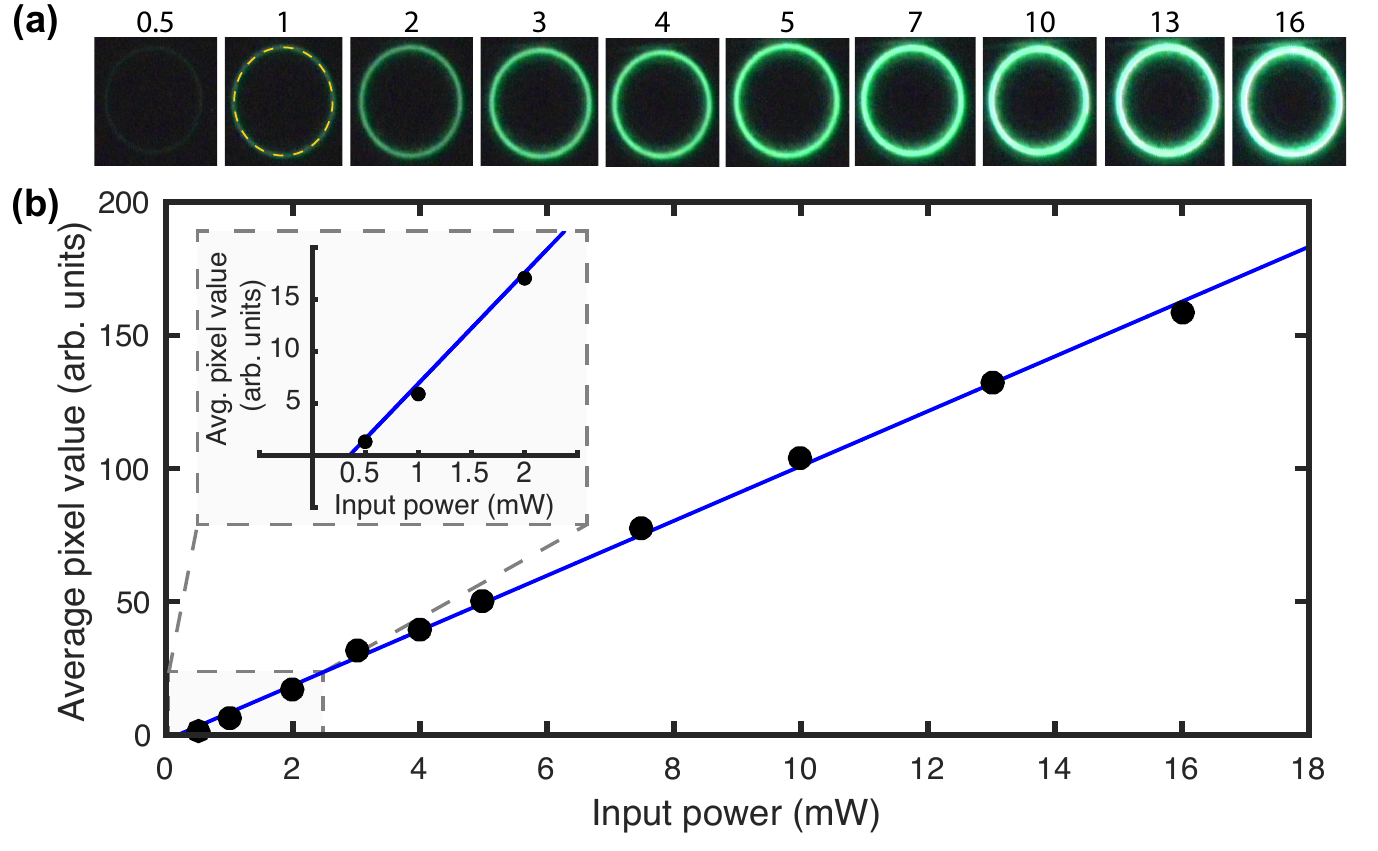}
\caption{\label{Fig:intensity_vs_power} Intensity of green optical emission as a function of input laser power, demonstrating green upconversion lasing. (a) Microscope images of emitting devices with increasing pump power. The region over which the optical intensity is recorded is  denoted by a dashed yellow ring in the second image (`1'). (b) Photoluminescence intensity in the recorded region for each image, extracted using ImageJ. For this plot, the average of the red component in the RGB-intensity of the pixels in the recorded ring is measured, in order to avoid artefacts from the saturation of the camera present in the green component, see Appendix \ref{Sec:Saturation}. Assuming an approximately constant wavelength as a function of input power, this is directly proportional to the total intensity of the emitted light. Variations in recorded intensities between pixels were found to result in a less than 2\% relative standard error in the total measured intensity. This is insignificant compared to the uncertainty in the pump power, which we believe accounts for the majority of the measurement uncertainty.}
\end{figure}

The parameters of the fiber taper and optical fiber used in our experiment are chosen for guiding light in the telecommunications window. This means that the tapered fiber is not correctly phase-matched for coupling green light from the cavity. As such, the green emission from the cavity was not guided by the output optical fiber, and no signal could be detected on the OSA in this wavelength range. It is in principle possible to achieve simultaneous coupling to such far separated wavelengths, for example by using two tapered fibers in an add-drop configuration~\cite{cai_fiber-optic_1999,monifi_robust_2012, murugan_optical_2010}, one for each wavelength, or a chaotic cavity~\cite{jiang_chaos-assisted_2017} which can be phase matched across a large wavelength range. For simplicity, however, here direct observation with a CCD microscope camera is used. This is a common technique for photoluminescence measurements in the visible and NIR wavelength ranges~\cite{mitchell_photoluminescence_2016,sun_camera-based_2015,sun_camera-based_2016,kasuya_resonant_1997,doke_photoluminescence_2013}. Observation with the CCD microscope camera reveals that the emission is concentrated around the outer perimeter of the device, and as far as can be ascertained is emitted isotropically. Top-view images of the device are taken with the microscope camera (Fig. \ref{Fig:intensity_vs_power}(a)), and post-processed to get a measure of the functional dependence of the intensity of the emission with pump power.

To measure the intensity of the green emission, the software package ImageJ~\cite{collins_imagej_2007} is used to extract the intensity of pixels along a ring within the optical mode area in each image. This region is shown as a yellow dashed line in Fig. \ref{Fig:intensity_vs_power}(a). The responsivity of the pixels in the CCD microscope camera is a function of the frequency of the green emission. However, even at the maximum pump powers used in our experiments, the relative shift in the optical mode frequencies is less than 0.1\%. Consequently, it is reasonable to approximate the responsivity as constant. The green component of the captured intensity saturates in the detector at input powers above 5~mW. Using the red component allows the intensity of the emission to be determined without reaching saturation (see Appendix \ref{Sec:Saturation} for further discussion).

In Fig. \ref{Fig:intensity_vs_power}(b) the emitted intensity is plotted as a function of optical input power, yielding a clear linear trend. A least-squares fit gives the equation for the trendline $I \propto (10.6\pm 0.2)P_\text{in} - (3.6\pm 0.8)$~mW where the uncertainties represent one standard deviation from the mean. This yields a threshold power of $P_\text{thresh} = 340 \pm 70 \; \mu$W. The existence of a threshold and linear dependence with pump power is consistent with green upconversion lasing~\cite{lu_-chip_2009}. Because the green light was not guided in the optical fibers used in the experimental setup, an optical spectrum of the lasing could not be obtained to determine whether the lasing is single- or multimode. 

It is important to note that the tuning mechanism demonstrated in Fig. \ref{Fig:tuningemission} applies to all of the cavity modes, giving confidence that the green lasing observed is also tunable. Further experiments could focus on suppressing green emission to produce lasing at telecommunications frequencies, or on capturing the green emission observed here in-fiber to directly observe tunable lasing, for example through use of an add-drop configuration~\cite{cai_fiber-optic_1999,monifi_robust_2012, murugan_optical_2010} or a chaotic cavity~\cite{jiang_chaos-assisted_2017}, as mentioned earlier.

\subsection{Concentration quenching}

The presence of green upconversion lasing in our devices, as opposed to lasing in the telecommunications band, can be explained through concentration-quenching mechanisms, in particular cooperative upconversion and excited-state absorption (ESA)~\cite{van_den_hoven_upconversion_1996,pollnau_time-resolved_1992}. Cooperative upconversion occurs between two neighboring excited erbium ions, where relaxation of one ion stimulates excitation of the other to a higher energy level~\cite{snoeks_cooperative_1995}, causing emission of a single higher-energy photon. ESA, on the other hand, occurs in a single ion as a two-step excitation process~\cite{kik_erbium-doped_1998}. Both processes can be considered concentration quenching, that is, effectively lowering the fraction of erbium in the first excited state at a given pump power~\cite{van_den_hoven_upconversion_1996} and therefore making lasing at telecommunication frequencies more difficult. This is especially the case for ESA, which can compete directly with stimulated emission if it occurs at the laser pumping frequency, effectively rendering lasing impossible in the 1550~nm band~\cite{pollnau_time-resolved_1992}. 

A further concentration quenching effect can arise from regions at the fringes of the optical mode, where the erbium ions are incompletely pumped. These regions give rise to a large absorption of pump power while not contributing to gain in the cavity, increasing the lasing threshold~\cite{kalkman_erbium-implanted_2006}. This effect can be avoided by selectively implanting regions of erbium where the optical modes will be situated through use of photoresist masks~\cite{johnson_method_1975}. 

In order to achieve lasing at telecommunications frequencies, the effect of the material matrix on the erbium ions' energy levels must be carefully considered. For example, it has been demonstrated that cooperative upconversion can be reduced by optimising the homogeneity of the ion implantation process~\cite{kik_erbium-doped_1998,kik_cooperative_2003}.

\begin{figure}[t]
\centering\includegraphics[width=\linewidth]{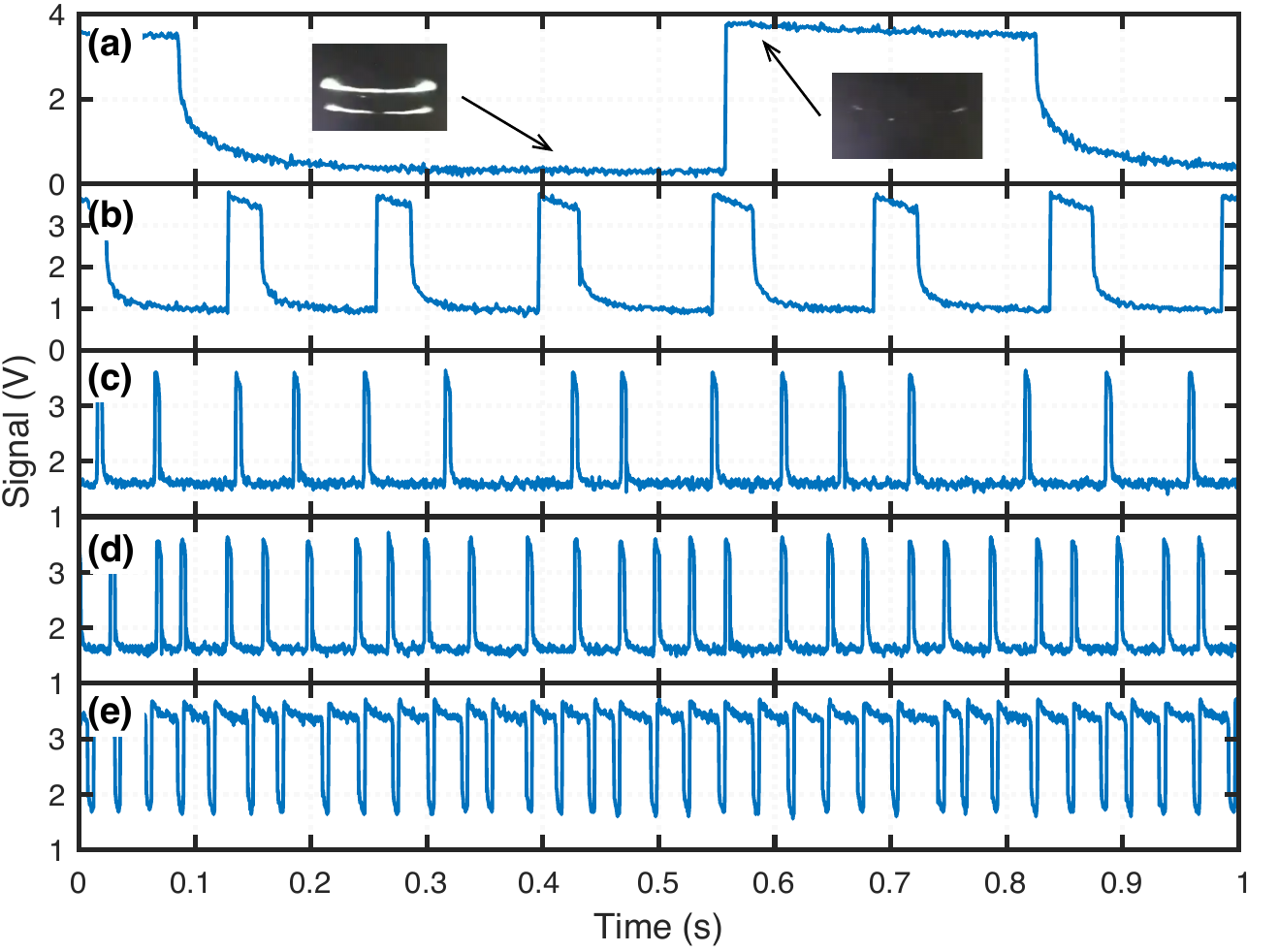}
\caption{\label{Fig:Blinking} Self-pulsing or `blinking' of green photoluminescence in double-disk cavities under constant laser input. A large output signal means no coupling to the device, and corresponds to no light on the microscope camera, whereas when the signal dips to low values, the light is on resonance with the cavity and the device lights up with bright green emission (see insets in (a), including reflection from the substrate). The rate of blinking can be controlled (a-d) through changes in the pumping power and detuning from resonance - in general, a higher power and smaller detuning lead to slower blinking rates, as also observed in~\cite{baker_optical_2012}. The style of the blinking can also be controlled in a similar manner, from rapid `dark' blinks for a mostly lit device (d) to the case where the device rapidly blinks `on' (e).}
\end{figure}

\section{Optomechanical instability and self-pulsing}

In addition to the phenomena described up to this point, we observe self-pulsing optomechanical effects in our double-disk devices (Fig. \ref{Fig:Blinking}). The existence of sets of forces of comparable magnitude acting in opposite directions and on different characteristic timescales can lead to instabilities in the optomechanical resonator. Such instabilities have been reported in a variety of systems, where the thermo-optic effect in the cavity is opposed by free carrier dynamics~\cite{johnson_self-induced_2006}, thermo-mechanical forces~\cite{baker_optical_2012,park_titanium_2014,diallo_giant_2015}
, gain dynamics in an external cavity~\cite{rowley_thermo-optical_2019} 
or the Kerr effect~\cite{park_regenerative_2007}. 
In our work, the fast driving is predominantly due to radiation pressure and the thermo-optic effect, while the slower relaxation is due to thermo-mechanical effects. Both the radiation pressure and thermo-optic effects here increase the effective refractive index of the cavity, leading to a red-shift of the resonance, while the thermo-mechanical force causes the disks to separate, blue-shifting the resonance. This latter force operates here on a slower timescale, gradually blue-shifting the WGM onto and past resonance, at which point the sign of radiation-pressure and thermo-optic feedback change from negative to positive, and the mode rapidly snaps out of resonance~\cite{baker_optical_2012}, as visible in Fig. \ref{Fig:tuningemission}(b). This eliminates further forcing from intracavity photons, until upon cooling, the mode is again brought into resonance, leading to a characteristic temporal transmission which results from forming a hysteresis loop around the bistable optical resonance~\cite{baker_optical_2012}. Because in this regime the thermo-optic tuning acts in the same direction as the radiation pressure tuning, the relative magnitude of these may be difficult to distinguish. We note however that the radiation pressure force in double-disks can be either attractive or repulsive depending on the spatial distribution of the optical mode~\cite{wiederhecker_controlling_2009}, which results respectively in a red- or blue-shifting of the cavity resonance. The direction of the thermo-optic and thermo-mechanical tuning, on the other hand, are dependent on geometric and material characteristics and are set for a given device. The fact that we have observed, on the same device, a reversal from a net red-shifting to a net blue-shifting effect for different WGMs indicates that the radiation pressure-based tuning is at least comparable in magnitude to the contribution from thermal effects.

Because the effects occur with different characteristic time constants, the steady state of the device is oscillatory, with the period of oscillation determined by the slow response.

The self-pulsing of the device can be controlled either through changing the detuning of the laser from optical resonance or the coupling rate to the tapered fiber. Several experimental measurements with different laser detuning are shown in Fig. \ref{Fig:Blinking}(a-e), corresponding to different periods and duty cycles. These are similar to the self-pulsing effects observed in other systems~\cite{baker_optical_2012, park_titanium_2014,deng_thermo-optomechanical_2013}. 

In order to better understand and explain the mechanism behind self-pulsing in this device, a theoretical model based on the work in Ref.~\cite{baker_optical_2012} (and including the radiation-pressure force) was developed and is discussed in Appendix~\ref{Sec:Model}. The good qualitative agreement between the model and experimental observations corroborates the explanation of self-pulsing given above, and provides an avenue for further theoretical exploration of this phenomenon.

It is important to note that this self-pulsing becomes much slower at higher optical powers due to mechanical stiffening of the device as the dropped power is increased \cite{rosenberg_static_2009}. Indeed, stable coupling to the side of resonance is achievable with sufficient optical power (as is the case in Fig. \ref{Fig:tuningemission}). This strong optical spring effect due to blue-detuned optical driving is not accompanied here by mechanical instabilities (i.e. regenerative oscillation arising due to dynamical back-action~\cite{bekker_injection_2017}) due to the very high squeeze-film damping of the double-disk resonator's vibrational modes in air~\cite{lin_mechanical_2009,rosenberg_static_2009}.

The combination of self-pulsing and a gain medium, in our case, allows the conversion of a continuous infrared pump into frequency-chirped pulsed signals, both as photoluminescence at infrared wavelengths and green lasing in the visible band. Self-pulsing behavior has been proposed as a method to sense fluctuations in humidity~\cite{deng_thermo-optomechanical_2013}, for biosensing~\cite{park_titanium_2014} and for all-optical control and synchronization~\cite{rowley_thermo-optical_2019}. An interesting question for future work is whether the additional optomechanical dynamics and loss channels due to the intracavity gain medium introduced here interacts with the self-pulsing instability, creating different forms of self-pulsing dynamics (e.g. Ref.~\cite{princepe_self-sustained_2018}).

\section{Conclusion}

In this work we have demonstrated optically tunable photoluminescence in the telecommunications band and green upconversion lasing emitted from chip-integrated silica double-disk optomechanical systems. The observation of radiation-pressure- and thermally-driven self-pulsing in our double-disk geometry provides the possibility to build pulsed light sources and explore different mechanisms of inducing optomechanical instability. 

Engineering the energy levels of the implanted erbium should allow for low-threshold lasing to also be achieved at 1550~nm, leading to applications of the system as an optically pumped, micron-scale, integrated laser light source for `lab on a chip' experiments or photonic integrated networks. Electronic tuning of the optical resonance frequencies can be achieved through capacitive actuation schemes, which has already been used to demonstrate super-FSR optical tuning ranges in similar devices~\cite{bekker_free_2018}, which opens up rich capabilities for fully integrated photonic-electronic hybrid technologies~\cite{atabaki_integrating_2018}.

\section*{Acknowledgments}

The ion implantation for these devices was performed at the NCRIS facilities (ANFF and the Heavy Ion Accelerator Capability) at the Australian National University. Fabrication was performed at the Queensland node of the Australian National Fabrication Facility (ANFF-Q) and the Microscopy Australia Facility at the Centre for Microscopy and Microanalysis, The University of Queensland. The authors acknowledge these facilities and the scientific and technical assistance of the staff. The authors thank E. Cheng and G. Harris for valuable insights and discussions and M. Lyu for annealing the chips. This work is funded by the Australian Research Council Centre of Excellence in Engineered Quantum Systems (EQUS) (CE170100009) and the Commonwealth of Australia as represented by the Defense Science and Technology Group at the Department of Defense. C.J.B. and C.G.B. acknowledge support through an Australian Government Research Training Program (RTP) Scholarship and an Australian Research Council Fellowship (DE190100318), respectively.

\appendix
\renewcommand\thesection{\Alph{section}}

\section{Device annealing} \label{Sec:Annealing}

\begin{figure} [h]
\medskip
\centering
\includegraphics[width=\linewidth]{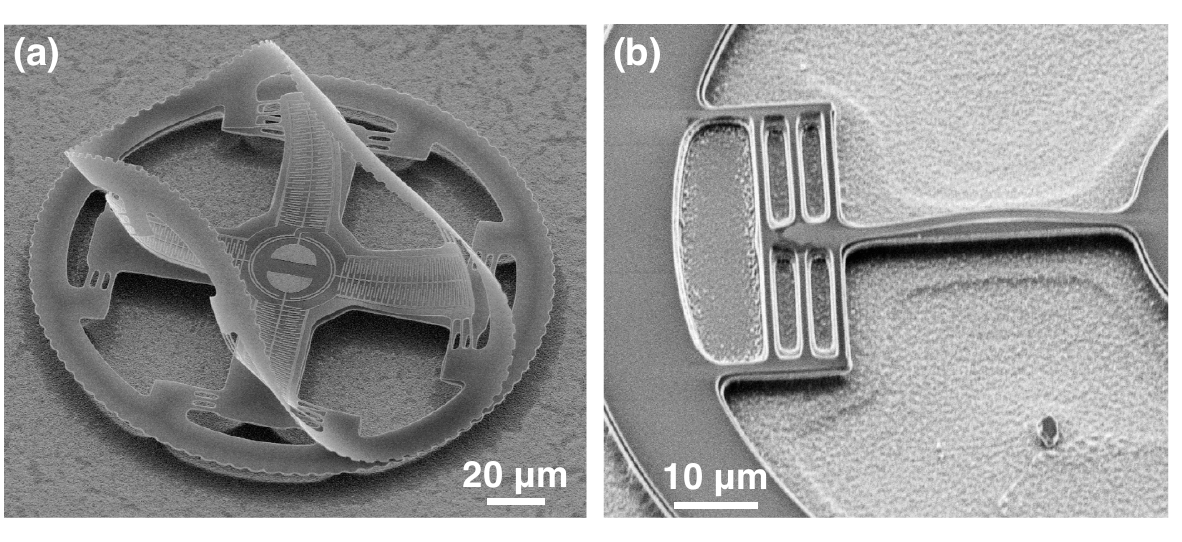}
\caption{\label{OCLFig:tacodisk}Electron micrographs showing warping in un-annealed devices with implanted erbium. This spectacular deformation arises from a stress gradient in the top disk induced by implantation damage, which must be repaired through an annealing step. The bottom disk had no implanted ions and displays minimal warping in comparison.}
\end{figure}

The first generation of devices were not annealed prior to fabrication. As shown in Fig. \ref{OCLFig:tacodisk}, the result was very large warping of the top silica disk. Even with just the tethers released (right-hand panel of Fig. \ref{OCLFig:tacodisk}), the warping is clearly visible. The reason for this is that during implantation, erbium ions dislocate other molecules from their places in the solid matrix both where the erbium is finally lodged and along its trajectory of travel. These dislocations cause not only large bulk stresses, but also a strong stress gradient in the material which enhances warping and distortion when the device is released. Here erbium was only implanted in the top disk, hence there was minimal damage to the bottom disk layer and consequently minimal warping as can be observed in Fig. \ref{OCLFig:tacodisk}.

In order to negate this warping the chips are annealed after ion implantation and prior to any further processing being done. At temperatures above 800$^\circ$C, the silica layers are allowed to relax, thereby `healing' the dislocations from the erbium implantation process~\cite{min_erbium-implanted_2004}. Though a few different parameters for annealing were used, it was found that sufficient layer repair is achieved by annealing at a temperature of 800$^\circ$C for one hour in a muffle furnace (atmospheric environment). However, to maximize the relaxation from the process, annealing for most chips is done at 900$^\circ$C for 5 hours. The inset of Figure \ref{Fig:Fig1}(b) shows an SEM of a fabricated annealed device, where the stress from ion implantation has been successfully removed.

\section{Saturation in the detection of green photoluminescence}
\label{Sec:Saturation}

In Section \ref{Sec:Green}, green photoluminescence of the device resulting from multi-photon absorption in the implanted erbium is presented. The green emission is captured using an optical CCD camera with a microscope objective to measure its dependence on the input power. The software package ImageJ is used to perform gamma correction on each image and to extract the individual R-, G- and B-component values as well as the weighted intensity of each pixel in the brightest region of emission (which corresponds to the centre of the optical mode). This data is then used to extract an average value and standard deviation for each of the four parameters at different optical pump powers. In this section, we discuss these measurements, and justify the use of the red component to display the linear relationship between the green emission intensity and pump power in the main text.

\begin{figure} [t]
\centering
\includegraphics[width=\linewidth]{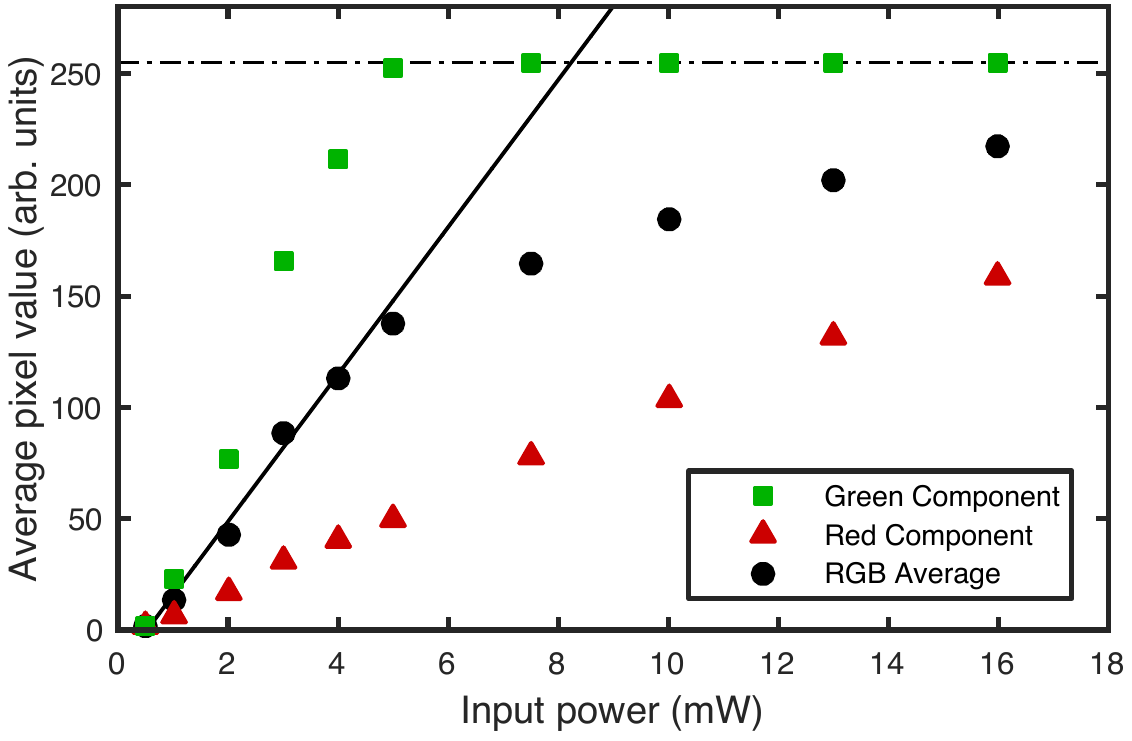}
\caption{\label{Fig:intensity_saturation} Saturation of optical intensity measurements for green emission. The pixel value of the green component of the emission (green squares, measured using RGB color intensities) saturates at the maximum detectable value of 255 at an input power of 5 mW, indicated by a dash-dotted horizontal line in the figure. Correspondingly the total measured signal (black squares), which consists of a weighted average of red, green and blue components, deviates from its linear trend at this value. as clearly seen in the figure. }
\end{figure} 

For the same optical images shown in Fig. \ref{Fig:intensity_vs_power}(a), Fig. \ref{Fig:intensity_saturation} shows the weighted average of the RGB pixels corresponding to total intensity (black circles) and green component of the pixel RGB values (green squares). As can be seen in the Figure, the green pixel component grows rapidly with increasing input power, and saturates at the maximum pixel value of 255 (corresponding to the maximum signal measurable by the camera) at a power of 5~mW. Because the total intensity of the measured emission takes into account all three RGB components, a corresponding saturation and deviation from the linear trend is clearly observable for the black circles in Fig. \ref{Fig:intensity_saturation} at pump powers larger than 5~mW. The fit to the first five points of the RGB average measurement (solid black line in Fig. \ref{Fig:intensity_saturation}) emphasises the linear trend before, and deviation after saturation.

Hence, neither the green component nor the RGB average measurements are accurate for input powers above 5 mW. In contrast, the red pixel measurement (red triangles in Fig. \ref{Fig:intensity_saturation}) does not approach saturation over the entire measured power range. This is because the wavelength of emission is in the green region of the visible spectrum, so that filtering out the `red' component of the emission wavelength results in a strong attenuation of the signal. This does not imply that there is red light in the emission, but that the red pixels in the CCD camera have some responsivity at green wavelengths. Therefore we can have confidence in the results obtained for this component, as presented in Fig. \ref{Fig:intensity_vs_power}(b), which indicates that the linear increase in the green emission intensity was maintained up to the highest pump power.

\section{Mechanism of optical resonance frequency tuning}
\label{Sec:Tuningmechanism}

Three forces interact in our system to tune the optical resonance frequencies of the device as a function of input laser power. In this section we model these forces to assess the mechanism behind the tuning in Fig. \ref{Fig:tuningemission}. 

Firstly, optical absorption in the silica disks causes heating and an increase of the material refractive index through the thermo-optic effect~\cite{almeida_optical_2004, parrain_origin_2015}. This is accompanied by thermal expansion of the heated material, which affects the spacing between the disks (this second effect is referred to here as the thermo-mechanical effect) and is opposed by mechanical rigidity. Lastly, optical power circulating in the cavity also alters the spacing between the disks through the radiation-pressure force. 

\begin{figure}[t]
\centering\includegraphics[width=\linewidth]{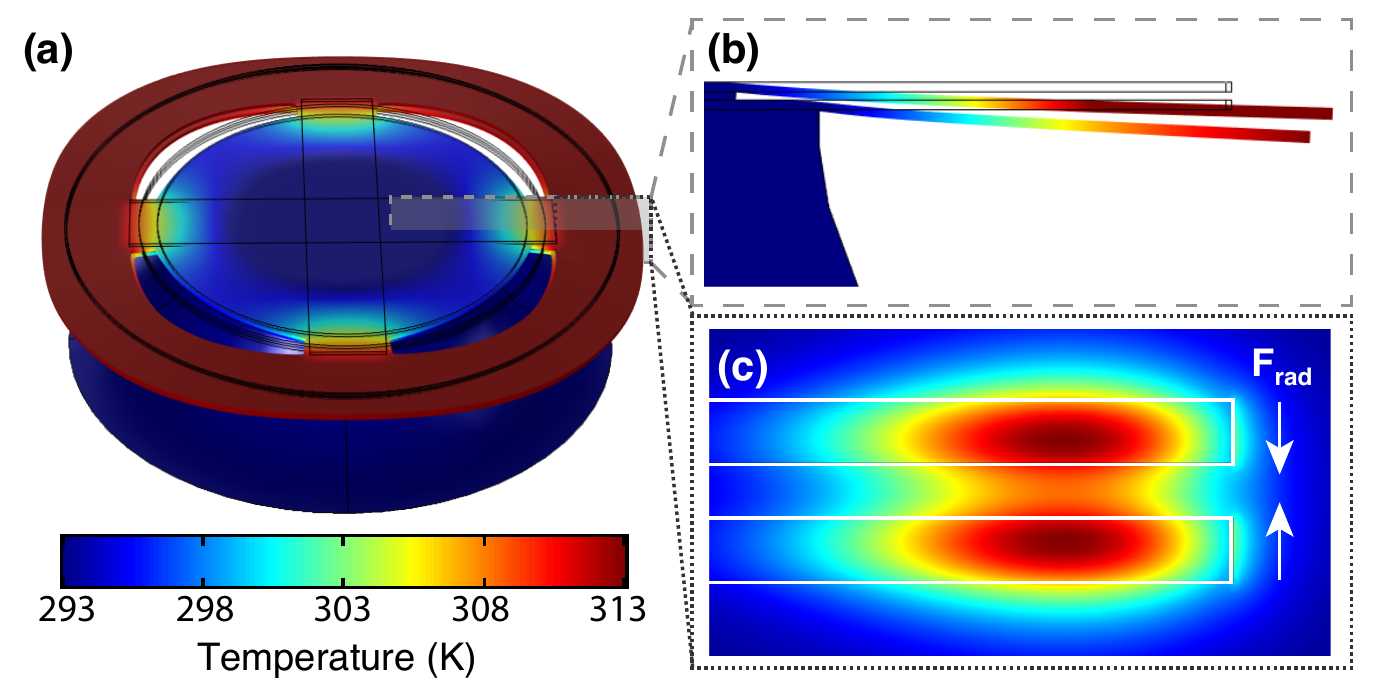}
\caption{\label{Fig:thermooptic} (a) 3D-view and (b) side-view of a COMSOL Multiphysics simulation of thermal expansion in the slotted double-disk geometry in response to an applied temperature difference of 20 K between the disk perimeter and substrate, where the color scheme denotes temperature of the device in both cases. The thermal bottleneck in the system is at the anchor points, creating a thermal gradient which differentially displaces the disks in the vertical direction, corresponding to a shift in the optical resonance frequency. (c) Simulation of electromagnetic field strength (color) in a bonding optical whispering gallery mode hosted by the geometry, which imparts a vertical force on the disks due to radiation pressure.}
\end{figure}

The magnitude of the thermo-optic and thermo-mechanical effects can be estimated through use of finite element modelling. In Fig. \ref{Fig:thermooptic}(a) and (b), a COMSOL Multiphysics simulation of the thermal distribution in a slotted double-disk is shown. The tethers present a bottleneck to thermal dissipation, so that the temperature is mostly homogeneous in the outer annulus. This results primarily in in-plane expansion of the disks (which does not affect the disk separation), with out-of-plane deflection arising due to the differences in thermal expansion coefficients between Si and SiO2 at the points where the top and bottom disks connect to the $\alpha$-Si sacrificial layer and Si pedestal, respectively. 

While the intrinsic absorption of telecom photons in silica is very low, in practice the optical heating of the devices will be dominated by external parameters such as fabrication-induced defects or the presence of adsorbed water on the dielectric~\cite{vernooy_high-q_1998}. Hence the power dissipated as heat cannot be determined solely by analytical means. However, to investigate the potential of the thermo-optic and thermo-mechanical effects to act as the mechanism for tuning in this work, we will make the assumption that 1\% of the optical power is converted to heat, for reasons outlined below (i.e. $\alpha_\text{abs} = 0.01$). For this work, the injected power of the pump laser is set to $P_\text{in}=10$ mW. The contrast of the mode is 84\% (or $T = 0.16$) giving a maximum dropped power on resonance of $P_d = 8.4$ mW.  Therefore the power dissipated as heat is predicted to be $P_\text{heat} = P_d\times0.01 = 84 \; \mu$W. The effective thermal resistance of the device was found from simulations to be $R_\theta \simeq 6.5\times 10^5$ K/W, yielding a temperature change in the material of $\Delta T = P_d R_\theta \simeq 55$ K. This corresponds to a heating per milliwatt dropped power of 6.5~K/mW, consistent with the result of 10~K/mW dropped power reported  for silica double-disks with longer spokes, and hence larger thermal resistance, in Ref.~\cite{rosenberg_static_2009}.

\subsection{Thermo-optic effect}

The thermo-optic coefficient of silica at room temperature is $\frac{\text{d} n}{\text{d} T} \simeq 1 \times 10^{-5} \; \text{K}^{-1}$~\cite{gao_investigation_2018}. Therefore with the assumption of $\alpha_\text{abs} = 0.01$ and $P_d = 8.4$ mW as above, a refractive index change in the silica disks of $\Delta n_\text{SiO2}= \frac{dn}{dT} \Delta T = 5\times 10^{-4}$ is expected. This does not directly correspond to a shift in the optical resonance through Eq. \ref{Eq:Resonancecondition}, since a proportion of the optical mode resides in the space between the disks, which does not experience a refractive index change. Instead, simulations provide a value for the shift of the optical resonance as a function of refractive index $\text{d} \lambda/\text{d} n_\text{SiO2} \simeq 700$ nm\footnote{This is approximately linear for small changes in refractive index and is simulated using the correct parameters for our geometry, not taking into account stress-induced warping.}, yielding an estimate of the thermo-optic shift in the optical resonance frequency of $\Delta \lambda_\text{TO} = 0.35$ nm.

\begin{table*}[t!]
    \begin{tabular}{l c c c c }
    \hline
      Parameter   &  Symbol & Value & Unit & Source\\
      \hline
      WGM intrinsic energy decay rate & $\kappa_i$ & $2\pi \times 2\times10^{9}$ & rad/s & measurement \\
      WGM extrinsic energy decay rate & $\kappa_{\mathrm{ex}} $ & $2\pi \times 2\times10^{9}$ & rad/s & measurement \\
      Double disk flapping mode effective mass & $m_{\mathrm{eff}} $ & $5.2\times 10^{-13}$ & kg & FEM\\
      Mechanical mode frequency & $\Omega_M/2\pi$ & 700 & kHz & FEM\\
      Mechanical mode decay rate & $\Gamma/2\pi$ & $175$ & kHz & \cite{lin_mechanical_2009}\\
      Fraction of intrinsic losses dissipated as heat   & $\alpha_{abs}$ & 0.01 & - & fit\\
      Optomechanical coupling rate & $G/2\pi$ & $-10$ & GHz/nm & FEM~\cite{bekker_free_2018}\\
      Double disk spring constant & $k$ & 10 & N/m & FEM\\
      Silica thermo-optic coefficient & $\frac{dn}{dT}$ & $1\times10^{-5}$ & K$^{-1}$ &\cite{noauthor_comsol_nodate}\\
     Resonator thermal conductance & $G_{\mathrm{th}}$ & $1.54 \times 10^{-6}$ & W/K & FEM\\
      Resonator thermal mass & $m$ & $4.2\times 10^{-11}$ & kg & FEM \\
      Thermo-mechanical coefficient & $\alpha_{\mathrm{mech}}$ & $225$ & pm/K & fit\\
         \hline
    \end{tabular}
    \caption{Physical parameters used in the simulation. FEM:Finite Element modelling. Mechanical decay rate $\Gamma$ is chosen to give a squeeze-film limited quality factor of $\sim 4$, see \cite{lin_mechanical_2009}. }
    \label{tableODEparameters}
\end{table*}

\subsection{Radiation pressure force}

The tuning range of the optical resonance frequency due to radiation pressure can be calculated using the equation $\Delta \omega = G \Delta x$, where $G/2\pi \simeq 10$~GHz/nm~\cite{bekker_free_2018} is the optomechanical coupling strength for an double-disk device with an air gap of 600 nm (obtained through COMSOL Multiphysics simulations, see Fig. \ref{Fig:thermooptic}(c)) and $\Delta x$ is the mechanical out-of-plane displacement. From Hooke's law $\Delta x = F_\text{rad}/k$, where $k$ is the effective spring constant of the double-disk system and the radiation pressure force $F_\text{rad} = \hbar G N_\text{cav}$. Finally we substitute the equation for the intracavity photon number $N_\text{cav} = \frac{Q_i P_d}{\hbar \omega_0^2}$, where $Q_i$ denotes the intrinsic quality factor of the optical mode, $\omega_0$ is the natural optomechanical resonance frequency and $P_d$ is the power dropped into the cavity, as defined in previous sections. Therefore the equation for the maximal frequency shift due to radiation pressure becomes:

\begin{equation}
\Delta \omega = - \frac{Q_i G^2}{\omega_0^2 k} P_d
\end{equation}

 The spring constant is obtained via finite-element simulations (COMSOL), through applying a force $F_\text{rad}$ to each disk (in opposite directions), and measuring the corresponding change in the spacing between the disks. Note that this measure for $k$, as opposed to the deformation in a single disk due to an applied force, is the origin of the factor of 2 difference between the equation presented here and that found in the literature~\cite{wiederhecker_broadband_2011, rosenberg_static_2009}. For out-of-plane bending, the simulations for this geometry yield $k=10$ N/m. Further we set $Q_i\simeq 1\times10^5$ and $\omega_0/2\pi = 1.947\times 10^{14}$ Hz to reflect a typical set of values from experiments. Therefore the predicted frequency tuning range through the radiation pressure force is $\Delta \omega \simeq 220$ GHz. Alternatively, this can be expressed in terms of wavelength tuning as $\Delta \lambda_\text{RP} \simeq 1.7$ nm.
 
Comparing the predicted optical resonance frequency tuning ranges due to the thermo-optic and radiation-pressure effects ($\Delta \lambda_\text{TO} \simeq 0.35$ nm and $\Delta \lambda_\text{RP} \simeq 1.7$ nm respectively) we find that they are are similar and within the same order of magnitude as the observed optical tuning. Due to the large uncertainty in the parameters for calculating these values, in particular the absorbed power $\alpha_\text{abs}$ and optomechanical coupling strength $G$, it is difficult to pinpoint whether one effect is dominant for a given device, and it is possible for this to vary between devices.

\subsection{Thermo-mechanical effect}

The thermal expansion of the double disks leads to a temperature-dependent change in the disk spacing, which in turn affects the effective refractive index of the optical mode. The displacement of the disks can be converted to a shift in the resonant frequency through $\Delta \omega = G \Delta x$, where $G = 10$~GHz is the optomechanical coupling strength and $\Delta x$ is the change in gap spacing, as defined earlier. In order to find $\Delta x$, the therm-mechanical displacement coefficient $\alpha_\text{mech} = \text{d}x/\text{d}T$ has to be found, however simulations in COMSOL Multiphysics yield widely varying results for different geometric parameters. Further, simulations cannot take into account warping effects in the disks, increasing the uncertainty in this parameter. Therefore a predicted parameter for the thermo-mechanical tuning range could not be determined.

Instead, we observe that for some devices self-pulsing behavior is observed while both the thermo-optic effect and radiation pressure force are red-shifting. This implies that the blue-shifting effect, namely the thermo-mechanical force, must be comparable to the other effects in order for self-pulsing to occur. Therefore in our model of this effect we designate the thermo-mechanical coefficient $\alpha_\text{mech}$ as a fitting parameter to successfully simulate self-pulsing. This is discussed further in Section \ref{Sec:Model}.

\begin{figure*}[t]
    \centering
    \includegraphics[width= 0.8\linewidth]{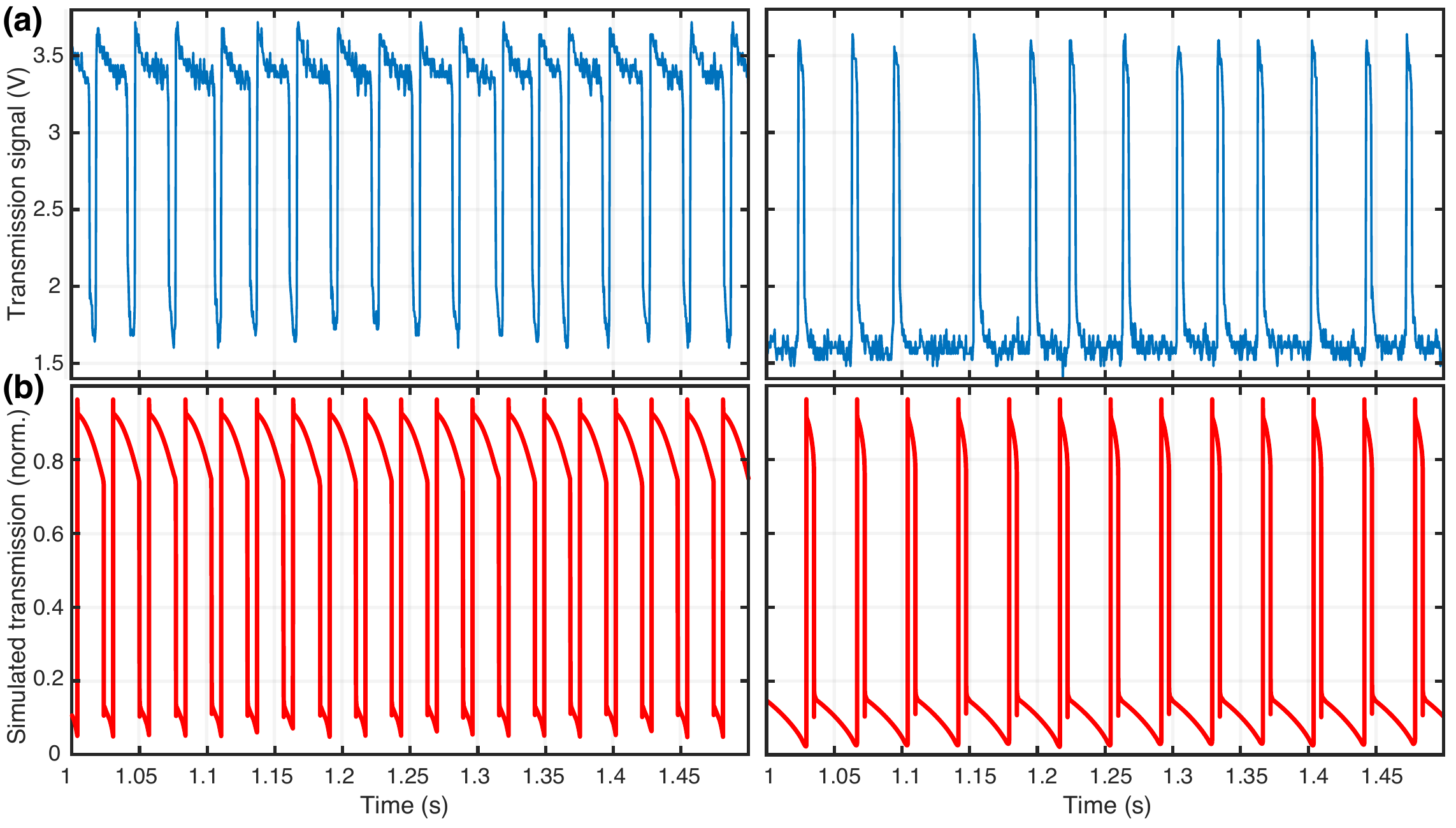}
    \caption{Comparison between the (a) experimental self-pulsing time traces (blue) and (b) numerical simulations (red), showing good qualitative agreement in terms of the dependency of frequency and duty cycle with laser power and detuning. The numerical script used to generate the red traces is available in the Supplementary Material~\cite{supplement}. Laser power P=2 mW. Detuning $\Delta \omega_0/2\pi$ respectively 8 GHz (left) and -31 GHz (right).}
    \label{fig:comparisonExperimentODE}
\end{figure*}

\section{Self-pulsing model}
\label{Sec:Model}

The dynamical behavior of the devices may be described by three coupled differential equations relating to the intracavity photon number $n_c$, the mechanical separation of the disks $x$ and their temperature $T_D$. Each of these parameters respectively responds on a characteristic timescale of $1/\kappa$; $1/\Gamma$  and $\tau_{th}$ (which are respectively here on the order of nanoseconds, microseconds and milliseconds). Since the optical decay rate $\kappa$ is much larger than all other decay rates, we consider that the intracavity photon number $n_c$ reacts instantaneously to any changes in the cavity, such that it takes the steady-state form~\cite{aspelmeyer_cavity_2014}: 
\begin{equation}
    n_c=\frac{\kappa_{\mathrm{ex}}}{\Delta^2+\left(\kappa/2\right)^2}\,\frac{P}{\hbar \, \omega_l},
\end{equation}
where $\Delta(T_D,x)=\omega_l-\omega_0(T_D,x)=\Delta_0(T_0,0)-G x + \frac{\partial \omega_0}{\partial n} \frac{d n}{d T} (T_D-T_0)$ is the temperature and position-dependent laser detuning and $T_0$ the background temperature. The dynamics can thus be reduced to the two following coupled equations of motion:

\begin{equation}
    m_{\mathrm{eff}}\,\ddot{x}+k \,x+\Gamma\, m_{\mathrm{eff}}\, \dot{x}=F_{\mathrm{th}} + F_{\mathrm{rp}},
\end{equation}
(where $m_{\mathrm{eff}}$ corresponds to the double disk fundamental flapping mode's effective mass, $F_{\mathrm{rp}}=-n_c \hbar \lvert G\rvert$ the attractive radiation pressure force acting upon the disks~\cite{rosenberg_static_2009,baker_photoelastic_2014,aspelmeyer_cavity_2014} and $F_{\mathrm{th}}=\alpha_{\mathrm{mech}}\,k\,\left(T_D-T_0\right)$ the thermo-mechanical force where $\alpha_{\mathrm{mech}}$ describes the temperature induced disk deflection in meters per Kelvin), and:
\begin{equation}
    \frac{dT_D}{dt}=\frac{n_c\,\hbar\, \omega_l\, \kappa_i\, \alpha_{\mathrm{abs}}}{m\,c}-\frac{G_{\mathrm{th}}\left(T_D-T_0\right)}{m\,c}.
\end{equation}
Here, $\alpha_{\mathrm{abs}}\in [0,1]$ corresponds to the fraction of the intrinsic losses dissipated as heat in the resonator, $m$ to the resonator's thermal mass and $c$ its specific heat, and $G_{\mathrm{th}}=\frac{m\,c}{\tau_{\mathrm{th}}}$ the resonator's thermal conductance to the substrate~\cite{baker_optical_2012}. In this model, thermo-optic and thermo-mechanical effects are thus effectively lumped together (unlike in Ref. \cite{baker_optical_2012}), resulting in an effective temperature-dependent frequency shift evolving with a characteristic timescale $\tau_{\mathrm{th}}$. The magnitude of the thermo-mechanical effect being larger than the thermo-optic contribution, this results in an effective blue-shifting contribution, which competes with the red-shifting radiation-pressure contribution acting on a faster timescale. Solving these coupled equations of motion using an ODE solver (see Supplementary Material~\cite{supplement}) and the parameters outlined in Table \ref{tableODEparameters}, we obtain the red time traces shown in red in Fig. \ref{fig:comparisonExperimentODE}. These qualitatively reproduce the observed self-pulsing behavior, as well as its frequency and duty cycle dependence with laser power and laser detuning.


%


\begin{thebibliography}{97}%
\makeatletter
\providecommand \@ifxundefined [1]{%
 \@ifx{#1\undefined}
}%
\providecommand \@ifnum [1]{%
 \ifnum #1\expandafter \@firstoftwo
 \else \expandafter \@secondoftwo
 \fi
}%
\providecommand \@ifx [1]{%
 \ifx #1\expandafter \@firstoftwo
 \else \expandafter \@secondoftwo
 \fi
}%
\providecommand \natexlab [1]{#1}%
\providecommand \enquote  [1]{``#1''}%
\providecommand \bibnamefont  [1]{#1}%
\providecommand \bibfnamefont [1]{#1}%
\providecommand \citenamefont [1]{#1}%
\providecommand \href@noop [0]{\@secondoftwo}%
\providecommand \href [0]{\begingroup \@sanitize@url \@href}%
\providecommand \@href[1]{\@@startlink{#1}\@@href}%
\providecommand \@@href[1]{\endgroup#1\@@endlink}%
\providecommand \@sanitize@url [0]{\catcode `\\12\catcode `\$12\catcode
  `\&12\catcode `\#12\catcode `\^12\catcode `\_12\catcode `\%12\relax}%
\providecommand \@@startlink[1]{}%
\providecommand \@@endlink[0]{}%
\providecommand \url  [0]{\begingroup\@sanitize@url \@url }%
\providecommand \@url [1]{\endgroup\@href {#1}{\urlprefix }}%
\providecommand \urlprefix  [0]{URL }%
\providecommand \Eprint [0]{\href }%
\providecommand \doibase [0]{https://doi.org/}%
\providecommand \selectlanguage [0]{\@gobble}%
\providecommand \bibinfo  [0]{\@secondoftwo}%
\providecommand \bibfield  [0]{\@secondoftwo}%
\providecommand \translation [1]{[#1]}%
\providecommand \BibitemOpen [0]{}%
\providecommand \bibitemStop [0]{}%
\providecommand \bibitemNoStop [0]{.\EOS\space}%
\providecommand \EOS [0]{\spacefactor3000\relax}%
\providecommand \BibitemShut  [1]{\csname bibitem#1\endcsname}%
\let\auto@bib@innerbib\@empty
\bibitem [{\citenamefont {Atabaki}\ \emph {et~al.}(2018)\citenamefont
  {Atabaki}, \citenamefont {Moazeni}, \citenamefont {Pavanello}, \citenamefont
  {Gevorgyan}, \citenamefont {Notaros}, \citenamefont {Alloatti}, \citenamefont
  {Wade}, \citenamefont {Sun}, \citenamefont {Kruger}, \citenamefont {Meng},
  \citenamefont {Qubaisi}, \citenamefont {Wang}, \citenamefont {Zhang},
  \citenamefont {Khilo}, \citenamefont {Baiocco}, \citenamefont {Popović},
  \citenamefont {Stojanović},\ and\ \citenamefont
  {Ram}}]{atabaki_integrating_2018}%
  \BibitemOpen
  \bibfield  {author} {\bibinfo {author} {\bibfnamefont {A.~H.}\ \bibnamefont
  {Atabaki}}, \bibinfo {author} {\bibfnamefont {S.}~\bibnamefont {Moazeni}},
  \bibinfo {author} {\bibfnamefont {F.}~\bibnamefont {Pavanello}}, \bibinfo
  {author} {\bibfnamefont {H.}~\bibnamefont {Gevorgyan}}, \bibinfo {author}
  {\bibfnamefont {J.}~\bibnamefont {Notaros}}, \bibinfo {author} {\bibfnamefont
  {L.}~\bibnamefont {Alloatti}}, \bibinfo {author} {\bibfnamefont {M.~T.}\
  \bibnamefont {Wade}}, \bibinfo {author} {\bibfnamefont {C.}~\bibnamefont
  {Sun}}, \bibinfo {author} {\bibfnamefont {S.~A.}\ \bibnamefont {Kruger}},
  \bibinfo {author} {\bibfnamefont {H.}~\bibnamefont {Meng}}, \bibinfo {author}
  {\bibfnamefont {K.~A.}\ \bibnamefont {Qubaisi}}, \bibinfo {author}
  {\bibfnamefont {I.}~\bibnamefont {Wang}}, \bibinfo {author} {\bibfnamefont
  {B.}~\bibnamefont {Zhang}}, \bibinfo {author} {\bibfnamefont
  {A.}~\bibnamefont {Khilo}}, \bibinfo {author} {\bibfnamefont {C.~V.}\
  \bibnamefont {Baiocco}}, \bibinfo {author} {\bibfnamefont {M.~A.}\
  \bibnamefont {Popović}}, \bibinfo {author} {\bibfnamefont {V.~M.}\
  \bibnamefont {Stojanović}},\ and\ \bibinfo {author} {\bibfnamefont {R.~J.}\
  \bibnamefont {Ram}},\ }\bibfield  {title} {{\bibinfo
  {title} {Integrating photonics with silicon nanoelectronics for the next
  generation of systems on a chip}},\ }\href
  {https://doi.org/10.1038/s41586-018-0028-z} {\bibfield  {journal} {\bibinfo
  {journal} {Nature}\ }\textbf {\bibinfo {volume} {556}},\ \bibinfo {pages}
  {349} (\bibinfo {year} {2018})}\BibitemShut {NoStop}%
\bibitem [{\citenamefont {Elshaari}\ \emph {et~al.}(2017)\citenamefont
  {Elshaari}, \citenamefont {Zadeh}, \citenamefont {Fognini}, \citenamefont
  {Reimer}, \citenamefont {Dalacu}, \citenamefont {Poole}, \citenamefont
  {Zwiller},\ and\ \citenamefont {Jöns}}]{elshaari_-chip_2017}%
  \BibitemOpen
  \bibfield  {author} {\bibinfo {author} {\bibfnamefont {A.~W.}\ \bibnamefont
  {Elshaari}}, \bibinfo {author} {\bibfnamefont {I.~E.}\ \bibnamefont {Zadeh}},
  \bibinfo {author} {\bibfnamefont {A.}~\bibnamefont {Fognini}}, \bibinfo
  {author} {\bibfnamefont {M.~E.}\ \bibnamefont {Reimer}}, \bibinfo {author}
  {\bibfnamefont {D.}~\bibnamefont {Dalacu}}, \bibinfo {author} {\bibfnamefont
  {P.~J.}\ \bibnamefont {Poole}}, \bibinfo {author} {\bibfnamefont
  {V.}~\bibnamefont {Zwiller}},\ and\ \bibinfo {author} {\bibfnamefont {K.~D.}\
  \bibnamefont {Jöns}},\ }\bibfield  {title} {{\bibinfo
  {title} {On-chip single photon filtering and multiplexing in hybrid quantum
  photonic circuits}},\ }\href {https://doi.org/10.1038/s41467-017-00486-8}
  {\bibfield  {journal} {\bibinfo  {journal} {Nature Communications}\ }\textbf
  {\bibinfo {volume} {8}},\ \bibinfo {pages} {379} (\bibinfo {year}
  {2017})}\BibitemShut {NoStop}%
\bibitem [{\citenamefont {Fujita}\ and\ \citenamefont
  {Baba}(2002)}]{fujita_microgear_2002}%
  \BibitemOpen
  \bibfield  {author} {\bibinfo {author} {\bibfnamefont {M.}~\bibnamefont
  {Fujita}}\ and\ \bibinfo {author} {\bibfnamefont {T.}~\bibnamefont {Baba}},\
  }\bibfield  {title} {\bibinfo {title} {Microgear laser},\ }\href
  {https://doi.org/10.1063/1.1462867} {\bibfield  {journal} {\bibinfo
  {journal} {Applied Physics Letters}\ }\textbf {\bibinfo {volume} {80}},\
  \bibinfo {pages} {2051} (\bibinfo {year} {2002})}\BibitemShut {NoStop}%
\bibitem [{\citenamefont {Yang}\ \emph {et~al.}(2003)\citenamefont {Yang},
  \citenamefont {Armani},\ and\ \citenamefont
  {Vahala}}]{yang_fiber-coupled_2003}%
  \BibitemOpen
  \bibfield  {author} {\bibinfo {author} {\bibfnamefont {L.}~\bibnamefont
  {Yang}}, \bibinfo {author} {\bibfnamefont {D.~K.}\ \bibnamefont {Armani}},\
  and\ \bibinfo {author} {\bibfnamefont {K.~J.}\ \bibnamefont {Vahala}},\
  }\bibfield  {title} {\bibinfo {title} {Fiber-coupled erbium microlasers on a
  chip},\ }\href {https://doi.org/10.1063/1.1598623} {\bibfield  {journal}
  {\bibinfo  {journal} {Applied Physics Letters}\ }\textbf {\bibinfo {volume}
  {83}},\ \bibinfo {pages} {825} (\bibinfo {year} {2003})}\BibitemShut
  {NoStop}%
\bibitem [{\citenamefont {Lu}\ \emph {et~al.}(2009)\citenamefont {Lu},
  \citenamefont {Yang}, \citenamefont {van Loon}, \citenamefont {Polman},\ and\
  \citenamefont {Vahala}}]{lu_-chip_2009}%
  \BibitemOpen
  \bibfield  {author} {\bibinfo {author} {\bibfnamefont {T.}~\bibnamefont
  {Lu}}, \bibinfo {author} {\bibfnamefont {L.}~\bibnamefont {Yang}}, \bibinfo
  {author} {\bibfnamefont {R.~V.~A.}\ \bibnamefont {van Loon}}, \bibinfo
  {author} {\bibfnamefont {A.}~\bibnamefont {Polman}},\ and\ \bibinfo {author}
  {\bibfnamefont {K.~J.}\ \bibnamefont {Vahala}},\ }\bibfield  {title}
  {{\bibinfo {title} {On-chip green silica upconversion
  microlaser}},\ }\href {https://doi.org/10.1364/OL.34.000482} {\bibfield
  {journal} {\bibinfo  {journal} {Optics Letters}\ }\textbf {\bibinfo {volume}
  {34}},\ \bibinfo {pages} {482} (\bibinfo {year} {2009})}\BibitemShut
  {NoStop}%
\bibitem [{\citenamefont {Mehrabani}\ and\ \citenamefont
  {Armani}(2013)}]{mehrabani_blue_2013}%
  \BibitemOpen
  \bibfield  {author} {\bibinfo {author} {\bibfnamefont {S.}~\bibnamefont
  {Mehrabani}}\ and\ \bibinfo {author} {\bibfnamefont {A.~M.}\ \bibnamefont
  {Armani}},\ }\bibfield  {title} {{\bibinfo {title} {Blue
  upconversion laser based on thulium-doped silica m icrocavity}},\ }\href
  {https://doi.org/10.1364/OL.38.004346} {\bibfield  {journal} {\bibinfo
  {journal} {Optics Letters}\ }\textbf {\bibinfo {volume} {38}},\ \bibinfo
  {pages} {4346} (\bibinfo {year} {2013})}\BibitemShut {NoStop}%
\bibitem [{\citenamefont {Park}\ \emph {et~al.}(2014)\citenamefont {Park},
  \citenamefont {Ozdemir}, \citenamefont {Monifi}, \citenamefont {Chadha},
  \citenamefont {Huang}, \citenamefont {Biswas},\ and\ \citenamefont
  {Yang}}]{park_titanium_2014}%
  \BibitemOpen
  \bibfield  {author} {\bibinfo {author} {\bibfnamefont {J.}~\bibnamefont
  {Park}}, \bibinfo {author} {\bibfnamefont {S.~K.}\ \bibnamefont {Ozdemir}},
  \bibinfo {author} {\bibfnamefont {F.}~\bibnamefont {Monifi}}, \bibinfo
  {author} {\bibfnamefont {T.}~\bibnamefont {Chadha}}, \bibinfo {author}
  {\bibfnamefont {S.~H.}\ \bibnamefont {Huang}}, \bibinfo {author}
  {\bibfnamefont {P.}~\bibnamefont {Biswas}},\ and\ \bibinfo {author}
  {\bibfnamefont {L.}~\bibnamefont {Yang}},\ }\bibfield  {title} {\bibinfo
  {title} {Titanium {Dioxide} {Whispering} {Gallery} {Microcavities}},\ }\href
  {https://doi.org/10.1002/adom.201400107} {\bibfield  {journal} {\bibinfo
  {journal} {Advanced Optical Materials}\ }\textbf {\bibinfo {volume} {2}},\
  \bibinfo {pages} {711} (\bibinfo {year} {2014})}\BibitemShut {NoStop}%
\bibitem [{\citenamefont {Yang}\ \emph {et~al.}(2017)\citenamefont {Yang},
  \citenamefont {Lei}, \citenamefont {Kasumie}, \citenamefont {Xu},
  \citenamefont {Ward}, \citenamefont {Yang},\ and\ \citenamefont
  {Chormaic}}]{yang_tunable_2017}%
  \BibitemOpen
  \bibfield  {author} {\bibinfo {author} {\bibfnamefont {Y.}~\bibnamefont
  {Yang}}, \bibinfo {author} {\bibfnamefont {F.}~\bibnamefont {Lei}}, \bibinfo
  {author} {\bibfnamefont {S.}~\bibnamefont {Kasumie}}, \bibinfo {author}
  {\bibfnamefont {L.}~\bibnamefont {Xu}}, \bibinfo {author} {\bibfnamefont
  {J.~M.}\ \bibnamefont {Ward}}, \bibinfo {author} {\bibfnamefont
  {L.}~\bibnamefont {Yang}},\ and\ \bibinfo {author} {\bibfnamefont {S.~N.}\
  \bibnamefont {Chormaic}},\ }\bibfield  {title} {{\bibinfo
  {title} {Tunable erbium-doped microbubble laser fabricated by sol-gel
  coating}},\ }\href {https://doi.org/10.1364/OE.25.001308} {\bibfield
  {journal} {\bibinfo  {journal} {Optics Express}\ }\textbf {\bibinfo {volume}
  {25}},\ \bibinfo {pages} {1308} (\bibinfo {year} {2017})}\BibitemShut
  {NoStop}%
\bibitem [{\citenamefont {Verbert}\ \emph {et~al.}(2005)\citenamefont
  {Verbert}, \citenamefont {Mazen}, \citenamefont {Charvolin}, \citenamefont
  {Picard}, \citenamefont {Calvo}, \citenamefont {Noé}, \citenamefont
  {Gérard},\ and\ \citenamefont {Hadji}}]{verbert_efficient_2005}%
  \BibitemOpen
  \bibfield  {author} {\bibinfo {author} {\bibfnamefont {J.}~\bibnamefont
  {Verbert}}, \bibinfo {author} {\bibfnamefont {F.}~\bibnamefont {Mazen}},
  \bibinfo {author} {\bibfnamefont {T.}~\bibnamefont {Charvolin}}, \bibinfo
  {author} {\bibfnamefont {E.}~\bibnamefont {Picard}}, \bibinfo {author}
  {\bibfnamefont {V.}~\bibnamefont {Calvo}}, \bibinfo {author} {\bibfnamefont
  {P.}~\bibnamefont {Noé}}, \bibinfo {author} {\bibfnamefont {J.-M.}\
  \bibnamefont {Gérard}},\ and\ \bibinfo {author} {\bibfnamefont
  {E.}~\bibnamefont {Hadji}},\ }\bibfield  {title} {\bibinfo {title} {Efficient
  coupling of {Er}-doped silicon-rich oxide to microdisk whispering gallery
  modes},\ }\href {https://doi.org/10.1063/1.1883331} {\bibfield  {journal}
  {\bibinfo  {journal} {Applied Physics Letters}\ }\textbf {\bibinfo {volume}
  {86}},\ \bibinfo {pages} {111117} (\bibinfo {year} {2005})}\BibitemShut
  {NoStop}%
\bibitem [{\citenamefont {Polman}\ \emph {et~al.}(2004)\citenamefont {Polman},
  \citenamefont {Min}, \citenamefont {Kalkman}, \citenamefont {Kippenberg},\
  and\ \citenamefont {Vahala}}]{polman_ultralow-threshold_2004}%
  \BibitemOpen
  \bibfield  {author} {\bibinfo {author} {\bibfnamefont {A.}~\bibnamefont
  {Polman}}, \bibinfo {author} {\bibfnamefont {B.}~\bibnamefont {Min}},
  \bibinfo {author} {\bibfnamefont {J.}~\bibnamefont {Kalkman}}, \bibinfo
  {author} {\bibfnamefont {T.~J.}\ \bibnamefont {Kippenberg}},\ and\ \bibinfo
  {author} {\bibfnamefont {K.~J.}\ \bibnamefont {Vahala}},\ }\bibfield  {title}
  {{\bibinfo {title} {Ultralow-threshold erbium-implanted
  toroidal microlaser on silicon}},\ }\href {https://doi.org/10.1063/1.1646748}
  {\bibfield  {journal} {\bibinfo  {journal} {Applied Physics Letters}\
  }\textbf {\bibinfo {volume} {84}},\ \bibinfo {pages} {1037} (\bibinfo {year}
  {2004})}\BibitemShut {NoStop}%
\bibitem [{\citenamefont {Kalkman}\ \emph {et~al.}(2006)\citenamefont
  {Kalkman}, \citenamefont {Polman}, \citenamefont {Kippenberg}, \citenamefont
  {Vahala},\ and\ \citenamefont {Brongersma}}]{kalkman_erbium-implanted_2006}%
  \BibitemOpen
  \bibfield  {author} {\bibinfo {author} {\bibfnamefont {J.}~\bibnamefont
  {Kalkman}}, \bibinfo {author} {\bibfnamefont {A.}~\bibnamefont {Polman}},
  \bibinfo {author} {\bibfnamefont {T.}~\bibnamefont {Kippenberg}}, \bibinfo
  {author} {\bibfnamefont {K.}~\bibnamefont {Vahala}},\ and\ \bibinfo {author}
  {\bibfnamefont {M.~L.}\ \bibnamefont {Brongersma}},\ }\bibfield  {title}
  {{\bibinfo {title} {Erbium-implanted silica microsphere
  laser}},\ }\href {https://doi.org/10.1016/j.nimb.2005.08.160} {\bibfield
  {journal} {\bibinfo  {journal} {Nuclear Instruments and Methods in Physics
  Research Section B: Beam Interactions with Materials and Atoms}\ }\textbf
  {\bibinfo {volume} {242}},\ \bibinfo {pages} {182} (\bibinfo {year}
  {2006})}\BibitemShut {NoStop}%
\bibitem [{\citenamefont {Kippenberg}\ \emph {et~al.}(2006)\citenamefont
  {Kippenberg}, \citenamefont {Kalkman}, \citenamefont {Polman},\ and\
  \citenamefont {Vahala}}]{kippenberg_demonstration_2006}%
  \BibitemOpen
  \bibfield  {author} {\bibinfo {author} {\bibfnamefont {T.~J.}\ \bibnamefont
  {Kippenberg}}, \bibinfo {author} {\bibfnamefont {J.}~\bibnamefont {Kalkman}},
  \bibinfo {author} {\bibfnamefont {A.}~\bibnamefont {Polman}},\ and\ \bibinfo
  {author} {\bibfnamefont {K.~J.}\ \bibnamefont {Vahala}},\ }\bibfield  {title}
  {\bibinfo {title} {Demonstration of an erbium-doped microdisk laser on a
  silicon chip},\ }\href {https://doi.org/10.1103/PhysRevA.74.051802}
  {\bibfield  {journal} {\bibinfo  {journal} {Physical Review A}\ }\textbf
  {\bibinfo {volume} {74}},\ \bibinfo {pages} {051802(R)} (\bibinfo {year}
  {2006})}\BibitemShut {NoStop}%
\bibitem [{\citenamefont {Min}\ \emph {et~al.}(2004)\citenamefont {Min},
  \citenamefont {Kippenberg}, \citenamefont {Yang}, \citenamefont {Vahala},
  \citenamefont {Kalkman},\ and\ \citenamefont
  {Polman}}]{min_erbium-implanted_2004}%
  \BibitemOpen
  \bibfield  {author} {\bibinfo {author} {\bibfnamefont {B.}~\bibnamefont
  {Min}}, \bibinfo {author} {\bibfnamefont {T.~J.}\ \bibnamefont {Kippenberg}},
  \bibinfo {author} {\bibfnamefont {L.}~\bibnamefont {Yang}}, \bibinfo {author}
  {\bibfnamefont {K.~J.}\ \bibnamefont {Vahala}}, \bibinfo {author}
  {\bibfnamefont {J.}~\bibnamefont {Kalkman}},\ and\ \bibinfo {author}
  {\bibfnamefont {A.}~\bibnamefont {Polman}},\ }\bibfield  {title}
  {{\bibinfo {title} {Erbium-implanted high- {Q} silica
  toroidal microcavity laser on a silicon chip}},\ }\href
  {https://link.aps.org/doi/10.1103/PhysRevA.70.033803} {\bibfield  {journal}
  {\bibinfo  {journal} {Physical Review A}\ }\textbf {\bibinfo {volume} {70}},\
  \bibinfo {pages} {033803} (\bibinfo {year} {2004})}\BibitemShut {NoStop}%
\bibitem [{\citenamefont {Sandoghdar}\ \emph {et~al.}(1996)\citenamefont
  {Sandoghdar}, \citenamefont {Treussart}, \citenamefont {Hare}, \citenamefont
  {Lefevre-Seguin}, \citenamefont {Raimond},\ and\ \citenamefont
  {Haroche}}]{sandoghdar_very_1996}%
  \BibitemOpen
  \bibfield  {author} {\bibinfo {author} {\bibfnamefont {V.}~\bibnamefont
  {Sandoghdar}}, \bibinfo {author} {\bibfnamefont {F.}~\bibnamefont
  {Treussart}}, \bibinfo {author} {\bibfnamefont {J.}~\bibnamefont {Hare}},
  \bibinfo {author} {\bibfnamefont {V.}~\bibnamefont {Lefèvre-Seguin}},
  \bibinfo {author} {\bibfnamefont {J.~M.}\ \bibnamefont {Raimond}},\ and\
  \bibinfo {author} {\bibfnamefont {S.}~\bibnamefont {Haroche}},\ }\bibfield
  {title} {{\bibinfo {title} {Very low threshold
  whispering-gallery-mode microsphere laser}},\ }\href
  {https://doi.org/10.1103/PhysRevA.54.R1777} {\bibfield  {journal} {\bibinfo
  {journal} {Physical Review A}\ }\textbf {\bibinfo {volume} {54}},\ \bibinfo
  {pages} {R1777} (\bibinfo {year} {1996})}\BibitemShut {NoStop}%
\bibitem [{\citenamefont {von Klitzing}\ \emph {et~al.}(2000)\citenamefont {von
  Klitzing}, \citenamefont {Jahier}, \citenamefont {Long}, \citenamefont
  {Lissillour}, \citenamefont {Lefèvre-Seguin}, \citenamefont {Hare},
  \citenamefont {Raimond},\ and\ \citenamefont
  {Haroche}}]{von_klitzing_very_2000}%
  \BibitemOpen
  \bibfield  {author} {\bibinfo {author} {\bibfnamefont {W.}~\bibnamefont {von
  Klitzing}}, \bibinfo {author} {\bibfnamefont {E.}~\bibnamefont {Jahier}},
  \bibinfo {author} {\bibfnamefont {R.}~\bibnamefont {Long}}, \bibinfo {author}
  {\bibfnamefont {F.}~\bibnamefont {Lissillour}}, \bibinfo {author}
  {\bibfnamefont {V.}~\bibnamefont {Lefèvre-Seguin}}, \bibinfo {author}
  {\bibfnamefont {J.}~\bibnamefont {Hare}}, \bibinfo {author} {\bibfnamefont
  {J.-M.}\ \bibnamefont {Raimond}},\ and\ \bibinfo {author} {\bibfnamefont
  {S.}~\bibnamefont {Haroche}},\ }\bibfield  {title} {{\bibinfo {title} {Very low threshold green lasing in microspheres by
  up-conversion of {IR} photons}},\ }\href
  {https://doi.org/10.1088/1464-4266/2/2/324} {\bibfield  {journal} {\bibinfo
  {journal} {Journal of Optics B: Quantum and Semiclassical Optics}\ }\textbf
  {\bibinfo {volume} {2}},\ \bibinfo {pages} {204} (\bibinfo {year}
  {2000})}\BibitemShut {NoStop}%
\bibitem [{\citenamefont {Zhu}\ \emph {et~al.}(2018)\citenamefont {Zhu},
  \citenamefont {Shi}, \citenamefont {Xiao}, \citenamefont {Zhang},\ and\
  \citenamefont {Fan}}]{zhu_all-optical_2018}%
  \BibitemOpen
  \bibfield  {author} {\bibinfo {author} {\bibfnamefont {S.}~\bibnamefont
  {Zhu}}, \bibinfo {author} {\bibfnamefont {L.}~\bibnamefont {Shi}}, \bibinfo
  {author} {\bibfnamefont {B.}~\bibnamefont {Xiao}}, \bibinfo {author}
  {\bibfnamefont {X.}~\bibnamefont {Zhang}},\ and\ \bibinfo {author}
  {\bibfnamefont {X.}~\bibnamefont {Fan}},\ }\bibfield  {title} {\bibinfo
  {title} {All-{Optical} {Tunable} {Microlaser} {Based} on an {Ultrahigh}-{Q}
  {Erbium}-{Doped} {Hybrid} {Microbottle} {Cavity}},\ }\href
  {https://doi.org/10.1021/acsphotonics.8b00838} {\bibfield  {journal}
  {\bibinfo  {journal} {ACS Photonics}\ }\textbf {\bibinfo {volume} {5}},\
  \bibinfo {pages} {3794} (\bibinfo {year} {2018})}\BibitemShut {NoStop}%
\bibitem [{\citenamefont {McCall}\ \emph {et~al.}(1998)\citenamefont {McCall},
  \citenamefont {Levi}, \citenamefont {Slusher}, \citenamefont {Pearton},\ and\
  \citenamefont {Logan}}]{mccall_whisperinggallery_1998}%
  \BibitemOpen
  \bibfield  {author} {\bibinfo {author} {\bibfnamefont {S.~L.}\ \bibnamefont
  {McCall}}, \bibinfo {author} {\bibfnamefont {A.~F.~J.}\ \bibnamefont {Levi}},
  \bibinfo {author} {\bibfnamefont {R.~E.}\ \bibnamefont {Slusher}}, \bibinfo
  {author} {\bibfnamefont {S.~J.}\ \bibnamefont {Pearton}},\ and\ \bibinfo
  {author} {\bibfnamefont {R.~A.}\ \bibnamefont {Logan}},\ }\bibfield  {title}
  {\bibinfo {title} {Whispering‐gallery mode microdisk lasers},\ }\href
  {https://doi.org/10.1063/1.106688@apl.2019.APLCLASS2019.issue-1} {\bibfield
  {journal} {\bibinfo  {journal} {Applied Physics Letters}\ }\textbf {\bibinfo
  {volume} {60}},\ \bibinfo {pages} {289} (\bibinfo {year} {1998})}\BibitemShut
  {NoStop}%
\bibitem [{\citenamefont {Princepe}\ \emph {et~al.}(2018)\citenamefont
  {Princepe}, \citenamefont {Wiederhecker}, \citenamefont {Favero},\ and\
  \citenamefont {Frateschi}}]{princepe_self-sustained_2018}%
  \BibitemOpen
  \bibfield  {author} {\bibinfo {author} {\bibfnamefont {D.}~\bibnamefont
  {Princepe}}, \bibinfo {author} {\bibfnamefont {G.~S.}\ \bibnamefont
  {Wiederhecker}}, \bibinfo {author} {\bibfnamefont {I.}~\bibnamefont
  {Favero}},\ and\ \bibinfo {author} {\bibfnamefont {N.~C.}\ \bibnamefont
  {Frateschi}},\ }\bibfield  {title} {\bibinfo {title} {Self-{Sustained}
  {Laser} {Pulsation} in {Active} {Optomechanical} {Devices}},\ }\href
  {https://doi.org/10.1109/JPHOT.2018.2831001} {\bibfield  {journal} {\bibinfo
  {journal} {IEEE Photonics Journal}\ }\textbf {\bibinfo {volume} {10}},\
  \bibinfo {pages} {1} (\bibinfo {year} {2018})}\BibitemShut {NoStop}%
\bibitem [{\citenamefont {Lin}\ \emph {et~al.}(1986)\citenamefont {Lin},
  \citenamefont {Huston}, \citenamefont {Justus},\ and\ \citenamefont
  {Campillo}}]{lin_characteristics_1986}%
  \BibitemOpen
  \bibfield  {author} {\bibinfo {author} {\bibfnamefont {H.-B.}\ \bibnamefont
  {Lin}}, \bibinfo {author} {\bibfnamefont {A.~L.}\ \bibnamefont {Huston}},
  \bibinfo {author} {\bibfnamefont {B.~L.}\ \bibnamefont {Justus}},\ and\
  \bibinfo {author} {\bibfnamefont {A.~J.}\ \bibnamefont {Campillo}},\
  }\bibfield  {title} {{\bibinfo {title} {Some
  characteristics of a droplet whispering-gallery-mode laser}},\ }\href
  {https://doi.org/10.1364/OL.11.000614} {\bibfield  {journal} {\bibinfo
  {journal} {Optics Letters}\ }\textbf {\bibinfo {volume} {11}},\ \bibinfo
  {pages} {614} (\bibinfo {year} {1986})}\BibitemShut {NoStop}%
\bibitem [{\citenamefont {Kuwata-Gonokami}\ \emph {et~al.}(1995)\citenamefont
  {Kuwata-Gonokami}, \citenamefont {Jordan}, \citenamefont {Dodabalapur},
  \citenamefont {Katz}, \citenamefont {Schilling}, \citenamefont {Slusher},\
  and\ \citenamefont {Ozawa}}]{kuwata-gonokami_polymer_1995}%
  \BibitemOpen
  \bibfield  {author} {\bibinfo {author} {\bibfnamefont {M.}~\bibnamefont
  {Kuwata-Gonokami}}, \bibinfo {author} {\bibfnamefont {R.~H.}\ \bibnamefont
  {Jordan}}, \bibinfo {author} {\bibfnamefont {A.}~\bibnamefont {Dodabalapur}},
  \bibinfo {author} {\bibfnamefont {H.~E.}\ \bibnamefont {Katz}}, \bibinfo
  {author} {\bibfnamefont {M.~L.}\ \bibnamefont {Schilling}}, \bibinfo {author}
  {\bibfnamefont {R.~E.}\ \bibnamefont {Slusher}},\ and\ \bibinfo {author}
  {\bibfnamefont {S.}~\bibnamefont {Ozawa}},\ }\bibfield  {title}
  {{\bibinfo {title} {Polymer microdisk and microring
  lasers}},\ }\href {https://doi.org/10.1364/OL.20.002093} {\bibfield
  {journal} {\bibinfo  {journal} {Optics Letters}\ }\textbf {\bibinfo {volume}
  {20}},\ \bibinfo {pages} {2093} (\bibinfo {year} {1995})}\BibitemShut
  {NoStop}%
\bibitem [{\citenamefont {Knight}\ \emph {et~al.}(1992)\citenamefont {Knight},
  \citenamefont {Driver}, \citenamefont {Hutcheon},\ and\ \citenamefont
  {Robertson}}]{knight_core-resonance_1992}%
  \BibitemOpen
  \bibfield  {author} {\bibinfo {author} {\bibfnamefont {J.~C.}\ \bibnamefont
  {Knight}}, \bibinfo {author} {\bibfnamefont {H.~S.~T.}\ \bibnamefont
  {Driver}}, \bibinfo {author} {\bibfnamefont {R.~J.}\ \bibnamefont
  {Hutcheon}},\ and\ \bibinfo {author} {\bibfnamefont {G.~N.}\ \bibnamefont
  {Robertson}},\ }\bibfield  {title} {{\bibinfo {title}
  {Core-resonance capillary-fiber whispering-gallery-mode laser}},\ }\href
  {https://doi.org/10.1364/OL.17.001280} {\bibfield  {journal} {\bibinfo
  {journal} {Optics Letters}\ }\textbf {\bibinfo {volume} {17}},\ \bibinfo
  {pages} {1280} (\bibinfo {year} {1992})}\BibitemShut {NoStop}%
\bibitem [{\citenamefont {Siegle}\ \emph {et~al.}(2017)\citenamefont {Siegle},
  \citenamefont {Remmel}, \citenamefont {Krämmer},\ and\ \citenamefont
  {Kalt}}]{siegle_split-disk_2017}%
  \BibitemOpen
  \bibfield  {author} {\bibinfo {author} {\bibfnamefont {T.}~\bibnamefont
  {Siegle}}, \bibinfo {author} {\bibfnamefont {M.}~\bibnamefont {Remmel}},
  \bibinfo {author} {\bibfnamefont {S.}~\bibnamefont {Krämmer}},\ and\
  \bibinfo {author} {\bibfnamefont {H.}~\bibnamefont {Kalt}},\ }\bibfield
  {title} {\bibinfo {title} {Split-disk micro-lasers: Tunable whispering
  gallery mode cavities},\ }\href {https://doi.org/10.1063/1.4985766}
  {\bibfield  {journal} {\bibinfo  {journal} {APL Photonics}\ }\textbf
  {\bibinfo {volume} {2}},\ \bibinfo {pages} {096103} (\bibinfo {year}
  {2017})}\BibitemShut {NoStop}%
\bibitem [{\citenamefont {Ward}\ \emph {et~al.}(2016)\citenamefont {Ward},
  \citenamefont {Yang},\ and\ \citenamefont
  {Nic~Chormaic}}]{ward_glass--glass_2016}%
  \BibitemOpen
  \bibfield  {author} {\bibinfo {author} {\bibfnamefont {J.~M.}\ \bibnamefont
  {Ward}}, \bibinfo {author} {\bibfnamefont {Y.}~\bibnamefont {Yang}},\ and\
  \bibinfo {author} {\bibfnamefont {S.}~\bibnamefont {Nic~Chormaic}},\
  }\bibfield  {title} {{\bibinfo {title} {Glass-on-{Glass}
  {Fabrication} of {Bottle}-{Shaped} {Tunable} {Microlasers} and their
  {Applications}}},\ }\href {https://doi.org/10.1038/srep25152} {\bibfield
  {journal} {\bibinfo  {journal} {Scientific Reports}\ }\textbf {\bibinfo
  {volume} {6}},\ \bibinfo {pages} {25152} (\bibinfo {year} {2016})},\ \bibinfo
  {note} {number: 1 Publisher: Nature Publishing Group}\BibitemShut {NoStop}%
\bibitem [{\citenamefont {Zhang}\ \emph {et~al.}(2013)\citenamefont {Zhang},
  \citenamefont {Meng}, \citenamefont {Guo}, \citenamefont {Wang},\ and\
  \citenamefont {He}}]{zhang_simple_2013}%
  \BibitemOpen
  \bibfield  {author} {\bibinfo {author} {\bibfnamefont {S.}~\bibnamefont
  {Zhang}}, \bibinfo {author} {\bibfnamefont {J.}~\bibnamefont {Meng}},
  \bibinfo {author} {\bibfnamefont {S.}~\bibnamefont {Guo}}, \bibinfo {author}
  {\bibfnamefont {L.}~\bibnamefont {Wang}},\ and\ \bibinfo {author}
  {\bibfnamefont {J.-J.}\ \bibnamefont {He}},\ }\bibfield  {title}
  {{\bibinfo {title} {Simple and compact {V}-cavity
  semiconductor laser with 50 x100 {GHz} wavelength tuning}},\ }\href
  {https://doi.org/10.1364/OE.21.013564} {\bibfield  {journal} {\bibinfo
  {journal} {Optics Express}\ }\textbf {\bibinfo {volume} {21}},\ \bibinfo
  {pages} {13564} (\bibinfo {year} {2013})}\BibitemShut {NoStop}%
\bibitem [{\citenamefont {Perahia}\ \emph {et~al.}(2010)\citenamefont
  {Perahia}, \citenamefont {Cohen}, \citenamefont {Meenehan}, \citenamefont
  {Alegre},\ and\ \citenamefont {Painter}}]{perahia_electrostatically_2010}%
  \BibitemOpen
  \bibfield  {author} {\bibinfo {author} {\bibfnamefont {R.}~\bibnamefont
  {Perahia}}, \bibinfo {author} {\bibfnamefont {J.~D.}\ \bibnamefont {Cohen}},
  \bibinfo {author} {\bibfnamefont {S.}~\bibnamefont {Meenehan}}, \bibinfo
  {author} {\bibfnamefont {T.~P.~M.}\ \bibnamefont {Alegre}},\ and\ \bibinfo
  {author} {\bibfnamefont {O.}~\bibnamefont {Painter}},\ }\bibfield  {title}
  {\bibinfo {title} {Electrostatically tunable optomechanical “zipper”
  cavity laser},\ }\href {https://doi.org/10.1063/1.3515296} {\bibfield
  {journal} {\bibinfo  {journal} {Applied Physics Letters}\ }\textbf {\bibinfo
  {volume} {97}},\ \bibinfo {pages} {191112} (\bibinfo {year}
  {2010})}\BibitemShut {NoStop}%
\bibitem [{\citenamefont {Weiß}\ \emph {et~al.}(2016)\citenamefont {Weiß},
  \citenamefont {Kapfinger}, \citenamefont {Reichert}, \citenamefont {Finley},
  \citenamefont {Wixforth}, \citenamefont {Kaniber},\ and\ \citenamefont
  {Krenner}}]{weis_surface_2016}%
  \BibitemOpen
  \bibfield  {author} {\bibinfo {author} {\bibfnamefont {M.}~\bibnamefont
  {Weiß}}, \bibinfo {author} {\bibfnamefont {S.}~\bibnamefont {Kapfinger}},
  \bibinfo {author} {\bibfnamefont {T.}~\bibnamefont {Reichert}}, \bibinfo
  {author} {\bibfnamefont {J.~J.}\ \bibnamefont {Finley}}, \bibinfo {author}
  {\bibfnamefont {A.}~\bibnamefont {Wixforth}}, \bibinfo {author}
  {\bibfnamefont {M.}~\bibnamefont {Kaniber}},\ and\ \bibinfo {author}
  {\bibfnamefont {H.~J.}\ \bibnamefont {Krenner}},\ }\bibfield  {title}
  {\bibinfo {title} {Surface acoustic wave regulated single photon emission
  from a coupled quantum dot–nanocavity system},\ }\href
  {https://doi.org/10.1063/1.4959079} {\bibfield  {journal} {\bibinfo
  {journal} {Applied Physics Letters}\ }\textbf {\bibinfo {volume} {109}},\
  \bibinfo {pages} {033105} (\bibinfo {year} {2016})},\ \bibinfo {note}
  {publisher: American Institute of Physics}\BibitemShut {NoStop}%
\bibitem [{\citenamefont {Sun}\ \emph {et~al.}(2013)\citenamefont {Sun},
  \citenamefont {Kim}, \citenamefont {Solomon},\ and\ \citenamefont
  {Waks}}]{sun_strain_2013}%
  \BibitemOpen
  \bibfield  {author} {\bibinfo {author} {\bibfnamefont {S.}~\bibnamefont
  {Sun}}, \bibinfo {author} {\bibfnamefont {H.}~\bibnamefont {Kim}}, \bibinfo
  {author} {\bibfnamefont {G.~S.}\ \bibnamefont {Solomon}},\ and\ \bibinfo
  {author} {\bibfnamefont {E.}~\bibnamefont {Waks}},\ }\bibfield  {title}
  {\bibinfo {title} {Strain tuning of a quantum dot strongly coupled to a
  photonic crystal cavity},\ }\href {https://doi.org/10.1063/1.4824712}
  {\bibfield  {journal} {\bibinfo  {journal} {Applied Physics Letters}\
  }\textbf {\bibinfo {volume} {103}},\ \bibinfo {pages} {151102} (\bibinfo
  {year} {2013})},\ \bibinfo {note} {publisher: American Institute of
  Physics}\BibitemShut {NoStop}%
\bibitem [{\citenamefont {Moczała-Dusanowska}\ \emph
  {et~al.}(2019)\citenamefont {Moczała-Dusanowska}, \citenamefont
  {Dusanowski}, \citenamefont {Gerhardt}, \citenamefont {He}, \citenamefont
  {Reindl}, \citenamefont {Rastelli}, \citenamefont {Trotta}, \citenamefont
  {Gregersen}, \citenamefont {Höfling},\ and\ \citenamefont
  {Schneider}}]{moczala-dusanowska_strain-tunable_2019}%
  \BibitemOpen
  \bibfield  {author} {\bibinfo {author} {\bibfnamefont {M.}~\bibnamefont
  {Moczała-Dusanowska}}, \bibinfo {author} {\bibfnamefont {u.}~\bibnamefont
  {Dusanowski}}, \bibinfo {author} {\bibfnamefont {S.}~\bibnamefont
  {Gerhardt}}, \bibinfo {author} {\bibfnamefont {Y.~M.}\ \bibnamefont {He}},
  \bibinfo {author} {\bibfnamefont {M.}~\bibnamefont {Reindl}}, \bibinfo
  {author} {\bibfnamefont {A.}~\bibnamefont {Rastelli}}, \bibinfo {author}
  {\bibfnamefont {R.}~\bibnamefont {Trotta}}, \bibinfo {author} {\bibfnamefont
  {N.}~\bibnamefont {Gregersen}}, \bibinfo {author} {\bibfnamefont
  {S.}~\bibnamefont {Höfling}},\ and\ \bibinfo {author} {\bibfnamefont
  {C.}~\bibnamefont {Schneider}},\ }\bibfield  {title} {\bibinfo {title}
  {Strain-{Tunable} {Single}-{Photon} {Source} {Based} on a {Quantum}
  {Dot}–{Micropillar} {System}},\ }\href
  {https://doi.org/10.1021/acsphotonics.9b00481} {\bibfield  {journal}
  {\bibinfo  {journal} {ACS Photonics}\ }\textbf {\bibinfo {volume} {6}},\
  \bibinfo {pages} {2025} (\bibinfo {year} {2019})},\ \bibinfo {note}
  {publisher: American Chemical Society}\BibitemShut {NoStop}%
\bibitem [{\citenamefont {Kremer}\ \emph {et~al.}(2014)\citenamefont {Kremer},
  \citenamefont {Dada}, \citenamefont {Kumar}, \citenamefont {Ma},
  \citenamefont {Kumar}, \citenamefont {Clarke},\ and\ \citenamefont
  {Gerardot}}]{kremer_strain-tunable_2014}%
  \BibitemOpen
  \bibfield  {author} {\bibinfo {author} {\bibfnamefont {P.~E.}\ \bibnamefont
  {Kremer}}, \bibinfo {author} {\bibfnamefont {A.~C.}\ \bibnamefont {Dada}},
  \bibinfo {author} {\bibfnamefont {P.}~\bibnamefont {Kumar}}, \bibinfo
  {author} {\bibfnamefont {Y.}~\bibnamefont {Ma}}, \bibinfo {author}
  {\bibfnamefont {S.}~\bibnamefont {Kumar}}, \bibinfo {author} {\bibfnamefont
  {E.}~\bibnamefont {Clarke}},\ and\ \bibinfo {author} {\bibfnamefont {B.~D.}\
  \bibnamefont {Gerardot}},\ }\bibfield  {title} {{\bibinfo
  {title} {Strain-tunable quantum dot embedded in a nanowire antenna}},\ }\href
  {https://doi.org/10.1103/PhysRevB.90.201408} {\bibfield  {journal} {\bibinfo
  {journal} {Physical Review B}\ }\textbf {\bibinfo {volume} {90}},\ \bibinfo
  {pages} {201408(R)} (\bibinfo {year} {2014})}\BibitemShut {NoStop}%
\bibitem [{\citenamefont {Alegre}\ \emph {et~al.}(2010)\citenamefont {Alegre},
  \citenamefont {Perahia},\ and\ \citenamefont
  {Painter}}]{alegre_optomechanical_2010}%
  \BibitemOpen
  \bibfield  {author} {\bibinfo {author} {\bibfnamefont {T.~P.~M.}\
  \bibnamefont {Alegre}}, \bibinfo {author} {\bibfnamefont {R.}~\bibnamefont
  {Perahia}},\ and\ \bibinfo {author} {\bibfnamefont {O.}~\bibnamefont
  {Painter}},\ }\bibfield  {title} {{\bibinfo {title}
  {Optomechanical zipper cavity lasers: theoretical analysis of tuning range
  and stability}},\ }\href {https://doi.org/10.1364/OE.18.007872} {\bibfield
  {journal} {\bibinfo  {journal} {Optics Express}\ }\textbf {\bibinfo {volume}
  {18}},\ \bibinfo {pages} {7872} (\bibinfo {year} {2010})}\BibitemShut
  {NoStop}%
\bibitem [{\citenamefont {Paterno}\ \emph {et~al.}(2007)\citenamefont
  {Paterno}, \citenamefont {Haramoni}, \citenamefont {Silva},\ and\
  \citenamefont {Kalinowski}}]{paterno_highly_2007}%
  \BibitemOpen
  \bibfield  {author} {\bibinfo {author} {\bibfnamefont {A.~S.}\ \bibnamefont
  {Paterno}}, \bibinfo {author} {\bibfnamefont {N.}~\bibnamefont {Haramoni}},
  \bibinfo {author} {\bibfnamefont {J.~C.~C.}\ \bibnamefont {Silva}},\ and\
  \bibinfo {author} {\bibfnamefont {H.~J.}\ \bibnamefont {Kalinowski}},\
  }\bibfield  {title} {{\bibinfo {title} {Highly reliable
  strain-tuning of an {Erbium}-doped fiber laser for the interrogation of
  multiplexed {Bragg} grating sensors}},\ }\href
  {https://doi.org/10.1016/j.optcom.2006.12.004} {\bibfield  {journal}
  {\bibinfo  {journal} {Optics Communications}\ }\textbf {\bibinfo {volume}
  {273}},\ \bibinfo {pages} {187} (\bibinfo {year} {2007})}\BibitemShut
  {NoStop}%
\bibitem [{\citenamefont {Ball}\ and\ \citenamefont
  {Morey}(1992)}]{ball_continuously_1992}%
  \BibitemOpen
  \bibfield  {author} {\bibinfo {author} {\bibfnamefont {G.~A.}\ \bibnamefont
  {Ball}}\ and\ \bibinfo {author} {\bibfnamefont {W.~W.}\ \bibnamefont
  {Morey}},\ }\bibfield  {title} {{\bibinfo {title}
  {Continuously tunable single-mode erbium fiber laser}},\ }\href
  {https://doi.org/10.1364/OL.17.000420} {\bibfield  {journal} {\bibinfo
  {journal} {Optics Letters}\ }\textbf {\bibinfo {volume} {17}},\ \bibinfo
  {pages} {420} (\bibinfo {year} {1992})},\ \bibinfo {note} {publisher: Optical
  Society of America}\BibitemShut {NoStop}%
\bibitem [{\citenamefont {Cheng}\ \emph {et~al.}(2008)\citenamefont {Cheng},
  \citenamefont {Shum}, \citenamefont {Tse}, \citenamefont {Zhou},
  \citenamefont {Tang}, \citenamefont {Tan}, \citenamefont {Wu},\ and\
  \citenamefont {Zhang}}]{cheng_single-longitudinal-mode_2008}%
  \BibitemOpen
  \bibfield  {author} {\bibinfo {author} {\bibfnamefont {X.~P.}\ \bibnamefont
  {Cheng}}, \bibinfo {author} {\bibfnamefont {P.}~\bibnamefont {Shum}},
  \bibinfo {author} {\bibfnamefont {C.~H.}\ \bibnamefont {Tse}}, \bibinfo
  {author} {\bibfnamefont {J.~L.}\ \bibnamefont {Zhou}}, \bibinfo {author}
  {\bibfnamefont {M.}~\bibnamefont {Tang}}, \bibinfo {author} {\bibfnamefont
  {W.~C.}\ \bibnamefont {Tan}}, \bibinfo {author} {\bibfnamefont {R.~F.}\
  \bibnamefont {Wu}},\ and\ \bibinfo {author} {\bibfnamefont {J.}~\bibnamefont
  {Zhang}},\ }\bibfield  {title} {\bibinfo {title}
  {Single-{Longitudinal}-{Mode} {Erbium}-{Doped} {Fiber} {Ring} {Laser} {Based}
  on {High} {Finesse} {Fiber} {Bragg} {Grating} {Fabry}–{PÉrot} {Etalon}},\
  }\href {https://doi.org/10.1109/LPT.2008.922974} {\bibfield  {journal}
  {\bibinfo  {journal} {IEEE Photonics Technology Letters}\ }\textbf {\bibinfo
  {volume} {20}},\ \bibinfo {pages} {976} (\bibinfo {year} {2008})},\ \bibinfo
  {note} {conference Name: IEEE Photonics Technology Letters}\BibitemShut
  {NoStop}%
\bibitem [{\citenamefont {Baker}\ \emph {et~al.}(2016)\citenamefont {Baker},
  \citenamefont {Bekker}, \citenamefont {McAuslan}, \citenamefont {Sheridan},\
  and\ \citenamefont {Bowen}}]{baker_high_2016}%
  \BibitemOpen
  \bibfield  {author} {\bibinfo {author} {\bibfnamefont {C.~G.}\ \bibnamefont
  {Baker}}, \bibinfo {author} {\bibfnamefont {C.}~\bibnamefont {Bekker}},
  \bibinfo {author} {\bibfnamefont {D.~L.}\ \bibnamefont {McAuslan}}, \bibinfo
  {author} {\bibfnamefont {E.}~\bibnamefont {Sheridan}},\ and\ \bibinfo
  {author} {\bibfnamefont {W.~P.}\ \bibnamefont {Bowen}},\ }\bibfield  {title}
  {{\bibinfo {title} {High bandwidth on-chip capacitive
  tuning of microtoroid resonators}},\ }\href
  {https://doi.org/10.1364/OE.24.020400} {\bibfield  {journal} {\bibinfo
  {journal} {Optics Express}\ }\textbf {\bibinfo {volume} {24}},\ \bibinfo
  {pages} {20400} (\bibinfo {year} {2016})}\BibitemShut {NoStop}%
\bibitem [{\citenamefont {Bekker}\ \emph {et~al.}(2018)\citenamefont {Bekker},
  \citenamefont {Baker}, \citenamefont {Kalra}, \citenamefont {Cheng},
  \citenamefont {Li}, \citenamefont {Prakash},\ and\ \citenamefont
  {Bowen}}]{bekker_free_2018}%
  \BibitemOpen
  \bibfield  {author} {\bibinfo {author} {\bibfnamefont {C.}~\bibnamefont
  {Bekker}}, \bibinfo {author} {\bibfnamefont {C.~G.}\ \bibnamefont {Baker}},
  \bibinfo {author} {\bibfnamefont {R.}~\bibnamefont {Kalra}}, \bibinfo
  {author} {\bibfnamefont {H.-H.}\ \bibnamefont {Cheng}}, \bibinfo {author}
  {\bibfnamefont {B.-B.}\ \bibnamefont {Li}}, \bibinfo {author} {\bibfnamefont
  {V.}~\bibnamefont {Prakash}},\ and\ \bibinfo {author} {\bibfnamefont {W.~P.}\
  \bibnamefont {Bowen}},\ }\bibfield  {title} {{\bibinfo
  {title} {Free spectral range electrical tuning of a high quality on-chip
  microcavity}},\ }\href {https://doi.org/10.1364/OE.26.033649} {\bibfield
  {journal} {\bibinfo  {journal} {Optics Express}\ }\textbf {\bibinfo {volume}
  {26}},\ \bibinfo {pages} {33649} (\bibinfo {year} {2018})}\BibitemShut
  {NoStop}%
\bibitem [{\citenamefont {Minzioni}\ \emph {et~al.}(2019)\citenamefont
  {Minzioni}, \citenamefont {Lacava}, \citenamefont {Tanabe}, \citenamefont
  {Dong}, \citenamefont {Hu}, \citenamefont {Csaba}, \citenamefont {Porod},
  \citenamefont {Singh}, \citenamefont {Willner}, \citenamefont {Almaiman},
  \citenamefont {Torres-Company}, \citenamefont {Schröder}, \citenamefont
  {Peacock}, \citenamefont {Strain}, \citenamefont {Parmigiani}, \citenamefont
  {Contestabile}, \citenamefont {Marpaung}, \citenamefont {Liu}, \citenamefont
  {Bowers}, \citenamefont {Chang}, \citenamefont {Fabbri}, \citenamefont
  {Vázquez}, \citenamefont {Bharadwaj}, \citenamefont {Eaton}, \citenamefont
  {Lodahl}, \citenamefont {Zhang}, \citenamefont {Eggleton}, \citenamefont
  {Munro}, \citenamefont {Nemoto}, \citenamefont {Morin}, \citenamefont
  {Laurat},\ and\ \citenamefont {Nunn}}]{minzioni_roadmap_2019}%
  \BibitemOpen
  \bibfield  {author} {\bibinfo {author} {\bibfnamefont {P.}~\bibnamefont
  {Minzioni}}, \bibinfo {author} {\bibfnamefont {C.}~\bibnamefont {Lacava}},
  \bibinfo {author} {\bibfnamefont {T.}~\bibnamefont {Tanabe}}, \bibinfo
  {author} {\bibfnamefont {J.}~\bibnamefont {Dong}}, \bibinfo {author}
  {\bibfnamefont {X.}~\bibnamefont {Hu}}, \bibinfo {author} {\bibfnamefont
  {G.}~\bibnamefont {Csaba}}, \bibinfo {author} {\bibfnamefont
  {W.}~\bibnamefont {Porod}}, \bibinfo {author} {\bibfnamefont
  {G.}~\bibnamefont {Singh}}, \bibinfo {author} {\bibfnamefont {A.~E.}\
  \bibnamefont {Willner}}, \bibinfo {author} {\bibfnamefont {A.}~\bibnamefont
  {Almaiman}}, \bibinfo {author} {\bibfnamefont {V.}~\bibnamefont
  {Torres-Company}}, \bibinfo {author} {\bibfnamefont {J.}~\bibnamefont
  {Schröder}}, \bibinfo {author} {\bibfnamefont {A.~C.}\ \bibnamefont
  {Peacock}}, \bibinfo {author} {\bibfnamefont {M.~J.}\ \bibnamefont {Strain}},
  \bibinfo {author} {\bibfnamefont {F.}~\bibnamefont {Parmigiani}}, \bibinfo
  {author} {\bibfnamefont {G.}~\bibnamefont {Contestabile}}, \bibinfo {author}
  {\bibfnamefont {D.}~\bibnamefont {Marpaung}}, \bibinfo {author}
  {\bibfnamefont {Z.}~\bibnamefont {Liu}}, \bibinfo {author} {\bibfnamefont
  {J.~E.}\ \bibnamefont {Bowers}}, \bibinfo {author} {\bibfnamefont
  {L.}~\bibnamefont {Chang}}, \bibinfo {author} {\bibfnamefont
  {S.}~\bibnamefont {Fabbri}}, \bibinfo {author} {\bibfnamefont {M.~R.}\
  \bibnamefont {Vázquez}}, \bibinfo {author} {\bibfnamefont {V.}~\bibnamefont
  {Bharadwaj}}, \bibinfo {author} {\bibfnamefont {S.~M.}\ \bibnamefont
  {Eaton}}, \bibinfo {author} {\bibfnamefont {P.}~\bibnamefont {Lodahl}},
  \bibinfo {author} {\bibfnamefont {X.}~\bibnamefont {Zhang}}, \bibinfo
  {author} {\bibfnamefont {B.~J.}\ \bibnamefont {Eggleton}}, \bibinfo {author}
  {\bibfnamefont {W.~J.}\ \bibnamefont {Munro}}, \bibinfo {author}
  {\bibfnamefont {K.}~\bibnamefont {Nemoto}}, \bibinfo {author} {\bibfnamefont
  {O.}~\bibnamefont {Morin}}, \bibinfo {author} {\bibfnamefont
  {J.}~\bibnamefont {Laurat}},\ and\ \bibinfo {author} {\bibfnamefont
  {J.}~\bibnamefont {Nunn}},\ }\bibfield  {title} {{\bibinfo {title} {Roadmap on all-optical processing}},\ }\href
  {https://doi.org/10.1088/2040-8986/ab0e66} {\bibfield  {journal} {\bibinfo
  {journal} {Journal of Optics}\ }\textbf {\bibinfo {volume} {21}},\ \bibinfo
  {pages} {063001} (\bibinfo {year} {2019})},\ \bibinfo {note} {publisher: IOP
  Publishing}\BibitemShut {NoStop}%
\bibitem [{\citenamefont {Thomson}\ \emph {et~al.}(2016)\citenamefont
  {Thomson}, \citenamefont {Zilkie}, \citenamefont {Bowers}, \citenamefont
  {Komljenovic}, \citenamefont {Reed}, \citenamefont {Vivien}, \citenamefont
  {Marris-Morini}, \citenamefont {Cassan}, \citenamefont {Virot}, \citenamefont
  {Fédéli}, \citenamefont {Hartmann}, \citenamefont {Schmid}, \citenamefont
  {Xu}, \citenamefont {Boeuf}, \citenamefont {O’Brien}, \citenamefont
  {Mashanovich},\ and\ \citenamefont {Nedeljkovic}}]{thomson_roadmap_2016}%
  \BibitemOpen
  \bibfield  {author} {\bibinfo {author} {\bibfnamefont {D.}~\bibnamefont
  {Thomson}}, \bibinfo {author} {\bibfnamefont {A.}~\bibnamefont {Zilkie}},
  \bibinfo {author} {\bibfnamefont {J.~E.}\ \bibnamefont {Bowers}}, \bibinfo
  {author} {\bibfnamefont {T.}~\bibnamefont {Komljenovic}}, \bibinfo {author}
  {\bibfnamefont {G.~T.}\ \bibnamefont {Reed}}, \bibinfo {author}
  {\bibfnamefont {L.}~\bibnamefont {Vivien}}, \bibinfo {author} {\bibfnamefont
  {D.}~\bibnamefont {Marris-Morini}}, \bibinfo {author} {\bibfnamefont
  {E.}~\bibnamefont {Cassan}}, \bibinfo {author} {\bibfnamefont
  {L.}~\bibnamefont {Virot}}, \bibinfo {author} {\bibfnamefont {J.-M.}\
  \bibnamefont {Fédéli}}, \bibinfo {author} {\bibfnamefont {J.-M.}\
  \bibnamefont {Hartmann}}, \bibinfo {author} {\bibfnamefont {J.~H.}\
  \bibnamefont {Schmid}}, \bibinfo {author} {\bibfnamefont {D.-X.}\
  \bibnamefont {Xu}}, \bibinfo {author} {\bibfnamefont {F.}~\bibnamefont
  {Boeuf}}, \bibinfo {author} {\bibfnamefont {P.}~\bibnamefont {O’Brien}},
  \bibinfo {author} {\bibfnamefont {G.~Z.}\ \bibnamefont {Mashanovich}},\ and\
  \bibinfo {author} {\bibfnamefont {M.}~\bibnamefont {Nedeljkovic}},\
  }\bibfield  {title} {{\bibinfo {title} {Roadmap on
  silicon photonics}},\ }\href {https://doi.org/10.1088/2040-8978/18/7/073003}
  {\bibfield  {journal} {\bibinfo  {journal} {Journal of Optics}\ }\textbf
  {\bibinfo {volume} {18}},\ \bibinfo {pages} {073003} (\bibinfo {year}
  {2016})}\BibitemShut {NoStop}%
\bibitem [{\citenamefont {Luo}\ \emph {et~al.}(2016)\citenamefont {Luo},
  \citenamefont {Ruan}, \citenamefont {Zhong}, \citenamefont {Cheng},
  \citenamefont {Yang}, \citenamefont {Xu}, \citenamefont {Xu},\ and\
  \citenamefont {Cai}}]{luo_compact_2016}%
  \BibitemOpen
  \bibfield  {author} {\bibinfo {author} {\bibfnamefont {Z.}~\bibnamefont
  {Luo}}, \bibinfo {author} {\bibfnamefont {Q.}~\bibnamefont {Ruan}}, \bibinfo
  {author} {\bibfnamefont {M.}~\bibnamefont {Zhong}}, \bibinfo {author}
  {\bibfnamefont {Y.}~\bibnamefont {Cheng}}, \bibinfo {author} {\bibfnamefont
  {R.}~\bibnamefont {Yang}}, \bibinfo {author} {\bibfnamefont {B.}~\bibnamefont
  {Xu}}, \bibinfo {author} {\bibfnamefont {H.}~\bibnamefont {Xu}},\ and\
  \bibinfo {author} {\bibfnamefont {Z.}~\bibnamefont {Cai}},\ }\bibfield
  {title} {{\bibinfo {title} {Compact self-{Q}-switched
  green upconversion {Er}:{ZBLAN} all-fiber laser operating at 5434 nm}},\
  }\href {https://doi.org/10.1364/OL.41.002258} {\bibfield  {journal} {\bibinfo
   {journal} {Optics Letters}\ }\textbf {\bibinfo {volume} {41}},\ \bibinfo
  {pages} {2258} (\bibinfo {year} {2016})}\BibitemShut {NoStop}%
\bibitem [{\citenamefont {Iqbal}\ \emph {et~al.}(2010)\citenamefont {Iqbal},
  \citenamefont {Gleeson}, \citenamefont {Spaugh}, \citenamefont {Tybor},
  \citenamefont {Gunn}, \citenamefont {Hochberg}, \citenamefont {Baehr-Jones},
  \citenamefont {Bailey},\ and\ \citenamefont {Gunn}}]{iqbal_label-free_2010}%
  \BibitemOpen
  \bibfield  {author} {\bibinfo {author} {\bibfnamefont {M.}~\bibnamefont
  {Iqbal}}, \bibinfo {author} {\bibfnamefont {M.~A.}\ \bibnamefont {Gleeson}},
  \bibinfo {author} {\bibfnamefont {B.}~\bibnamefont {Spaugh}}, \bibinfo
  {author} {\bibfnamefont {F.}~\bibnamefont {Tybor}}, \bibinfo {author}
  {\bibfnamefont {W.~G.}\ \bibnamefont {Gunn}}, \bibinfo {author}
  {\bibfnamefont {M.}~\bibnamefont {Hochberg}}, \bibinfo {author}
  {\bibfnamefont {T.}~\bibnamefont {Baehr-Jones}}, \bibinfo {author}
  {\bibfnamefont {R.~C.}\ \bibnamefont {Bailey}},\ and\ \bibinfo {author}
  {\bibfnamefont {L.~C.}\ \bibnamefont {Gunn}},\ }\bibfield  {title} {\bibinfo
  {title} {Label-{Free} {Biosensor} {Arrays} {Based} on {Silicon} {Ring}
  {Resonators} and {High}-{Speed} {Optical} {Scanning} {Instrumentation}},\
  }\href {https://doi.org/10.1109/JSTQE.2009.2032510} {\bibfield  {journal}
  {\bibinfo  {journal} {IEEE Journal of Selected Topics in Quantum
  Electronics}\ }\textbf {\bibinfo {volume} {16}},\ \bibinfo {pages} {654}
  (\bibinfo {year} {2010})}\BibitemShut {NoStop}%
\bibitem [{\citenamefont {Washburn}\ \emph {et~al.}(2016)\citenamefont
  {Washburn}, \citenamefont {Shia}, \citenamefont {Lenkeit}, \citenamefont
  {Lee},\ and\ \citenamefont {Bailey}}]{washburn_multiplexed_2016}%
  \BibitemOpen
  \bibfield  {author} {\bibinfo {author} {\bibfnamefont {A.}~\bibnamefont
  {Washburn}}, \bibinfo {author} {\bibfnamefont {W.}~\bibnamefont {Shia}},
  \bibinfo {author} {\bibfnamefont {K.}~\bibnamefont {Lenkeit}}, \bibinfo
  {author} {\bibfnamefont {S.-H.}\ \bibnamefont {Lee}},\ and\ \bibinfo {author}
  {\bibfnamefont {R.}~\bibnamefont {Bailey}},\ }\bibfield  {title}
  {{\bibinfo {title} {Multiplexed cancer biomarker
  detection using chip-integrated silicon photonic sensor arrays}},\ }\href
  {https://doi.org/10.1039/C6AN01076H} {\bibfield  {journal} {\bibinfo
  {journal} {Analyst}\ }\textbf {\bibinfo {volume} {141}},\ \bibinfo {pages}
  {5358} (\bibinfo {year} {2016})}\BibitemShut {NoStop}%
\bibitem [{\citenamefont {Stern}\ \emph {et~al.}(2018)\citenamefont {Stern},
  \citenamefont {Ji}, \citenamefont {Okawachi}, \citenamefont {Gaeta},\ and\
  \citenamefont {Lipson}}]{stern_battery-operated_2018}%
  \BibitemOpen
  \bibfield  {author} {\bibinfo {author} {\bibfnamefont {B.}~\bibnamefont
  {Stern}}, \bibinfo {author} {\bibfnamefont {X.}~\bibnamefont {Ji}}, \bibinfo
  {author} {\bibfnamefont {Y.}~\bibnamefont {Okawachi}}, \bibinfo {author}
  {\bibfnamefont {A.~L.}\ \bibnamefont {Gaeta}},\ and\ \bibinfo {author}
  {\bibfnamefont {M.}~\bibnamefont {Lipson}},\ }\bibfield  {title}
  {{\bibinfo {title} {Battery-operated integrated frequency
  comb generator}},\ }\href {https://doi.org/10.1038/s41586-018-0598-9}
  {\bibfield  {journal} {\bibinfo  {journal} {Nature}\ }\textbf {\bibinfo
  {volume} {562}},\ \bibinfo {pages} {401} (\bibinfo {year}
  {2018})}\BibitemShut {NoStop}%
\bibitem [{\citenamefont {Del’Haye}\ \emph {et~al.}(2007)\citenamefont
  {Del’Haye}, \citenamefont {Schliesser}, \citenamefont {Arcizet},
  \citenamefont {Wilken}, \citenamefont {Holzwarth},\ and\ \citenamefont
  {Kippenberg}}]{delhaye_optical_2007}%
  \BibitemOpen
  \bibfield  {author} {\bibinfo {author} {\bibfnamefont {P.}~\bibnamefont
  {Del’Haye}}, \bibinfo {author} {\bibfnamefont {A.}~\bibnamefont
  {Schliesser}}, \bibinfo {author} {\bibfnamefont {O.}~\bibnamefont {Arcizet}},
  \bibinfo {author} {\bibfnamefont {T.}~\bibnamefont {Wilken}}, \bibinfo
  {author} {\bibfnamefont {R.}~\bibnamefont {Holzwarth}},\ and\ \bibinfo
  {author} {\bibfnamefont {T.~J.}\ \bibnamefont {Kippenberg}},\ }\bibfield
  {title} {{\bibinfo {title} {Optical frequency comb
  generation from a monolithic microresonator}},\ }\href
  {https://doi.org/10.1038/nature06401} {\bibfield  {journal} {\bibinfo
  {journal} {Nature}\ }\textbf {\bibinfo {volume} {450}},\ \bibinfo {pages}
  {1214} (\bibinfo {year} {2007})}\BibitemShut {NoStop}%
\bibitem [{\citenamefont {Kuramochi}\ \emph {et~al.}(2014)\citenamefont
  {Kuramochi}, \citenamefont {Nozaki}, \citenamefont {Shinya}, \citenamefont
  {Takeda}, \citenamefont {Sato}, \citenamefont {Matsuo}, \citenamefont
  {Taniyama}, \citenamefont {Sumikura},\ and\ \citenamefont
  {Notomi}}]{kuramochi_large-scale_2014}%
  \BibitemOpen
  \bibfield  {author} {\bibinfo {author} {\bibfnamefont {E.}~\bibnamefont
  {Kuramochi}}, \bibinfo {author} {\bibfnamefont {K.}~\bibnamefont {Nozaki}},
  \bibinfo {author} {\bibfnamefont {A.}~\bibnamefont {Shinya}}, \bibinfo
  {author} {\bibfnamefont {K.}~\bibnamefont {Takeda}}, \bibinfo {author}
  {\bibfnamefont {T.}~\bibnamefont {Sato}}, \bibinfo {author} {\bibfnamefont
  {S.}~\bibnamefont {Matsuo}}, \bibinfo {author} {\bibfnamefont
  {H.}~\bibnamefont {Taniyama}}, \bibinfo {author} {\bibfnamefont
  {H.}~\bibnamefont {Sumikura}},\ and\ \bibinfo {author} {\bibfnamefont
  {M.}~\bibnamefont {Notomi}},\ }\bibfield  {title} {{\bibinfo {title} {Large-scale integration of wavelength-addressable
  all-optical memories on a photonic crystal chip}},\ }\href
  {https://doi.org/10.1038/nphoton.2014.93} {\bibfield  {journal} {\bibinfo
  {journal} {Nature Photonics}\ }\textbf {\bibinfo {volume} {8}},\ \bibinfo
  {pages} {474} (\bibinfo {year} {2014})}\BibitemShut {NoStop}%
\bibitem [{\citenamefont {Polman}(2001)}]{polman_erbium_2001}%
  \BibitemOpen
  \bibfield  {author} {\bibinfo {author} {\bibfnamefont {A.}~\bibnamefont
  {Polman}},\ }\bibfield  {title} {{\\bibinfo {title}
  {Erbium as a probe of everything?}},\ }\href
  {https://doi.org/10.1016/S0921-4526(01)00573-7} {\bibfield  {journal}
  {\bibinfo  {journal} {Physica B: Condensed Matter}\ }\textbf {\bibinfo
  {volume} {300}},\ \bibinfo {pages} {78} (\bibinfo {year} {2001})}\BibitemShut
  {NoStop}%
\bibitem [{\citenamefont {Kik}\ and\ \citenamefont
  {Polman}(1998)}]{kik_erbium-doped_1998}%
  \BibitemOpen
  \bibfield  {author} {\bibinfo {author} {\bibfnamefont {P.}~\bibnamefont
  {Kik}}\ and\ \bibinfo {author} {\bibfnamefont {A.}~\bibnamefont {Polman}},\
  }\bibfield  {title} {{\bibinfo {title} {Erbium-{Doped}
  {Optical}-{Waveguide} {Amplifiers} on {Silicon}}},\ }\href
  {https://doi.org/10.1557/S0883769400030268} {\bibfield  {journal} {\bibinfo
  {journal} {MRS Bulletin}\ }\textbf {\bibinfo {volume} {23}},\ \bibinfo
  {pages} {48} (\bibinfo {year} {1998})}\BibitemShut {NoStop}%
\bibitem [{\citenamefont {Becker}\ \emph {et~al.}(1999)\citenamefont {Becker},
  \citenamefont {Olsson},\ and\ \citenamefont
  {Simpson}}]{becker_erbium-doped_1999}%
  \BibitemOpen
  \bibfield  {author} {\bibinfo {author} {\bibfnamefont {P.~M.}\ \bibnamefont
  {Becker}}, \bibinfo {author} {\bibfnamefont {A.~A.}\ \bibnamefont {Olsson}},\
  and\ \bibinfo {author} {\bibfnamefont {J.~R.}\ \bibnamefont {Simpson}},\
  }\href@noop {} {{\emph {\bibinfo {title} {Erbium-{Doped}
  {Fiber} {Amplifiers}: {Fundamentals} and {Technology}}}}}\ (\bibinfo
  {publisher} {Elsevier},\ \bibinfo {year} {1999})\BibitemShut {NoStop}%
\bibitem [{\citenamefont {Lin}\ \emph {et~al.}(2009)\citenamefont {Lin},
  \citenamefont {Rosenberg}, \citenamefont {Jiang}, \citenamefont {Vahala},\
  and\ \citenamefont {Painter}}]{lin_mechanical_2009}%
  \BibitemOpen
  \bibfield  {author} {\bibinfo {author} {\bibfnamefont {Q.}~\bibnamefont
  {Lin}}, \bibinfo {author} {\bibfnamefont {J.}~\bibnamefont {Rosenberg}},
  \bibinfo {author} {\bibfnamefont {X.}~\bibnamefont {Jiang}}, \bibinfo
  {author} {\bibfnamefont {K.~J.}\ \bibnamefont {Vahala}},\ and\ \bibinfo
  {author} {\bibfnamefont {O.}~\bibnamefont {Painter}},\ }\bibfield  {title}
  {\bibinfo {title} {Mechanical {Oscillation} and {Cooling} {Actuated} by the
  {Optical} {Gradient} {Force}},\ }\href
  {https://doi.org/10.1103/PhysRevLett.103.103601} {\bibfield  {journal}
  {\bibinfo  {journal} {Physical Review Letters}\ }\textbf {\bibinfo {volume}
  {103}},\ \bibinfo {pages} {103601} (\bibinfo {year} {2009})}\BibitemShut
  {NoStop}%
\bibitem [{\citenamefont {Jiang}\ \emph {et~al.}(2009)\citenamefont {Jiang},
  \citenamefont {Lin}, \citenamefont {Rosenberg}, \citenamefont {Vahala},\ and\
  \citenamefont {Painter}}]{jiang_high-q_2009}%
  \BibitemOpen
  \bibfield  {author} {\bibinfo {author} {\bibfnamefont {X.}~\bibnamefont
  {Jiang}}, \bibinfo {author} {\bibfnamefont {Q.}~\bibnamefont {Lin}}, \bibinfo
  {author} {\bibfnamefont {J.}~\bibnamefont {Rosenberg}}, \bibinfo {author}
  {\bibfnamefont {K.}~\bibnamefont {Vahala}},\ and\ \bibinfo {author}
  {\bibfnamefont {O.}~\bibnamefont {Painter}},\ }\bibfield  {title}
  {{\bibinfo {title} {High-{Q} double-disk microcavities
  for cavity optomechanics}},\ }\href {https://doi.org/10.1364/OE.17.020911}
  {\bibfield  {journal} {\bibinfo  {journal} {Optics Express}\ }\textbf
  {\bibinfo {volume} {17}},\ \bibinfo {pages} {20911} (\bibinfo {year}
  {2009})}\BibitemShut {NoStop}%
\bibitem [{\citenamefont {Rosenberg}\ \emph {et~al.}(2009)\citenamefont
  {Rosenberg}, \citenamefont {Lin},\ and\ \citenamefont
  {Painter}}]{rosenberg_static_2009}%
  \BibitemOpen
  \bibfield  {author} {\bibinfo {author} {\bibfnamefont {J.}~\bibnamefont
  {Rosenberg}}, \bibinfo {author} {\bibfnamefont {Q.}~\bibnamefont {Lin}},\
  and\ \bibinfo {author} {\bibfnamefont {O.}~\bibnamefont {Painter}},\
  }\bibfield  {title} {{\bibinfo {title} {Static and
  dynamic wavelength routing via the gradient optical force}},\ }\href
  {https://doi.org/10.1038/nphoton.2009.137} {\bibfield  {journal} {\bibinfo
  {journal} {Nature Photonics}\ }\textbf {\bibinfo {volume} {3}},\ \bibinfo
  {pages} {478} (\bibinfo {year} {2009})}\BibitemShut {NoStop}%
\bibitem [{\citenamefont {Lee}\ \emph {et~al.}(2010)\citenamefont {Lee},
  \citenamefont {Eom}, \citenamefont {Chang}, \citenamefont {Huh},
  \citenamefont {Sung},\ and\ \citenamefont {Shin}}]{lee_silicon_2010}%
  \BibitemOpen
  \bibfield  {author} {\bibinfo {author} {\bibfnamefont {S.}~\bibnamefont
  {Lee}}, \bibinfo {author} {\bibfnamefont {S.~C.}\ \bibnamefont {Eom}},
  \bibinfo {author} {\bibfnamefont {J.~S.}\ \bibnamefont {Chang}}, \bibinfo
  {author} {\bibfnamefont {C.}~\bibnamefont {Huh}}, \bibinfo {author}
  {\bibfnamefont {G.~Y.}\ \bibnamefont {Sung}},\ and\ \bibinfo {author}
  {\bibfnamefont {J.~H.}\ \bibnamefont {Shin}},\ }\bibfield  {title}
  {{\bibinfo {title} {A silicon nitride microdisk resonator
  with a 40-nm-thin horizontal air slot}},\ }\href
  {https://doi.org/10.1364/OE.18.011209} {\bibfield  {journal} {\bibinfo
  {journal} {Optics Express}\ }\textbf {\bibinfo {volume} {18}},\ \bibinfo
  {pages} {11209} (\bibinfo {year} {2010})}\BibitemShut {NoStop}%
\bibitem [{\citenamefont {Wiederhecker}\ \emph {et~al.}(2009)\citenamefont
  {Wiederhecker}, \citenamefont {Chen}, \citenamefont {Gondarenko},\ and\
  \citenamefont {Lipson}}]{wiederhecker_controlling_2009}%
  \BibitemOpen
  \bibfield  {author} {\bibinfo {author} {\bibfnamefont {G.~S.}\ \bibnamefont
  {Wiederhecker}}, \bibinfo {author} {\bibfnamefont {L.}~\bibnamefont {Chen}},
  \bibinfo {author} {\bibfnamefont {A.}~\bibnamefont {Gondarenko}},\ and\
  \bibinfo {author} {\bibfnamefont {M.}~\bibnamefont {Lipson}},\ }\bibfield
  {title} {{\bibinfo {title} {Controlling photonic
  structures using optical forces}},\ }\href
  {https://doi.org/10.1038/nature08584} {\bibfield  {journal} {\bibinfo
  {journal} {Nature}\ }\textbf {\bibinfo {volume} {462}},\ \bibinfo {pages}
  {633} (\bibinfo {year} {2009})}\BibitemShut {NoStop}%
\bibitem [{\citenamefont {Wiederhecker}\ \emph {et~al.}(2011)\citenamefont
  {Wiederhecker}, \citenamefont {Manipatruni}, \citenamefont {Lee},\ and\
  \citenamefont {Lipson}}]{wiederhecker_broadband_2011}%
  \BibitemOpen
  \bibfield  {author} {\bibinfo {author} {\bibfnamefont {G.~S.}\ \bibnamefont
  {Wiederhecker}}, \bibinfo {author} {\bibfnamefont {S.}~\bibnamefont
  {Manipatruni}}, \bibinfo {author} {\bibfnamefont {S.}~\bibnamefont {Lee}},\
  and\ \bibinfo {author} {\bibfnamefont {M.}~\bibnamefont {Lipson}},\
  }\bibfield  {title} {\bibinfo {title} {Broadband tuning of optomechanical
  cavities},\ }\href
  {https://www.osapublishing.org/abstract.cfm?uri=oe-19-3-2782} {\bibfield
  {journal} {\bibinfo  {journal} {Optics Express}\ }\textbf {\bibinfo {volume}
  {19}},\ \bibinfo {pages} {2782} (\bibinfo {year} {2011})}\BibitemShut
  {NoStop}%
\bibitem [{\citenamefont {Lu}\ \emph {et~al.}(2019)\citenamefont {Lu},
  \citenamefont {Lee}, \citenamefont {Rogers},\ and\ \citenamefont
  {Lin}}]{lu_silicon_2019}%
  \BibitemOpen
  \bibfield  {author} {\bibinfo {author} {\bibfnamefont {X.}~\bibnamefont
  {Lu}}, \bibinfo {author} {\bibfnamefont {J.~Y.}\ \bibnamefont {Lee}},
  \bibinfo {author} {\bibfnamefont {S.~D.}\ \bibnamefont {Rogers}},\ and\
  \bibinfo {author} {\bibfnamefont {Q.}~\bibnamefont {Lin}},\ }\bibfield
  {title} {{\bibinfo {title} {Silicon carbide
  double-microdisk resonator}},\ }\href {https://doi.org/10.1364/OL.44.004295}
  {\bibfield  {journal} {\bibinfo  {journal} {Optics Letters}\ }\textbf
  {\bibinfo {volume} {44}},\ \bibinfo {pages} {4295} (\bibinfo {year}
  {2019})}\BibitemShut {NoStop}%
\bibitem [{\citenamefont {Zheng}\ \emph {et~al.}(2019)\citenamefont {Zheng},
  \citenamefont {Fang}, \citenamefont {Liu}, \citenamefont {Cheng},\ and\
  \citenamefont {Chen}}]{zheng_high-q_2019}%
  \BibitemOpen
  \bibfield  {author} {\bibinfo {author} {\bibfnamefont {Y.}~\bibnamefont
  {Zheng}}, \bibinfo {author} {\bibfnamefont {Z.}~\bibnamefont {Fang}},
  \bibinfo {author} {\bibfnamefont {S.}~\bibnamefont {Liu}}, \bibinfo {author}
  {\bibfnamefont {Y.}~\bibnamefont {Cheng}},\ and\ \bibinfo {author}
  {\bibfnamefont {X.}~\bibnamefont {Chen}},\ }\bibfield  {title} {\bibinfo
  {title} {High-{Q} {Exterior} {Whispering}-{Gallery} {Modes} in a
  {Double}-{Layer} {Crystalline} {Microdisk} {Resonator}},\ }\href
  {https://doi.org/10.1103/PhysRevLett.122.253902} {\bibfield  {journal}
  {\bibinfo  {journal} {Physical Review Letters}\ }\textbf {\bibinfo {volume}
  {122}},\ \bibinfo {pages} {253902} (\bibinfo {year} {2019})}\BibitemShut
  {NoStop}%
\bibitem [{\citenamefont {Fang}\ \emph {et~al.}(2017)\citenamefont {Fang},
  \citenamefont {Yao}, \citenamefont {Wang}, \citenamefont {Lin}, \citenamefont
  {Zhang}, \citenamefont {Wu}, \citenamefont {Qiao}, \citenamefont {Fang},
  \citenamefont {Lu},\ and\ \citenamefont {Cheng}}]{fang_fabrication_2017}%
  \BibitemOpen
  \bibfield  {author} {\bibinfo {author} {\bibfnamefont {Z.}~\bibnamefont
  {Fang}}, \bibinfo {author} {\bibfnamefont {N.}~\bibnamefont {Yao}}, \bibinfo
  {author} {\bibfnamefont {M.}~\bibnamefont {Wang}}, \bibinfo {author}
  {\bibfnamefont {J.}~\bibnamefont {Lin}}, \bibinfo {author} {\bibfnamefont
  {J.}~\bibnamefont {Zhang}}, \bibinfo {author} {\bibfnamefont
  {R.}~\bibnamefont {Wu}}, \bibinfo {author} {\bibfnamefont {L.}~\bibnamefont
  {Qiao}}, \bibinfo {author} {\bibfnamefont {W.}~\bibnamefont {Fang}}, \bibinfo
  {author} {\bibfnamefont {T.}~\bibnamefont {Lu}},\ and\ \bibinfo {author}
  {\bibfnamefont {Y.}~\bibnamefont {Cheng}},\ }\bibfield  {title} {\bibinfo
  {title} {Fabrication of high quality factor lithium niobate double-disk using
  a femtosecond laser},\ }\href {https://doi.org/10.1080/15599612.2017.1406024}
  {\bibfield  {journal} {\bibinfo  {journal} {International Journal of
  Optomechatronics}\ }\textbf {\bibinfo {volume} {11}},\ \bibinfo {pages} {47}
  (\bibinfo {year} {2017})}\BibitemShut {NoStop}%
\bibitem [{\citenamefont {Dehghannasiri}\ \emph {et~al.}(2018)\citenamefont
  {Dehghannasiri}, \citenamefont {Moradinejad}, \citenamefont {Fan},
  \citenamefont {Hosseinnia}, \citenamefont {Eftekhar},\ and\ \citenamefont
  {Adibi}}]{dehghannasiri_integrated_2018}%
  \BibitemOpen
  \bibfield  {author} {\bibinfo {author} {\bibfnamefont {R.}~\bibnamefont
  {Dehghannasiri}}, \bibinfo {author} {\bibfnamefont {H.}~\bibnamefont
  {Moradinejad}}, \bibinfo {author} {\bibfnamefont {T.}~\bibnamefont {Fan}},
  \bibinfo {author} {\bibfnamefont {A.~H.}\ \bibnamefont {Hosseinnia}},
  \bibinfo {author} {\bibfnamefont {A.~A.}\ \bibnamefont {Eftekhar}},\ and\
  \bibinfo {author} {\bibfnamefont {A.}~\bibnamefont {Adibi}},\ }\bibfield
  {title} {\bibinfo {title} {Integrated {Optomechanical} {Resonators} in
  {Double}-{Layer} {Crystalline} {Silicon} {Platforms}},\ }in\ \href
  {https://doi.org/10.1109/IPCon.2018.8527120} {\emph {\bibinfo {booktitle}
  {2018 {IEEE} {Photonics} {Conference} ({IPC})}}}\ (\bibinfo {year} {2018})\
  pp.\ \bibinfo {pages} {1--2}\BibitemShut {NoStop}%
\bibitem [{\citenamefont {Ding}\ \emph {et~al.}(2010)\citenamefont {Ding},
  \citenamefont {Baker}, \citenamefont {Senellart}, \citenamefont {Lemaitre},
  \citenamefont {Ducci}, \citenamefont {Leo},\ and\ \citenamefont
  {Favero}}]{ding_high_2010}%
  \BibitemOpen
  \bibfield  {author} {\bibinfo {author} {\bibfnamefont {L.}~\bibnamefont
  {Ding}}, \bibinfo {author} {\bibfnamefont {C.}~\bibnamefont {Baker}},
  \bibinfo {author} {\bibfnamefont {P.}~\bibnamefont {Senellart}}, \bibinfo
  {author} {\bibfnamefont {A.}~\bibnamefont {Lemaitre}}, \bibinfo {author}
  {\bibfnamefont {S.}~\bibnamefont {Ducci}}, \bibinfo {author} {\bibfnamefont
  {G.}~\bibnamefont {Leo}},\ and\ \bibinfo {author} {\bibfnamefont
  {I.}~\bibnamefont {Favero}},\ }\bibfield  {title} {\bibinfo {title} {High
  {Frequency} {GaAs} {Nano}-{Optomechanical} {Disk} {Resonator}},\ }\href
  {https://doi.org/10.1103/PhysRevLett.105.263903} {\bibfield  {journal}
  {\bibinfo  {journal} {Physical Review Letters}\ }\textbf {\bibinfo {volume}
  {105}},\ \bibinfo {pages} {263903} (\bibinfo {year} {2010})}\BibitemShut
  {NoStop}%
\bibitem [{\citenamefont {Polman}(1997)}]{polman_erbium_1997}%
  \BibitemOpen
  \bibfield  {author} {\bibinfo {author} {\bibfnamefont {A.}~\bibnamefont
  {Polman}},\ }\bibfield  {title} {{\bibinfo {title}
  {Erbium implanted thin film photonic materials}},\ }\href
  {https://doi.org/10.1063/1.366265} {\bibfield  {journal} {\bibinfo  {journal}
  {Journal of Applied Physics}\ }\textbf {\bibinfo {volume} {82}},\ \bibinfo
  {pages} {1} (\bibinfo {year} {1997})}\BibitemShut {NoStop}%
\bibitem [{\citenamefont {Romeira}\ and\ \citenamefont
  {Fiore}(2018)}]{romeira_purcell_2018}%
  \BibitemOpen
  \bibfield  {author} {\bibinfo {author} {\bibfnamefont {B.}~\bibnamefont
  {Romeira}}\ and\ \bibinfo {author} {\bibfnamefont {A.}~\bibnamefont
  {Fiore}},\ }\bibfield  {title} {\bibinfo {title} {Purcell {Effect} in the
  {Stimulated} and {Spontaneous} {Emission} {Rates} of {Nanoscale}
  {Semiconductor} {Lasers}},\ }\href {https://doi.org/10.1109/JQE.2018.2802464}
  {\bibfield  {journal} {\bibinfo  {journal} {IEEE Journal of Quantum
  Electronics}\ }\textbf {\bibinfo {volume} {54}},\ \bibinfo {pages} {1}
  (\bibinfo {year} {2018})}\BibitemShut {NoStop}%
\bibitem [{\citenamefont {Jager}\ \emph {et~al.}(2009)\citenamefont {Jager},
  \citenamefont {Noé}, \citenamefont {Picard}, \citenamefont {Calvo},
  \citenamefont {Delamadeleine},\ and\ \citenamefont
  {Hadji}}]{jager_whispering_2009}%
  \BibitemOpen
  \bibfield  {author} {\bibinfo {author} {\bibfnamefont {J.~B.}\ \bibnamefont
  {Jager}}, \bibinfo {author} {\bibfnamefont {P.}~\bibnamefont {Noé}},
  \bibinfo {author} {\bibfnamefont {E.}~\bibnamefont {Picard}}, \bibinfo
  {author} {\bibfnamefont {V.}~\bibnamefont {Calvo}}, \bibinfo {author}
  {\bibfnamefont {E.}~\bibnamefont {Delamadeleine}},\ and\ \bibinfo {author}
  {\bibfnamefont {E.}~\bibnamefont {Hadji}},\ }\bibfield  {title}
  {{\bibinfo {title} {Whispering gallery modes in
  {Er}-doped silicon-rich oxide toroidal microcavities on chip}},\ }\href
  {https://doi.org/10.1016/j.physe.2008.08.011} {\bibfield  {journal} {\bibinfo
   {journal} {Physica E: Low-dimensional Systems and Nanostructures}\ }\textbf
  {\bibinfo {volume} {41}},\ \bibinfo {pages} {1127} (\bibinfo {year}
  {2009})}\BibitemShut {NoStop}%
\bibitem [{\citenamefont {Hsu}\ \emph {et~al.}(2009)\citenamefont {Hsu},
  \citenamefont {Cai},\ and\ \citenamefont
  {Armani}}]{hsu_ultra-low-threshold_2009}%
  \BibitemOpen
  \bibfield  {author} {\bibinfo {author} {\bibfnamefont {H.-S.}\ \bibnamefont
  {Hsu}}, \bibinfo {author} {\bibfnamefont {C.}~\bibnamefont {Cai}},\ and\
  \bibinfo {author} {\bibfnamefont {A.~M.}\ \bibnamefont {Armani}},\ }\bibfield
   {title} {{\bibinfo {title} {Ultra-low-threshold
  {Er}:{Yb} sol-gel microlaser on silicon}},\ }\href
  {https://doi.org/10.1364/OE.17.023265} {\bibfield  {journal} {\bibinfo
  {journal} {Optics Express}\ }\textbf {\bibinfo {volume} {17}},\ \bibinfo
  {pages} {23265} (\bibinfo {year} {2009})}\BibitemShut {NoStop}%
\bibitem [{\citenamefont {Tao}\ \emph {et~al.}(2020)\citenamefont {Tao},
  \citenamefont {Wei}, \citenamefont {Li}, \citenamefont {Ou}, \citenamefont
  {Wang}, \citenamefont {Wang}, \citenamefont {Zhang}, \citenamefont {Zhang},
  \citenamefont {Gan},\ and\ \citenamefont {Ou}}]{tao_-chip_2020}%
  \BibitemOpen
  \bibfield  {author} {\bibinfo {author} {\bibfnamefont {L.}~\bibnamefont
  {Tao}}, \bibinfo {author} {\bibfnamefont {W.}~\bibnamefont {Wei}}, \bibinfo
  {author} {\bibfnamefont {Y.}~\bibnamefont {Li}}, \bibinfo {author}
  {\bibfnamefont {W.}~\bibnamefont {Ou}}, \bibinfo {author} {\bibfnamefont
  {T.}~\bibnamefont {Wang}}, \bibinfo {author} {\bibfnamefont {C.}~\bibnamefont
  {Wang}}, \bibinfo {author} {\bibfnamefont {J.}~\bibnamefont {Zhang}},
  \bibinfo {author} {\bibfnamefont {J.}~\bibnamefont {Zhang}}, \bibinfo
  {author} {\bibfnamefont {F.}~\bibnamefont {Gan}},\ and\ \bibinfo {author}
  {\bibfnamefont {X.}~\bibnamefont {Ou}},\ }\bibfield  {title} {\bibinfo
  {title} {On-{Chip} {Integration} of {Energy}-{Tunable} {Quantum} {Dot}
  {Based} {Single}-{Photon} {Sources} via {Strain} {Tuning} of {GaAs}
  {Waveguides}},\ }\href {https://doi.org/10.1021/acsphotonics.0c00748}
  {\bibfield  {journal} {\bibinfo  {journal} {ACS Photonics}\ }\textbf
  {\bibinfo {volume} {7}},\ \bibinfo {pages} {2723} (\bibinfo {year} {2020})},\
  \bibinfo {note} {publisher: American Chemical Society}\BibitemShut {NoStop}%
\bibitem [{\citenamefont {Lee}\ \emph {et~al.}(2016)\citenamefont {Lee},
  \citenamefont {Lee}, \citenamefont {Ovartchaiyapong}, \citenamefont
  {Minguzzi}, \citenamefont {Maze},\ and\ \citenamefont
  {Bleszynski~Jayich}}]{lee_strain_2016}%
  \BibitemOpen
  \bibfield  {author} {\bibinfo {author} {\bibfnamefont {K.~W.}\ \bibnamefont
  {Lee}}, \bibinfo {author} {\bibfnamefont {D.}~\bibnamefont {Lee}}, \bibinfo
  {author} {\bibfnamefont {P.}~\bibnamefont {Ovartchaiyapong}}, \bibinfo
  {author} {\bibfnamefont {J.}~\bibnamefont {Minguzzi}}, \bibinfo {author}
  {\bibfnamefont {J.~R.}\ \bibnamefont {Maze}},\ and\ \bibinfo {author}
  {\bibfnamefont {A.~C.}\ \bibnamefont {Bleszynski~Jayich}},\ }\bibfield
  {title} {\bibinfo {title} {Strain {Coupling} of a {Mechanical} {Resonator} to
  a {Single} {Quantum} {Emitter} in {Diamond}},\ }\href
  {https://doi.org/10.1103/PhysRevApplied.6.034005} {\bibfield  {journal}
  {\bibinfo  {journal} {Physical Review Applied}\ }\textbf {\bibinfo {volume}
  {6}},\ \bibinfo {pages} {034005} (\bibinfo {year} {2016})},\ \bibinfo {note}
  {publisher: American Physical Society}\BibitemShut {NoStop}%
\bibitem [{\citenamefont {Feng}\ \emph {et~al.}(2010)\citenamefont {Feng},
  \citenamefont {Li}, \citenamefont {Sedhain}, \citenamefont {Lin},
  \citenamefont {Jiang},\ and\ \citenamefont {Zavada}}]{feng_enhancing_2010}%
  \BibitemOpen
  \bibfield  {author} {\bibinfo {author} {\bibfnamefont {I.~W.}\ \bibnamefont
  {Feng}}, \bibinfo {author} {\bibfnamefont {J.}~\bibnamefont {Li}}, \bibinfo
  {author} {\bibfnamefont {A.}~\bibnamefont {Sedhain}}, \bibinfo {author}
  {\bibfnamefont {J.~Y.}\ \bibnamefont {Lin}}, \bibinfo {author} {\bibfnamefont
  {H.~X.}\ \bibnamefont {Jiang}},\ and\ \bibinfo {author} {\bibfnamefont
  {J.}~\bibnamefont {Zavada}},\ }\bibfield  {title} {\bibinfo {title}
  {Enhancing erbium emission by strain engineering in {GaN} heteroepitaxial
  layers},\ }\href {https://doi.org/10.1063/1.3295705} {\bibfield  {journal}
  {\bibinfo  {journal} {Applied Physics Letters}\ }\textbf {\bibinfo {volume}
  {96}},\ \bibinfo {pages} {031908} (\bibinfo {year} {2010})},\ \bibinfo {note}
  {publisher: American Institute of Physics}\BibitemShut {NoStop}%
\bibitem [{\citenamefont {Zhang}\ \emph {et~al.}(2019)\citenamefont {Zhang},
  \citenamefont {Hu}, \citenamefont {de~Boo}, \citenamefont {Rančić},
  \citenamefont {Johnson}, \citenamefont {McCallum}, \citenamefont {Du},
  \citenamefont {Sellars}, \citenamefont {Yin},\ and\ \citenamefont
  {Rogge}}]{zhang_single_2019}%
  \BibitemOpen
  \bibfield  {author} {\bibinfo {author} {\bibfnamefont {Q.}~\bibnamefont
  {Zhang}}, \bibinfo {author} {\bibfnamefont {G.}~\bibnamefont {Hu}}, \bibinfo
  {author} {\bibfnamefont {G.~G.}\ \bibnamefont {de~Boo}}, \bibinfo {author}
  {\bibfnamefont {M.}~\bibnamefont {Rančić}}, \bibinfo {author}
  {\bibfnamefont {B.~C.}\ \bibnamefont {Johnson}}, \bibinfo {author}
  {\bibfnamefont {J.~C.}\ \bibnamefont {McCallum}}, \bibinfo {author}
  {\bibfnamefont {J.}~\bibnamefont {Du}}, \bibinfo {author} {\bibfnamefont
  {M.~J.}\ \bibnamefont {Sellars}}, \bibinfo {author} {\bibfnamefont
  {C.}~\bibnamefont {Yin}},\ and\ \bibinfo {author} {\bibfnamefont
  {S.}~\bibnamefont {Rogge}},\ }\bibfield  {title} {\bibinfo {title} {Single
  {Rare}-{Earth} {Ions} as {Atomic}-{Scale} {Probes} in {Ultrascaled}
  {Transistors}},\ }\href {https://doi.org/10.1021/acs.nanolett.9b01281}
  {\bibfield  {journal} {\bibinfo  {journal} {Nano Letters}\ }\textbf {\bibinfo
  {volume} {19}},\ \bibinfo {pages} {5025} (\bibinfo {year} {2019})},\ \bibinfo
  {note} {publisher: American Chemical Society}\BibitemShut {NoStop}%
\bibitem [{\citenamefont {He}\ \emph {et~al.}(2013)\citenamefont {He},
  \citenamefont {\"{O}zdemir},\ and\ \citenamefont
  {Yang}}]{he_whispering_2013}%
  \BibitemOpen
  \bibfield  {author} {\bibinfo {author} {\bibfnamefont {L.}~\bibnamefont
  {He}}, \bibinfo {author} {\bibfnamefont {c.~K.}\ \bibnamefont
  {\"{O}zdemir}},\ and\ \bibinfo {author} {\bibfnamefont {L.}~\bibnamefont
  {Yang}},\ }\bibfield  {title} {\bibinfo {title} {Whispering gallery
  microcavity lasers},\ }\href {https://doi.org/10.1002/lpor.201100032}
  {\bibfield  {journal} {\bibinfo  {journal} {Laser \& Photonics Reviews}\
  }\textbf {\bibinfo {volume} {7}},\ \bibinfo {pages} {60} (\bibinfo {year}
  {2013})}\BibitemShut {NoStop}%
\bibitem [{\citenamefont {Cai}\ \emph {et~al.}(1999)\citenamefont {Cai},
  \citenamefont {Hunziker},\ and\ \citenamefont
  {Vahala}}]{cai_fiber-optic_1999}%
  \BibitemOpen
  \bibfield  {author} {\bibinfo {author} {\bibfnamefont {M.}~\bibnamefont
  {Cai}}, \bibinfo {author} {\bibfnamefont {G.}~\bibnamefont {Hunziker}},\ and\
  \bibinfo {author} {\bibfnamefont {K.}~\bibnamefont {Vahala}},\ }\bibfield
  {title} {\bibinfo {title} {Fiber-optic add-drop device based on a silica
  microsphere-whispering gallery mode system},\ }\href
  {https://doi.org/10.1109/68.766785} {\bibfield  {journal} {\bibinfo
  {journal} {IEEE Photonics Technology Letters}\ }\textbf {\bibinfo {volume}
  {11}},\ \bibinfo {pages} {686} (\bibinfo {year} {1999})},\ \bibinfo {note}
  {conference Name: IEEE Photonics Technology Letters}\BibitemShut {NoStop}%
\bibitem [{\citenamefont {Monifi}\ \emph {et~al.}(2012)\citenamefont {Monifi},
  \citenamefont {Friedlein}, \citenamefont {Ozdemir},\ and\ \citenamefont
  {Yang}}]{monifi_robust_2012}%
  \BibitemOpen
  \bibfield  {author} {\bibinfo {author} {\bibfnamefont {F.}~\bibnamefont
  {Monifi}}, \bibinfo {author} {\bibfnamefont {J.}~\bibnamefont {Friedlein}},
  \bibinfo {author} {\bibfnamefont {.~K.}\ \bibnamefont {Ozdemir}},\ and\
  \bibinfo {author} {\bibfnamefont {L.}~\bibnamefont {Yang}},\ }\bibfield
  {title} {\bibinfo {title} {A {Robust} and {Tunable} {Add}–{Drop} {Filter}
  {Using} {Whispering} {Gallery} {Mode} {Microtoroid} {Resonator}},\ }\href
  {https://doi.org/10.1109/JLT.2012.2214026} {\bibfield  {journal} {\bibinfo
  {journal} {Journal of Lightwave Technology}\ }\textbf {\bibinfo {volume}
  {30}},\ \bibinfo {pages} {3306} (\bibinfo {year} {2012})},\ \bibinfo {note}
  {conference Name: Journal of Lightwave Technology}\BibitemShut {NoStop}%
\bibitem [{\citenamefont {Murugan}\ \emph {et~al.}(2010)\citenamefont
  {Murugan}, \citenamefont {Wilkinson},\ and\ \citenamefont
  {Zervas}}]{murugan_optical_2010}%
  \BibitemOpen
  \bibfield  {author} {\bibinfo {author} {\bibfnamefont {G.~S.}\ \bibnamefont
  {Murugan}}, \bibinfo {author} {\bibfnamefont {J.~S.}\ \bibnamefont
  {Wilkinson}},\ and\ \bibinfo {author} {\bibfnamefont {M.~N.}\ \bibnamefont
  {Zervas}},\ }\bibfield  {title} {{\bibinfo {title}
  {Optical excitation and probing of whispering gallery modes in bottle
  microresonators: potential for all-fiber add-drop filters}},\ }\href
  {https://doi.org/10.1364/OL.35.001893} {\bibfield  {journal} {\bibinfo
  {journal} {Optics Letters}\ }\textbf {\bibinfo {volume} {35}},\ \bibinfo
  {pages} {1893} (\bibinfo {year} {2010})},\ \bibinfo {note} {publisher:
  Optical Society of America}\BibitemShut {NoStop}%
\bibitem [{\citenamefont {Jiang}\ \emph {et~al.}(2017)\citenamefont {Jiang},
  \citenamefont {Shao}, \citenamefont {Zhang}, \citenamefont {Yi},
  \citenamefont {Wiersig}, \citenamefont {Wang}, \citenamefont {Gong},
  \citenamefont {Lončar}, \citenamefont {Yang},\ and\ \citenamefont
  {Xiao}}]{jiang_chaos-assisted_2017}%
  \BibitemOpen
  \bibfield  {author} {\bibinfo {author} {\bibfnamefont {X.}~\bibnamefont
  {Jiang}}, \bibinfo {author} {\bibfnamefont {L.}~\bibnamefont {Shao}},
  \bibinfo {author} {\bibfnamefont {S.-X.}\ \bibnamefont {Zhang}}, \bibinfo
  {author} {\bibfnamefont {X.}~\bibnamefont {Yi}}, \bibinfo {author}
  {\bibfnamefont {J.}~\bibnamefont {Wiersig}}, \bibinfo {author} {\bibfnamefont
  {L.}~\bibnamefont {Wang}}, \bibinfo {author} {\bibfnamefont {Q.}~\bibnamefont
  {Gong}}, \bibinfo {author} {\bibfnamefont {M.}~\bibnamefont {Lončar}},
  \bibinfo {author} {\bibfnamefont {L.}~\bibnamefont {Yang}},\ and\ \bibinfo
  {author} {\bibfnamefont {Y.-F.}\ \bibnamefont {Xiao}},\ }\bibfield  {title}
  {{\bibinfo {title} {Chaos-assisted broadband momentum
  transformation in optical microresonators}},\ }\href
  {https://doi.org/10.1126/science.aao0763} {\bibfield  {journal} {\bibinfo
  {journal} {Science}\ }\textbf {\bibinfo {volume} {358}},\ \bibinfo {pages}
  {344} (\bibinfo {year} {2017})},\ \bibinfo {note} {publisher: American
  Association for the Advancement of Science Section: Report}\BibitemShut
  {NoStop}%
\bibitem [{\citenamefont {Mitchell}\ \emph {et~al.}(2016)\citenamefont
  {Mitchell}, \citenamefont {Chung},\ and\ \citenamefont
  {Teal}}]{mitchell_photoluminescence_2016}%
  \BibitemOpen
  \bibfield  {author} {\bibinfo {author} {\bibfnamefont {B.}~\bibnamefont
  {Mitchell}}, \bibinfo {author} {\bibfnamefont {D.}~\bibnamefont {Chung}},\
  and\ \bibinfo {author} {\bibfnamefont {A.}~\bibnamefont {Teal}},\ }\bibfield
  {title} {\bibinfo {title} {Photoluminescence {Imaging} {Using} {Silicon}
  {Line}-{Scanning} {Cameras}},\ }\href
  {https://doi.org/10.1109/JPHOTOV.2016.2557064} {\bibfield  {journal}
  {\bibinfo  {journal} {IEEE Journal of Photovoltaics}\ }\textbf {\bibinfo
  {volume} {6}},\ \bibinfo {pages} {967} (\bibinfo {year} {2016})},\ \bibinfo
  {note} {conference Name: IEEE Journal of Photovoltaics}\BibitemShut {NoStop}%
\bibitem [{\citenamefont {Sun}\ \emph {et~al.}(2015)\citenamefont {Sun},
  \citenamefont {Melnikov},\ and\ \citenamefont
  {Mandelis}}]{sun_camera-based_2015}%
  \BibitemOpen
  \bibfield  {author} {\bibinfo {author} {\bibfnamefont {Q.~M.}\ \bibnamefont
  {Sun}}, \bibinfo {author} {\bibfnamefont {A.}~\bibnamefont {Melnikov}},\ and\
  \bibinfo {author} {\bibfnamefont {A.}~\bibnamefont {Mandelis}},\ }\bibfield
  {title} {{\bibinfo {title} {Camera-{Based} {Lock}-in and
  {Heterodyne} {Carrierographic} {Photoluminescence} {Imaging} of {Crystalline}
  {Silicon} {Wafers}}},\ }\href {https://doi.org/10.1007/s10765-014-1599-z}
  {\bibfield  {journal} {\bibinfo  {journal} {International Journal of
  Thermophysics}\ }\textbf {\bibinfo {volume} {36}},\ \bibinfo {pages} {1274}
  (\bibinfo {year} {2015})}\BibitemShut {NoStop}%
\bibitem [{\citenamefont {Sun}\ \emph {et~al.}(2016)\citenamefont {Sun},
  \citenamefont {Melnikov},\ and\ \citenamefont
  {Mandelis}}]{sun_camera-based_2016}%
  \BibitemOpen
  \bibfield  {author} {\bibinfo {author} {\bibfnamefont {Q.}~\bibnamefont
  {Sun}}, \bibinfo {author} {\bibfnamefont {A.}~\bibnamefont {Melnikov}},\ and\
  \bibinfo {author} {\bibfnamefont {A.}~\bibnamefont {Mandelis}},\ }\bibfield
  {title} {{\bibinfo {title} {Camera-based high frequency
  heterodyne lock-in carrierographic (frequency-domain photoluminescence)
  imaging of crystalline silicon wafers}},\ }\href
  {https://doi.org/https://doi.org/10.1002/pssa.201532033} {\bibfield
  {journal} {\bibinfo  {journal} {physica status solidi (a)}\ }\textbf
  {\bibinfo {volume} {213}},\ \bibinfo {pages} {405} (\bibinfo {year}
  {2016})}\BibitemShut {NoStop}%
\bibitem [{\citenamefont {Kasuya}\ and\ \citenamefont
  {Suezawa}(1997)}]{kasuya_resonant_1997}%
  \BibitemOpen
  \bibfield  {author} {\bibinfo {author} {\bibfnamefont {A.}~\bibnamefont
  {Kasuya}}\ and\ \bibinfo {author} {\bibfnamefont {M.}~\bibnamefont
  {Suezawa}},\ }\bibfield  {title} {\bibinfo {title} {Resonant excitation of
  visible photoluminescence from an erbium-oxide overlayer on {Si}},\ }\href
  {https://doi.org/10.1063/1.120119} {\bibfield  {journal} {\bibinfo  {journal}
  {Applied Physics Letters}\ }\textbf {\bibinfo {volume} {71}},\ \bibinfo
  {pages} {2728} (\bibinfo {year} {1997})},\ \bibinfo {note} {publisher:
  American Institute of Physics}\BibitemShut {NoStop}%
\bibitem [{\citenamefont {Doke}\ \emph {et~al.}(2013)\citenamefont {Doke},
  \citenamefont {Sarakovskis}, \citenamefont {Grube},\ and\ \citenamefont
  {Springis}}]{doke_photoluminescence_2013}%
  \BibitemOpen
  \bibfield  {author} {\bibinfo {author} {\bibfnamefont {G.}~\bibnamefont
  {Doke}}, \bibinfo {author} {\bibfnamefont {A.}~\bibnamefont {Sarakovskis}},
  \bibinfo {author} {\bibfnamefont {J.}~\bibnamefont {Grube}},\ and\ \bibinfo
  {author} {\bibfnamefont {M.}~\bibnamefont {Springis}},\ }\bibfield  {title}
  {{\bibinfo {title} {Photoluminescence of neodymium and
  erbium doped {NaLaF4} material}},\ }\href
  {https://doi.org/10.1016/j.radmeas.2013.03.009} {\bibfield  {journal}
  {\bibinfo  {journal} {Radiation Measurements}\ }\bibinfo {series}
  {Proceedings of the 8th {International} {Conference} on {Luminescent}
  {Detectors} and {Transformers} of {Ionizing} {Radiation} ({LUMDETR} 2012)},\
  \textbf {\bibinfo {volume} {56}},\ \bibinfo {pages} {27} (\bibinfo {year}
  {2013})}\BibitemShut {NoStop}%
\bibitem [{\citenamefont {Collins}(2007)}]{collins_imagej_2007}%
  \BibitemOpen
  \bibfield  {author} {\bibinfo {author} {\bibfnamefont {T.~J.}\ \bibnamefont
  {Collins}},\ }\bibfield  {title} {\bibinfo {title} {{ImageJ} for
  microscopy},\ }\href {https://doi.org/10.2144/000112517} {\bibfield
  {journal} {\bibinfo  {journal} {BioTechniques}\ }\textbf {\bibinfo {volume}
  {43}},\ \bibinfo {pages} {S25} (\bibinfo {year} {2007})}\BibitemShut
  {NoStop}%
\bibitem [{\citenamefont {van~den Hoven}\ \emph {et~al.}(1996)\citenamefont
  {van~den Hoven}, \citenamefont {Snoeks}, \citenamefont {Polman},
  \citenamefont {van Dam}, \citenamefont {van Uffelen},\ and\ \citenamefont
  {Smit}}]{van_den_hoven_upconversion_1996}%
  \BibitemOpen
  \bibfield  {author} {\bibinfo {author} {\bibfnamefont {G.~N.}\ \bibnamefont
  {van~den Hoven}}, \bibinfo {author} {\bibfnamefont {E.}~\bibnamefont
  {Snoeks}}, \bibinfo {author} {\bibfnamefont {A.}~\bibnamefont {Polman}},
  \bibinfo {author} {\bibfnamefont {C.}~\bibnamefont {van Dam}}, \bibinfo
  {author} {\bibfnamefont {J.~W.~M.}\ \bibnamefont {van Uffelen}},\ and\
  \bibinfo {author} {\bibfnamefont {M.~K.}\ \bibnamefont {Smit}},\ }\bibfield
  {title} {{\bibinfo {title} {Upconversion in
  {Er}‐implanted {Al} $_{\textrm{2}}${O}$_{\textrm{3}}$ waveguides}},\ }\href
  {https://doi.org/10.1063/1.361020} {\bibfield  {journal} {\bibinfo  {journal}
  {Journal of Applied Physics}\ }\textbf {\bibinfo {volume} {79}},\ \bibinfo
  {pages} {1258} (\bibinfo {year} {1996})}\BibitemShut {NoStop}%
\bibitem [{\citenamefont {Pollnau}\ \emph {et~al.}(1992)\citenamefont
  {Pollnau}, \citenamefont {Heumann},\ and\ \citenamefont
  {Huber}}]{pollnau_time-resolved_1992}%
  \BibitemOpen
  \bibfield  {author} {\bibinfo {author} {\bibfnamefont {M.}~\bibnamefont
  {Pollnau}}, \bibinfo {author} {\bibfnamefont {E.}~\bibnamefont {Heumann}},\
  and\ \bibinfo {author} {\bibfnamefont {G.}~\bibnamefont {Huber}},\ }\bibfield
   {title} {{\bibinfo {title} {Time-resolved spectra of
  excited-state absorption in {Er}$^{3+}$ doped {YAlO}$_3$}},\ }\href
  {https://doi.org/10.1007/BF00324164} {\bibfield  {journal} {\bibinfo
  {journal} {Applied Physics A}\ }\textbf {\bibinfo {volume} {54}},\ \bibinfo
  {pages} {404} (\bibinfo {year} {1992})}\BibitemShut {NoStop}%
\bibitem [{\citenamefont {Snoeks}\ \emph {et~al.}(1995)\citenamefont {Snoeks},
  \citenamefont {Hoven}, \citenamefont {Polman}, \citenamefont {Hendriksen},
  \citenamefont {Diemeer},\ and\ \citenamefont
  {Priolo}}]{snoeks_cooperative_1995}%
  \BibitemOpen
  \bibfield  {author} {\bibinfo {author} {\bibfnamefont {E.}~\bibnamefont
  {Snoeks}}, \bibinfo {author} {\bibfnamefont {G.~N. v.~d.}\ \bibnamefont
  {Hoven}}, \bibinfo {author} {\bibfnamefont {A.}~\bibnamefont {Polman}},
  \bibinfo {author} {\bibfnamefont {B.}~\bibnamefont {Hendriksen}}, \bibinfo
  {author} {\bibfnamefont {M.~B.~J.}\ \bibnamefont {Diemeer}},\ and\ \bibinfo
  {author} {\bibfnamefont {F.}~\bibnamefont {Priolo}},\ }\bibfield  {title}
  {{\bibinfo {title} {Cooperative upconversion in
  erbium-implanted soda-lime silicate glass optical waveguides}},\ }\href
  {https://doi.org/10.1364/JOSAB.12.001468} {\bibfield  {journal} {\bibinfo
  {journal} {Journal of the Optical Society of America B}\ }\textbf {\bibinfo
  {volume} {12}},\ \bibinfo {pages} {1468} (\bibinfo {year}
  {1995})}\BibitemShut {NoStop}%
\bibitem [{\citenamefont {Johnson}\ \emph {et~al.}(1975)\citenamefont
  {Johnson}, \citenamefont {Ku}, \citenamefont {Lillja},\ and\ \citenamefont
  {Shih-To}}]{johnson_method_1975}%
  \BibitemOpen
  \bibfield  {author} {\bibinfo {author} {\bibfnamefont {J.~C.}\ \bibnamefont
  {Johnson}}, \bibinfo {author} {\bibfnamefont {S.-M.}\ \bibnamefont {Ku}},
  \bibinfo {author} {\bibfnamefont {H.~V.}\ \bibnamefont {Lillja}},\ and\
  \bibinfo {author} {\bibfnamefont {P.~E.}\ \bibnamefont {Shih-To}},\ }\href
  {https://patents.google.com/patent/US3920483A/en} {\bibinfo {title} {Method
  of ion implantation through a photoresist mask}} (\bibinfo {year} {1975}),\
  \bibinfo {note} {library Catalog: Google Patents}\BibitemShut {NoStop}%
\bibitem [{\citenamefont {Kik}\ and\ \citenamefont
  {Polman}(2003)}]{kik_cooperative_2003}%
  \BibitemOpen
  \bibfield  {author} {\bibinfo {author} {\bibfnamefont {P.~G.}\ \bibnamefont
  {Kik}}\ and\ \bibinfo {author} {\bibfnamefont {A.}~\bibnamefont {Polman}},\
  }\bibfield  {title} {{\bibinfo {title} {Cooperative
  upconversion as the gain-limiting factor in {Er} doped miniature {Al2O3}
  optical waveguide amplifiers}},\ }\href {https://doi.org/10.1063/1.1565697}
  {\bibfield  {journal} {\bibinfo  {journal} {Journal of Applied Physics}\
  }\textbf {\bibinfo {volume} {93}},\ \bibinfo {pages} {5008} (\bibinfo {year}
  {2003})}\BibitemShut {NoStop}%
\bibitem [{\citenamefont {Baker}\ \emph {et~al.}(2012)\citenamefont {Baker},
  \citenamefont {Stapfner}, \citenamefont {Parrain}, \citenamefont {Ducci},
  \citenamefont {Leo}, \citenamefont {Weig},\ and\ \citenamefont
  {Favero}}]{baker_optical_2012}%
  \BibitemOpen
  \bibfield  {author} {\bibinfo {author} {\bibfnamefont {C.}~\bibnamefont
  {Baker}}, \bibinfo {author} {\bibfnamefont {S.}~\bibnamefont {Stapfner}},
  \bibinfo {author} {\bibfnamefont {D.}~\bibnamefont {Parrain}}, \bibinfo
  {author} {\bibfnamefont {S.}~\bibnamefont {Ducci}}, \bibinfo {author}
  {\bibfnamefont {G.}~\bibnamefont {Leo}}, \bibinfo {author} {\bibfnamefont
  {E.~M.}\ \bibnamefont {Weig}},\ and\ \bibinfo {author} {\bibfnamefont
  {I.}~\bibnamefont {Favero}},\ }\bibfield  {title} {{\bibinfo {title} {Optical instability and self-pulsing in silicon nitride
  whispering gallery resonators}},\ }\href
  {https://doi.org/10.1364/OE.20.029076} {\bibfield  {journal} {\bibinfo
  {journal} {Optics Express}\ }\textbf {\bibinfo {volume} {20}},\ \bibinfo
  {pages} {29076} (\bibinfo {year} {2012})}\BibitemShut {NoStop}%
\bibitem [{\citenamefont {Johnson}\ \emph {et~al.}(2006)\citenamefont
  {Johnson}, \citenamefont {Borselli},\ and\ \citenamefont
  {Painter}}]{johnson_self-induced_2006}%
  \BibitemOpen
  \bibfield  {author} {\bibinfo {author} {\bibfnamefont {T.~J.}\ \bibnamefont
  {Johnson}}, \bibinfo {author} {\bibfnamefont {M.}~\bibnamefont {Borselli}},\
  and\ \bibinfo {author} {\bibfnamefont {O.}~\bibnamefont {Painter}},\
  }\bibfield  {title} {\bibinfo {title} {Self-induced optical modulation of the
  transmission through a high-{Q} silicon microdisk resonator},\ }\href
  {https://doi.org/10.1364/OPEX.14.000817} {\bibfield  {journal} {\bibinfo
  {journal} {Opt. Express}\ }\textbf {\bibinfo {volume} {14}},\ \bibinfo
  {pages} {817} (\bibinfo {year} {2006})}\BibitemShut {NoStop}%
\bibitem [{\citenamefont {Diallo}\ \emph {et~al.}(2015)\citenamefont {Diallo},
  \citenamefont {Lin},\ and\ \citenamefont {Chembo}}]{diallo_giant_2015}%
  \BibitemOpen
  \bibfield  {author} {\bibinfo {author} {\bibfnamefont {S.}~\bibnamefont
  {Diallo}}, \bibinfo {author} {\bibfnamefont {G.}~\bibnamefont {Lin}},\ and\
  \bibinfo {author} {\bibfnamefont {Y.~K.}\ \bibnamefont {Chembo}},\ }\bibfield
   {title} {{\bibinfo {title} {Giant thermo-optical
  relaxation oscillations in millimeter-size whispering gallery mode disk
  resonators}},\ }\href {https://doi.org/10.1364/OL.40.003834} {\bibfield
  {journal} {\bibinfo  {journal} {Optics Letters}\ }\textbf {\bibinfo {volume}
  {40}},\ \bibinfo {pages} {3834} (\bibinfo {year} {2015})}\BibitemShut
  {NoStop}%
\bibitem [{\citenamefont {Rowley}\ \emph {et~al.}(2019)\citenamefont {Rowley},
  \citenamefont {Wetzel}, \citenamefont {Lauro}, \citenamefont {Gongora},
  \citenamefont {Bao}, \citenamefont {Silver}, \citenamefont {Bino},
  \citenamefont {Haye}, \citenamefont {Peccianti},\ and\ \citenamefont
  {Pasquazi}}]{rowley_thermo-optical_2019}%
  \BibitemOpen
  \bibfield  {author} {\bibinfo {author} {\bibfnamefont {M.}~\bibnamefont
  {Rowley}}, \bibinfo {author} {\bibfnamefont {B.}~\bibnamefont {Wetzel}},
  \bibinfo {author} {\bibfnamefont {L.~D.}\ \bibnamefont {Lauro}}, \bibinfo
  {author} {\bibfnamefont {J.~S.~T.}\ \bibnamefont {Gongora}}, \bibinfo
  {author} {\bibfnamefont {H.}~\bibnamefont {Bao}}, \bibinfo {author}
  {\bibfnamefont {J.}~\bibnamefont {Silver}}, \bibinfo {author} {\bibfnamefont
  {L.~D.}\ \bibnamefont {Bino}}, \bibinfo {author} {\bibfnamefont {P.~D.}\
  \bibnamefont {Haye}}, \bibinfo {author} {\bibfnamefont {M.}~\bibnamefont
  {Peccianti}},\ and\ \bibinfo {author} {\bibfnamefont {A.}~\bibnamefont
  {Pasquazi}},\ }\bibfield  {title} {{\bibinfo {title}
  {Thermo-optical pulsing in a microresonator filtered fiber-laser: a route
  towards all-optical control and synchronization}},\ }\href
  {https://doi.org/10.1364/OE.27.019242} {\bibfield  {journal} {\bibinfo
  {journal} {Optics Express}\ }\textbf {\bibinfo {volume} {27}},\ \bibinfo
  {pages} {19242} (\bibinfo {year} {2019})}\BibitemShut {NoStop}%
\bibitem [{\citenamefont {Park}\ and\ \citenamefont
  {Wang}(2007)}]{park_regenerative_2007}%
  \BibitemOpen
  \bibfield  {author} {\bibinfo {author} {\bibfnamefont {Y.-S.}\ \bibnamefont
  {Park}}\ and\ \bibinfo {author} {\bibfnamefont {H.}~\bibnamefont {Wang}},\
  }\bibfield  {title} {{\bibinfo {title} {Regenerative
  pulsation in silica microspheres}},\ }\href
  {https://doi.org/10.1364/OL.32.003104} {\bibfield  {journal} {\bibinfo
  {journal} {Optics Letters}\ }\textbf {\bibinfo {volume} {32}},\ \bibinfo
  {pages} {3104} (\bibinfo {year} {2007})}\BibitemShut {NoStop}%
\bibitem [{\citenamefont {Deng}\ \emph {et~al.}(2013)\citenamefont {Deng},
  \citenamefont {Liu}, \citenamefont {Leseman},\ and\ \citenamefont
  {Hossein-Zadeh}}]{deng_thermo-optomechanical_2013}%
  \BibitemOpen
  \bibfield  {author} {\bibinfo {author} {\bibfnamefont {Y.}~\bibnamefont
  {Deng}}, \bibinfo {author} {\bibfnamefont {F.}~\bibnamefont {Liu}}, \bibinfo
  {author} {\bibfnamefont {Z.~C.}\ \bibnamefont {Leseman}},\ and\ \bibinfo
  {author} {\bibfnamefont {M.}~\bibnamefont {Hossein-Zadeh}},\ }\bibfield
  {title} {{\bibinfo {title} {Thermo-optomechanical
  oscillator for sensing applications}},\ }\href
  {https://doi.org/10.1364/OE.21.004653} {\bibfield  {journal} {\bibinfo
  {journal} {Optics Express}\ }\textbf {\bibinfo {volume} {21}},\ \bibinfo
  {pages} {4653} (\bibinfo {year} {2013})}\BibitemShut {NoStop}%
\bibitem [{\citenamefont {Bekker}\ \emph {et~al.}(2017)\citenamefont {Bekker},
  \citenamefont {Kalra}, \citenamefont {Baker},\ and\ \citenamefont
  {Bowen}}]{bekker_injection_2017}%
  \BibitemOpen
  \bibfield  {author} {\bibinfo {author} {\bibfnamefont {C.}~\bibnamefont
  {Bekker}}, \bibinfo {author} {\bibfnamefont {R.}~\bibnamefont {Kalra}},
  \bibinfo {author} {\bibfnamefont {C.}~\bibnamefont {Baker}},\ and\ \bibinfo
  {author} {\bibfnamefont {W.~P.}\ \bibnamefont {Bowen}},\ }\bibfield  {title}
  {{\bibinfo {title} {Injection locking of an
  electro-optomechanical device}},\ }\href
  {https://doi.org/10.1364/OPTICA.4.001196} {\bibfield  {journal} {\bibinfo
  {journal} {Optica}\ }\textbf {\bibinfo {volume} {4}},\ \bibinfo {pages}
  {1196} (\bibinfo {year} {2017})}\BibitemShut {NoStop}%
\bibitem [{\citenamefont {Almeida}\ and\ \citenamefont
  {Lipson}(2004)}]{almeida_optical_2004}%
  \BibitemOpen
  \bibfield  {author} {\bibinfo {author} {\bibfnamefont {V.~R.}\ \bibnamefont
  {Almeida}}\ and\ \bibinfo {author} {\bibfnamefont {M.}~\bibnamefont
  {Lipson}},\ }\bibfield  {title} {{\bibinfo {title}
  {Optical bistability on a silicon chip}},\ }\href
  {https://doi.org/10.1364/OL.29.002387} {\bibfield  {journal} {\bibinfo
  {journal} {Optics Letters}\ }\textbf {\bibinfo {volume} {29}},\ \bibinfo
  {pages} {2387} (\bibinfo {year} {2004})}\BibitemShut {NoStop}%
\bibitem [{\citenamefont {Parrain}\ \emph {et~al.}(2015)\citenamefont
  {Parrain}, \citenamefont {Baker}, \citenamefont {Wang}, \citenamefont {Guha},
  \citenamefont {Santos}, \citenamefont {Lemaitre}, \citenamefont {Senellart},
  \citenamefont {Leo}, \citenamefont {Ducci},\ and\ \citenamefont
  {Favero}}]{parrain_origin_2015}%
  \BibitemOpen
  \bibfield  {author} {\bibinfo {author} {\bibfnamefont {D.}~\bibnamefont
  {Parrain}}, \bibinfo {author} {\bibfnamefont {C.}~\bibnamefont {Baker}},
  \bibinfo {author} {\bibfnamefont {G.}~\bibnamefont {Wang}}, \bibinfo {author}
  {\bibfnamefont {B.}~\bibnamefont {Guha}}, \bibinfo {author} {\bibfnamefont
  {E.~G.}\ \bibnamefont {Santos}}, \bibinfo {author} {\bibfnamefont
  {A.}~\bibnamefont {Lemaitre}}, \bibinfo {author} {\bibfnamefont
  {P.}~\bibnamefont {Senellart}}, \bibinfo {author} {\bibfnamefont
  {G.}~\bibnamefont {Leo}}, \bibinfo {author} {\bibfnamefont {S.}~\bibnamefont
  {Ducci}},\ and\ \bibinfo {author} {\bibfnamefont {I.}~\bibnamefont
  {Favero}},\ }\bibfield  {title} {{\bibinfo {title}
  {Origin of optical losses in gallium arsenide disk whispering gallery
  resonators}},\ }\href {https://doi.org/10.1364/OE.23.019656} {\bibfield
  {journal} {\bibinfo  {journal} {Optics Express}\ }\textbf {\bibinfo {volume}
  {23}},\ \bibinfo {pages} {19656} (\bibinfo {year} {2015})}\BibitemShut
  {NoStop}%
\bibitem [{\citenamefont {Vernooy}\ \emph {et~al.}(1998)\citenamefont
  {Vernooy}, \citenamefont {Ilchenko}, \citenamefont {Mabuchi}, \citenamefont
  {Streed},\ and\ \citenamefont {Kimble}}]{vernooy_high-q_1998}%
  \BibitemOpen
  \bibfield  {author} {\bibinfo {author} {\bibfnamefont {D.~W.}\ \bibnamefont
  {Vernooy}}, \bibinfo {author} {\bibfnamefont {V.~S.}\ \bibnamefont
  {Ilchenko}}, \bibinfo {author} {\bibfnamefont {H.}~\bibnamefont {Mabuchi}},
  \bibinfo {author} {\bibfnamefont {E.~W.}\ \bibnamefont {Streed}},\ and\
  \bibinfo {author} {\bibfnamefont {H.~J.}\ \bibnamefont {Kimble}},\ }\bibfield
   {title} {{\bibinfo {title} {High-{Q} measurements of
  fused-silica microspheres in the near infrared}},\ }\href
  {https://doi.org/10.1364/OL.23.000247} {\bibfield  {journal} {\bibinfo
  {journal} {Optics Letters}\ }\textbf {\bibinfo {volume} {23}},\ \bibinfo
  {pages} {247} (\bibinfo {year} {1998})}\BibitemShut {NoStop}%
\bibitem [{\citenamefont {Gao}\ \emph {et~al.}(2018)\citenamefont {Gao},
  \citenamefont {Jiang}, \citenamefont {Cui}, \citenamefont {Zhang},
  \citenamefont {Jia},\ and\ \citenamefont {Jiang}}]{gao_investigation_2018}%
  \BibitemOpen
  \bibfield  {author} {\bibinfo {author} {\bibfnamefont {H.}~\bibnamefont
  {Gao}}, \bibinfo {author} {\bibfnamefont {Y.}~\bibnamefont {Jiang}}, \bibinfo
  {author} {\bibfnamefont {Y.}~\bibnamefont {Cui}}, \bibinfo {author}
  {\bibfnamefont {L.}~\bibnamefont {Zhang}}, \bibinfo {author} {\bibfnamefont
  {J.}~\bibnamefont {Jia}},\ and\ \bibinfo {author} {\bibfnamefont
  {L.}~\bibnamefont {Jiang}},\ }\bibfield  {title} {\bibinfo {title}
  {Investigation on the {Thermo}-{Optic} {Coefficient} of {Silica} {Fiber}
  {Within} a {Wide} {Temperature} {Range}},\ }\href
  {https://doi.org/10.1109/JLT.2018.2875941} {\bibfield  {journal} {\bibinfo
  {journal} {Journal of Lightwave Technology}\ }\textbf {\bibinfo {volume}
  {36}},\ \bibinfo {pages} {5881} (\bibinfo {year} {2018})}\BibitemShut
  {NoStop}%
\bibitem [{Note1()}]{Note1}%
  \BibitemOpen
  \bibinfo {note} {This is approximately linear for small changes in refractive
  index and is simulated using the correct parameters for our geometry, not
  taking into account stress-induced warping.}\BibitemShut {Stop}%
\bibitem [{noa()}]{noauthor_comsol_nodate}%
  \BibitemOpen
  \href@noop {} {\ }\bibinfo {note} {Comsol {Material} {Library}}\BibitemShut
  {NoStop}%
\bibitem [{\citenamefont {Aspelmeyer}\ \emph {et~al.}(2014)\citenamefont
  {Aspelmeyer}, \citenamefont {Kippenberg},\ and\ \citenamefont
  {Marquardt}}]{aspelmeyer_cavity_2014}%
  \BibitemOpen
  \bibfield  {author} {\bibinfo {author} {\bibfnamefont {M.}~\bibnamefont
  {Aspelmeyer}}, \bibinfo {author} {\bibfnamefont {T.~J.}\ \bibnamefont
  {Kippenberg}},\ and\ \bibinfo {author} {\bibfnamefont {F.}~\bibnamefont
  {Marquardt}},\ }\bibfield  {title} {\bibinfo {title} {Cavity optomechanics},\
  }\href {https://doi.org/10.1103/RevModPhys.86.1391} {\bibfield  {journal}
  {\bibinfo  {journal} {Reviews of Modern Physics}\ }\textbf {\bibinfo {volume}
  {86}},\ \bibinfo {pages} {1391} (\bibinfo {year} {2014})}\BibitemShut
  {NoStop}%
\bibitem [{\citenamefont {Baker}\ \emph {et~al.}(2014)\citenamefont {Baker},
  \citenamefont {Hease}, \citenamefont {Nguyen}, \citenamefont {Andronico},
  \citenamefont {Ducci}, \citenamefont {Leo},\ and\ \citenamefont
  {Favero}}]{baker_photoelastic_2014}%
  \BibitemOpen
  \bibfield  {author} {\bibinfo {author} {\bibfnamefont {C.}~\bibnamefont
  {Baker}}, \bibinfo {author} {\bibfnamefont {W.}~\bibnamefont {Hease}},
  \bibinfo {author} {\bibfnamefont {D.-T.}\ \bibnamefont {Nguyen}}, \bibinfo
  {author} {\bibfnamefont {A.}~\bibnamefont {Andronico}}, \bibinfo {author}
  {\bibfnamefont {S.}~\bibnamefont {Ducci}}, \bibinfo {author} {\bibfnamefont
  {G.}~\bibnamefont {Leo}},\ and\ \bibinfo {author} {\bibfnamefont
  {I.}~\bibnamefont {Favero}},\ }\bibfield  {title} {{\bibinfo {title} {Photoelastic coupling in gallium arsenide
  optomechanical disk resonators}},\ }\href
  {https://doi.org/10.1364/OE.22.014072} {\bibfield  {journal} {\bibinfo
  {journal} {Optics Express}\ }\textbf {\bibinfo {volume} {22}},\ \bibinfo
  {pages} {14072} (\bibinfo {year} {2014})}\BibitemShut {NoStop}%
\bibitem [{supplement()}]{supplement}%
  \BibitemOpen
  \bibinfo {note} {See {Supplemental Material} at [URL will be
  inserted by publisher] for the {MATLAB} script used to model self-pulsing behavior arising from radiation pressure and the thermo-optic and thermomechanical effects. In addition to the model as the main script, a secondary script providing the equations of motion of the system is supplied.}\BibitemShut {Stop}%
\end{thebibliography}
\end{document}